\documentclass[floats,floatfix,
amssymb,prd,onecolumn,notitlepage,
superscriptaddress,nofootinbib]{revtex4-1}

\usepackage{amssymb,amsmath,verbatim,mathtools,needspace,enumitem,etoolbox,graphicx,physics,microtype,afterpage,bm}
\usepackage[dvipsnames, usenames]{xcolor}
\definecolor{linkcolor}{rgb}{0.0,0.3,0.5}
\usepackage{booktabs}

\usepackage[T1]{fontenc}
\usepackage[utf8]{inputenc}
\usepackage{tabularx}
\usepackage{adjustbox}
\usepackage{float}
\usepackage{ulem}
\usepackage{xfrac}

\usepackage{slashed}
\usepackage{bm}
\usepackage{latexsym,amssymb,amsmath,float,url,mathrsfs}
\usepackage{latexsym}
\usepackage{graphicx}
\usepackage{epstopdf}
\usepackage{amsmath}
\usepackage{comment}
\usepackage{subfigure}
\usepackage{natbib}
\usepackage{hyperref}
\usepackage{pifont}
\usepackage{blindtext}
\usepackage{adjustbox}
\usepackage{multirow}
\usepackage{tabularx}
\usepackage{orcidlink}
\usepackage{tikz}
\usetikzlibrary{tikzmark}
\usepackage{fontawesome}

\usepackage{graphics,appendix,afterpage,makecell} 

\definecolor{verdes}{rgb}{0.1, 0.5, 0.1}%
\definecolor{cornellred}{rgb}{0.7, 0.11, 0.11}

\newcommand{\nn}{\nonumber}
\newcommand{\fd}[2]{\parbox{#1}{\includegraphics[width=#1]{#2}}}

\definecolor{VioletRed4}{rgb}{0.55, 0.13, .32}
\definecolor{cerisepink}{rgb}{0.93, 0.23, 0.51}
\definecolor{azure}{rgb}{0.0, 0.5, 1.0}

\hypersetup{
     colorlinks   = true,
     citecolor    = violet,
     urlcolor     = violet,
     linkcolor    = violet}

\definecolor{ForestGreen}{rgb}{0.13, 0.55, 0.13}

\begin{document}

\title{
Primordial black hole dark matter from inflation: the reverse engineering approach
% Observational signatures of ultra slow-roll: The reverse engineering approach
}

\author{Gabriele Franciolini\orcidlink{0000-0002-6892-9145}}
\thanks{{\scriptsize Email}: \href{mailto:gabriele.franciolini@uniroma1.it}{gabriele.franciolini@uniroma1.it}.}
\affiliation{Dipartimento di Fisica, ``Sapienza'' Universit\`a di Roma, Piazzale Aldo Moro 5, 00185, Roma, Italy}
\affiliation{INFN sezione di Roma, Piazzale Aldo Moro 5, 00185, Roma, Italy}

\author{Alfredo Urbano\orcidlink{0000-0002-0488-3256}}
\thanks{{\scriptsize Email}: \href{mailto:alfredo.urbano@uniroma1.it}{alfredo.urbano@uniroma1.it}.}
\affiliation{Dipartimento di Fisica, ``Sapienza'' Universit\`a di Roma, Piazzale Aldo Moro 5, 00185, Roma, Italy}
\affiliation{INFN sezione di Roma, Piazzale Aldo Moro 5, 00185, Roma, Italy}

\date{\today}

\begin{abstract}\noindent
Constraining the inflationary epoch is one of the aims of modern cosmology. 
In order to fully exploit current and future small-scale observations, 
it is necessary to devise tools to directly relate them to the early universes dynamics.
We present here a novel reverse engineer approach able to connect 
fundamental late-time observables to consistent inflationary dynamics and, 
eventually, to the inflaton potential. 
Employing this procedure, 
we are able to describe which conditions 
can give rise to a raised plateau in the power spectrum of curvature perturbations
at small scales, which are not constrained by CMB observations. 
Within this new phenomenologically-driven approach, 
we find that inflation can generate a raised plateau in the spectrum of curvature perturbations
that potentially connects 
three fundamental observables:
a dominant component of the dark matter in the form of 
asteroid-mass/atomic-size primordial black holes; 
detectable signals in stochastic gravitational waves and 
a subdominant fraction of stellar-mass primordial black holes mergers. 
\end{abstract}

\maketitle

{
  \hypersetup{linkcolor=black}
  \tableofcontents
}

%%%%%%%%%%%%%%%%%%%%%%%%%%%%%%%%%%%%%%%%%%%%%%%%
\noindent
\section{Introduction}
%%%%%%%%%%%%%%%%%%%%%%%%%%%%%%%%%%%%%%%%%%%%%%%%

{The increasing accuracy of observational data relating to measurements of the cosmic microwave background (CMB) anisotropy placed severe constraints on cosmic inflation. 
In the framework of standard single-field inflationary models with Einstein gravity, the latest results reported by the Planck and BICEP2/Keck collaborations \cite{Planck:2018jri,BICEP:2021xfz} imply that the predictions of slow-roll models with a concave potential are strongly  favoured by data and no evidence for dynamics beyond slow-roll was found. On the theoretical side, these constraints  have far-reaching implications. For instance, a simple inspection of the theoretical predictions regarding  the power spectra of scalar and tensor perturbations 
%spectral index of scalar perturbations %and the  tensor-to-scalar ratio 
leads to the conclusion that the standard version of natural inflation and the full class of monomial potentials $V(\phi) \sim \phi^n$ are now strongly disfavored\,\cite{Kallosh:2021mnu}.

However, it is important to keep in mind that the above discussion is limited only to a relatively short part of the inflationary dynamics, namely the one that took place at around 60 $e$-folds before the end of inflation when curvature perturbations with comoving wavenumber  in the range 
$0.005 \lesssim k\,[\textrm{Mpc}^{-1}] \lesssim 0.2$ exited the Hubble horizon. 
On smaller scales (larger $k$), the observational constraints are far weaker implying that deviations from the slow-roll paradigm are possible; consequently, claiming any theoretical control over the inflationary potential is, at these scales,  way more difficult. 
\begin{figure}[h]
\leavevmode
\centering
\includegraphics[width=.95\textwidth]{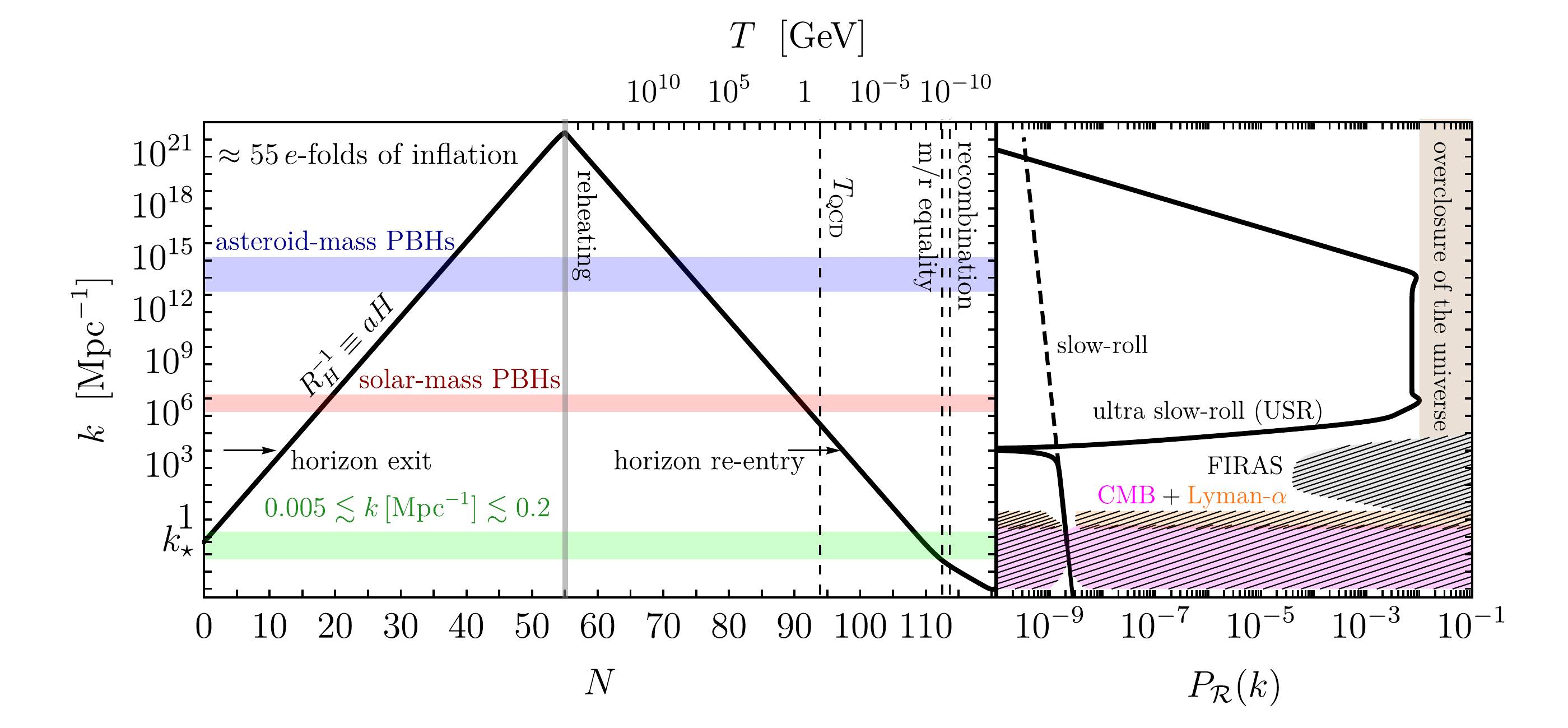}  
\caption{
Left part.
Time evolution (in terms of the number of $e$-folds $N$ defined by $dN = H dt$) of the inverse comoving Hubble horizon $R_H^{-1} \equiv aH$ throughout the history of our observable universe. We start from $N =0$, defined as the time at  which the CMB pivot scale $k_{\star} = 0.05$ Mpc$^{-1}$ crossed the Hubble horizon, $k_{\star} = a(0)H(0)$. 
We assume instantaneous reheating and, after inflation, standard $\Lambda$CDM cosmology. The three horizontal bands mark the milestones of our phenomenological analysis. 
The region shaded in green ($0.005 \lesssim k\,[\textrm{Mpc}^{-1}] \lesssim 0.2$) represents the range of comoving wavenumbers (horizontal lines with constant $k$ in the figure) constrained by CMB anisotropy measurements; the region shaded in red 
($1.7\times 10^6 \lesssim k\,[\textrm{Mpc}^{-1}] \lesssim 1.7\times 10^7$) corresponds to the range of comoving wavenumbers for which curvature perturbations, after re-entering the cosmological horizon, have the chance of generating solar-mass PBHs  ($1\lesssim M_{\rm PBH}\,[M_{\odot}] \lesssim  100$); the region shaded in blue 
($1.5\times 10^{13} \lesssim k\,[\textrm{Mpc}^{-1}] \lesssim 1.5\times 10^{14}$) is the same as the red band but corresponds to asteroid-mass PBHs ($10^{-16}\lesssim M_{\rm PBH}\,[M_{\odot}] \lesssim  10^{-12}$). 
Right part. We plot the power spectrum of scalar perturbations $P_{\mathcal{R}}(k)$ as a function of the comoving wavenumber $k$. The plot is rotated in such a way as to share the same $y$-axis with the left part of the figure. 
We plot the region excluded by CMB anisotropy measurements, ref.\,\cite{Planck:2018jri}, the FIRAS bound on CMB spectral distortions, ref.\,\cite{Fixsen:1996nj} and the bound obtained from Lyman-$\alpha$ forest data \cite{Bird:2010mp}. 
If $P_{\mathcal{R}}(k) \gtrsim 10^{-2}$ the abundance of PBHs overcloses the universe (that is, their abundance would be larger than the cold dark matter density of the universe). The  black dashed line is the typical prediction of slow-roll inflationary models. 
The solid black line, on the contrary, is characterized by the presence of an USR phase (concretely, it corresponds to model {\color{verdes}{(1)}} in section\,\ref{sec:SM:background} and ref.\,\cite{Franciolini:2022pav}).
}\label{fig:SummaryPlot}
\end{figure}

Deviations from slow-roll dynamics at small scales may have interesting consequences as far as the formation of primordial black holes (PBHs) is concerned \cite{Zeldovich:1967lct,Hawking:1974rv,Chapline:1975ojl,Carr:1975qj}.  
 In the inflationary picture, space-time fluctuates quantum mechanically around a background that is expanding
exponentially fast; after the end of inflation, these curvature fluctuations are transferred to the radiation field, creating
slightly over- and under-dense regions.  
Regions where the overdensity is large
enough, gravitationally collapse and form PBHs \cite{Ivanov:1994pa,GarciaBellido:1996qt,Ivanov:1997ia,Blinnikov:2016bxu}.
At the practical level, the implementation of this idea requires some mechanism that boosts, at scales relevant for PBH formation, the power spectrum 
of curvature fluctuations $P_{\mathcal{R}}(k)$ way above the value inferred from CMB observations (that is, $P_{\mathcal{R}}(k_{\star}) \approx 2\times 10^{-9}$ with $k_{\star} \equiv 0.05$ Mpc$^{-1}$ the CMB pivot scale) and necessarily breaks the slow-roll paradigm \cite{Motohashi:2017kbs}. 
A popular option is the introduction of an ultra slow-roll (USR) phase during the inflationary dynamics. At the  classical level, during USR the inflaton nearly stops its descent along the potential and remains for a long interval of time with almost zero velocity before re-accelerating towards 
the end of inflation. 
At the quantum level,  during USR comoving curvature perturbations on super-horizon scales are not conserved and are subject to exponential growth due to the presence of a negative friction term in their equation of motion. It is precisely this exponential enhancement that provides the above-mentioned boost in the power spectrum of scalar perturbations. 
The simplest option to get such dynamics is to consider an inflationary potential that features (after the first flattish region that ensures the slow-roll dynamics needed for the fit of CMB measurements) an approximate stationary inflection point. 
}

{The cosmological setup we have in mind is summarized in fig.\,\ref{fig:SummaryPlot} (see caption for details). 
Observational data force the curvature power spectrum (that we plot in the right panel of fig.\,\ref{fig:SummaryPlot}) to have, in the range 
$0.005 \lesssim k\,[\textrm{Mpc}^{-1}] \lesssim 0.2$, a power law functional form  of the type
$P_{\mathcal{R}}(k) = A_s (k/k_{\star})^{n_s -1}$, 
with amplitude $A_s \simeq 2.1 \times 10^{-9}$ and spectral index $n_s \simeq 0.965$, which fits extremely well the typical outcome of slow-roll inflationary models (black dashed line in the right panel of fig.\,\ref{fig:SummaryPlot}). 
However, if we consider larger $k$ an almost uncharted territory opens up, and huge deviations from the slow-roll paradigm are possible. 
The solid black line in the right panel of fig.\,\ref{fig:SummaryPlot} differs from the dashed line because of the presence of an USR phase. 
In this respect, fig.\,\ref{fig:SummaryPlot} summarizes the main objectives of the present work. 
We are interested in curvature power spectra that feature, because of USR, a raised plateau at small scales which are not
constrained by CMB observations. More in detail, we impose three phenomenological requirements.
\begin{itemize}
    \item[{\it i)}] The part of the power spectrum at large scales (that is, for comoving wavenumbers corresponding to the horizontal green band in the left panel of fig.\,\ref{fig:SummaryPlot}) must be consistent with CMB observations. 
    \item[{\it ii)}] The left-side edge of the plateau (that is, at small $k$) corresponds to values of $k$ for which curvature perturbations re-enter the cosmological horizon when the latter has a mass of the order of the solar mass (the horizontal red band in the left panel of fig.\,\ref{fig:SummaryPlot}). 
    This is to generate a sizable abundance of solar-mass PBHs.  
    This is an interesting phenomenological requirement since it implies the possibility that a fraction of merger events directly observed by the LIGO/Virgo/KAGRA collaboration (LVKC) is (or will be) ascribable to stellar-mass PBHs \cite{DeLuca:2021hde,Pujolas:2021yaw}.
    We remark that the red band re-enters the cosmological horizon when the temperature of the universe (labels on the upper $x$-axis) is of the order of the QCD quark-hadron phase transition (in fig.\,\ref{fig:SummaryPlot}  taken to be $T_{\rm QCD} = 0.1$ GeV).

    The right-side edge of the plateau (that is, at large $k$) corresponds to values of $k$ for which curvature perturbations re-enter the cosmological horizon when the latter has a mass of the order of the asteroid mass (the horizontal blue band in the left panel of fig.\,\ref{fig:SummaryPlot}).  This is to generate a sizable abundance of asteroid-mass PBHs.
    \item[{\it iii)}] We take the amplitude of the plateau to be as close as possible to the allowed upper limit, $P_{\mathcal{R}}(k) = O(10^{-2})$. 
    This is to generate an abundance of asteroid-mass PBHs compatible with the observed dark matter (DM) content of the universe. 
\end{itemize}
As well known, an interesting byproduct of {\it ii)} and {\it iii)} 
is the possibility to generate a stochastic signal of gravitational waves (GWs) that are induced, as a second-order effect, 
by curvature perturbations \cite{Tomita:1975kj,Matarrese:1993zf,Acquaviva:2002ud,Mollerach:2003nq,Ananda:2006af,Baumann:2007zm} (see ref.\,\cite{Domenech:2021ztg} for a recent review). 
The frequency $f$ is related to the comoving wavenumber $k$ by the relation 
$k \simeq 7\times 10^{14} (
f/{\rm Hz})\,{\rm Mpc}^{-1}$
so  that the two sides of the plateau in fig.\,\ref{fig:SummaryPlot} correspond to 
$f = O(0.1)$ Hz (the typical target of future space-based GW interferometers like LISA \cite{LISACosmologyWorkingGroup:2022jok,Kuns:2019upi,Sesana:2019vho})  and 
$f = O(10^{-9})$ Hz (the typical target of Pulsar Timing Array (PTA) experiments). Interestingly, 
the NANOGrav collaboration has recently published an analysis of 12.5 yrs of pulsar timing data reporting a strong evidence for a stochastic
 common process, potentially induced by a SGWB, at a frequency of $O(10^{-9}\,{\rm Hz})$ \,\cite{NANOGrav:2020bcs} (also independently supported other by PTA experiments \cite{Goncharov:2021oub,Chen:2021rqp,Antoniadis:2022pcn}). 
 
 The presence of the plateau in the power spectrum opens the possibility to connect all the above observables even if characterized by widely different scales.  In ref.\,\cite{DeLuca:2020agl} it was indeed shown that a 
broad power spectrum in the form of a simple double-Heaviside theta function 
$P_{\mathcal{R}}(k) = A\,\Theta(k - k_{\rm min})\,\Theta(k_{\rm max} - k)$ 
with endpoints $k_{\rm min} \simeq 10^{-9}k_{\rm max}$ and $k_{\rm max} \simeq 10^{15}$ Mpc$^{-1}$ 
 and amplitude $A = 5.8\times 10^{-3}$
has the 
chance to produce the observed abundance of DM in the form of PBHs and, at the same time, 
generate a GW signal compatible (in frequency and amplitude) 
with the NANOGrav signal.
In this paper, we will explicitly derive the inflationary dynamics required to realise an analogous power spectrum, revealing the much richer phenomenology associated with this scenario.

To be more specific, the solid black line in the  right panel of fig.\,\ref{fig:SummaryPlot} corresponds to one of the USR models recently constructed in ref.\,\cite{Franciolini:2022pav}.  
The analysis of ref.\,\cite{Franciolini:2022pav} is based on what is called a ``reverse engineering approach'' (see refs.\,\cite{Ragavendra:2020sop,Tasinato:2020vdk,Ng:2021hll,Karam:2022nym} for a similar viewpoint). The idea that lies at the heart of this approach is that the starting point of the analysis is not the inflaton potential but rather the inflationary dynamics. 
Let us motivate this change of perspective.  
As mentioned above, the presence of an USR phase in the inflationary dynamics can be obtained is one takes 
a scalar potential that features  an approximate stationary 
inflection point. The latter is usually controlled by a number of free parameters that need to be fine-tuned up to very special values in order to 
guarantee the desired enhancement in the power spectrum of curvature perturbations \cite{Inomata:2016rbd,Garcia-Bellido:2017mdw,Ballesteros:2017fsr,Hertzberg:2017dkh,Kannike:2017bxn,Dalianis:2018frf,Inomata:2018cht,Cheong:2019vzl,Bhaumik:2019tvl,Bhaumik:2020dor,Ballesteros:2020qam,Iacconi:2021ltm,Kawai:2021edk}. At the technical level, this operation is not very transparent in the sense that it is typically difficult to isolate which parameters in the scalar potential control some specific feature of the power spectrum. 
In the approach of ref.\,\cite{Franciolini:2022pav} the scalar potential is nothing but an outcome of the analysis which, on the contrary, puts in the foreground the inflationary  dynamics starting from an analytical {\it ansatz} for the Hubble parameter $\eta$. 
As a result, the construction of inflationary models that give curvature power spectra with features compatible with the phenomenological requirements enumerated above becomes, at the technical level, far way accessible and, from the point of view of the physics involved, more transparent.  

Let us stress the following important conceptual point. 
Applying the reverse engineering approach of ref.\,\cite{Franciolini:2022pav} would be almost meaningless if one were only interested in the part of the power spectrum constrained by CMB observations. 
The reason is that, as mentioned at the very beginning of this introductory discussion, at CMB scales the experimental constraints are so tight that they almost completely nail down, at the corresponding field values, the form of the inflationary potential and a detailed analysis of specific models is possible. This is not true, however, if one is interested in the behaviour of the power spectrum at much smaller scales where, as illustrated in  fig.\,\ref{fig:SummaryPlot}, observational constraints are weaker and deviations from the slow-roll paradigm possible. 
In this case, contrary to what  happens in the reverse engineering approach of ref.\,\cite{Franciolini:2022pav},  there is no clear mapping between the free parameters of the scalar potential and the phenomenological implications that the presence of an USR phase may have.

The purpose of this work is to deepen the analysis presented in ref.\,\cite{Franciolini:2022pav}, and we organize our material as follows. 
In section\,\ref{sec:SM:background} we clarify the details of the reverse engineering approach 
by carefully describing the steps used to compute the spectrum of perturbations starting from the inflationary dynamics with a special emphasis on explaining with analytical arguments the mechanism that generates the plateau in the power spectrum as well as the physical meaning of the free parameters describing the inflationary dynamics.
In section\,\ref{sec:fPBH} we review the computation of the PBH abundance.
In section\,\ref{sec:GW} we discuss the implications for the scalar-induced GW signal. 
In section\,\ref{sec:potential} we give more details about the profile of the reconstructed inflationary potential and its theoretical interpretation.
Finally, we conclude in section\,\ref{sec:Conclusions}.
}

Throughout this paper, we use natural units and set the reduced Planck mass to unity.

\color{black}
%%%%%%%%%%%%%%%%%%%%%%%%%%%%%%%%%%%%%%%%%%%%%%%%
\section{Background evolution and spectrum of curvature perturbations}
\label{sec:SM:background}
%%%%%%%%%%%%%%%%%%%%%%%%%%%%%%%%%%%%%%%%%%%%%%%%

In this section, we introduce the bases of our reverse engineer approach.
We start with a parameterised background Hubble evolution, 
followed by the computation of curvature perturbations 
and an in-depth discussion of the characteristic features 
leading to the important phenomenological 
signatures presented in the following sections.

\subsection{Background evolution}

The inflationary background can be
described by modelling the evolution of the Hubble rate $H$.
This is dictated by dynamical equations relating $H$ 
to the Hubble parameters, which are 
\begin{align}\label{eq:HubbleParameters}
\epsilon\equiv -\frac{\dot{H}}{H^2}\,,~~~~~~~~~
\eta \equiv -\frac{\ddot{H}}{2H\dot{H}} = \epsilon - \frac{1}{2}\frac{d\log\epsilon}{dN}\,,
\end{align} 
where $\dot{H} = dH/dt$ is the cosmic-time derivative of $H$ while
$N$, defined such as $dN = Hdt$, is the number of $e$-folds. 
One can notice that, if we assume $\epsilon$ to be small and $\eta$ constant, 
eq.\,(\ref{eq:HubbleParameters}) admits the 
solution $\epsilon(N) \propto e^{-2\eta N}$.
As we will see in the following, 
this behaviour leads to 
an exponential enhancement of the amplitude of perturbations 
when the dynamics is characterised by large and positive $\eta$, as it is the case in an USR phase.

\begin{figure}[t]
\leavevmode
\centering
\includegraphics[width=0.65\textwidth]{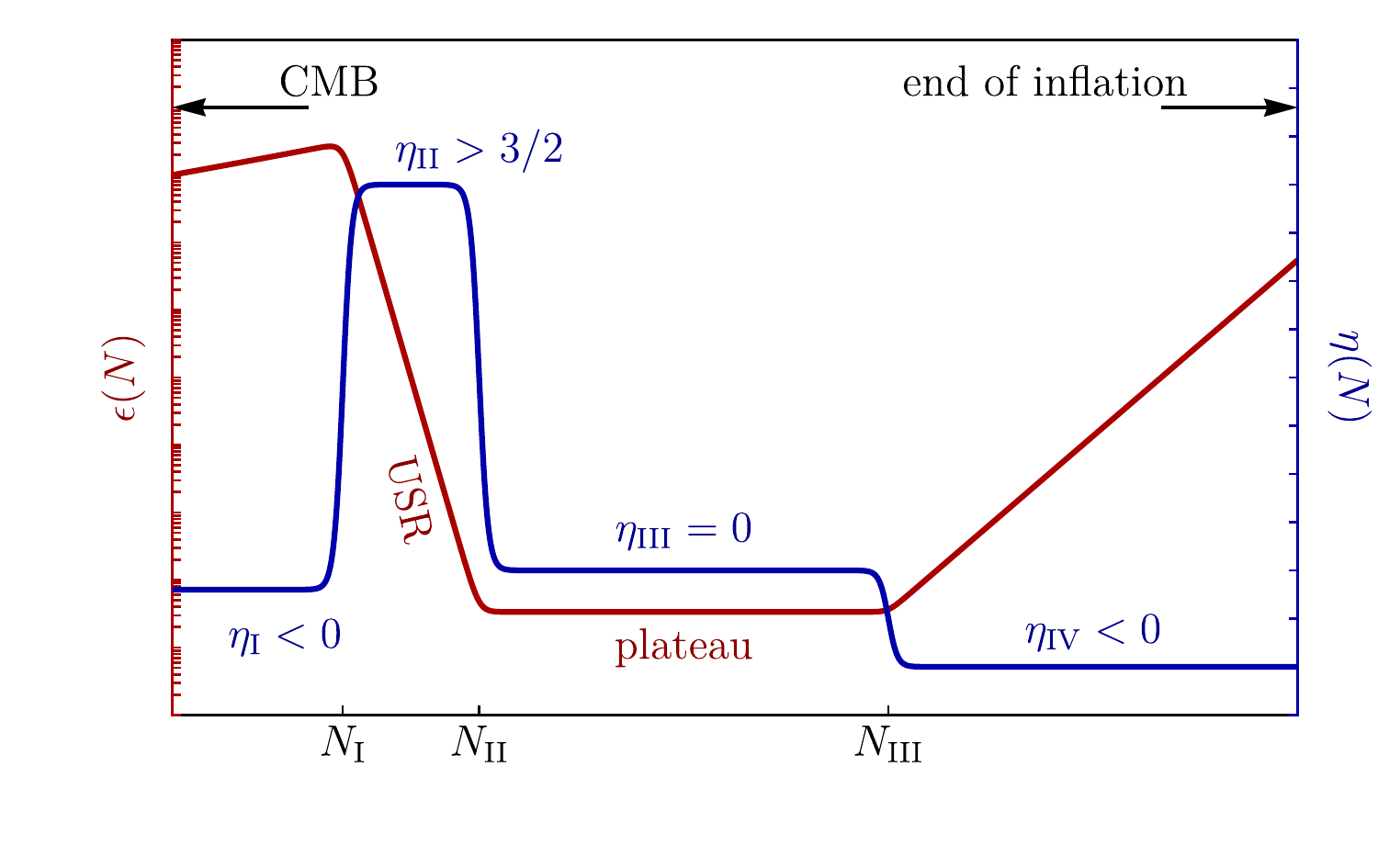}  
\caption{\label{fig:PlotEpsilonEta}
Schematic evolution of $\eta(N)$ and $\epsilon(N)$.
The former is given by our analytical ansatz in eq.\,(\ref{eq:MainEqEta}); the latter follows from  
eq.\,(\ref{eq:HubbleParameters}).
We label the USR phase characterised by a negative friction and 
the plateau in $\epsilon$ in the pahse of vanishing $\eta_{\rm III}$.}
\end{figure}

We base our construction on an analytical {\it ansatz} for the time-evolution of $\eta$ of the form
\begin{align}
\eta(N) = \frac{1}{2}\bigg\{&\left[
\eta_{\rm I} - \eta_{\rm II} + (\eta_{\rm II}-\eta_{\rm I})\tanh\left(\frac{N - N_{\rm I}}{\delta N_{\rm I}}\right)
\right] + 
\left[
\eta_{\rm II} + \eta_{\rm III} + (\eta_{\rm III}-\eta_{\rm II})\tanh\left(\frac{N - N_{\rm II}}{\delta N_{\rm II}}\right)
\right] + \nn \\
&\left[
\eta_{\rm IV}-\eta_{\rm III} + (\eta_{\rm IV}-\eta_{\rm III})\tanh\left(\frac{N - N_{\rm III}}{\delta N_{\rm III}}\right)
\right]\bigg\}\,.\label{eq:MainEqEta}
\end{align}
The consequent behaviour of $\epsilon(N)$ 
that follows from such ansatz
is derived by directly integrating the differential 
eq.\,(\ref{eq:HubbleParameters}). 
The reasons that leads to the specific ansatz presented in  eq.~\eqref{eq:MainEqEta} 
will become clear in the following. 
The inflationary dynamics can be divided into four subsequent stages, as we also show in fig.\,\ref{fig:PlotEpsilonEta}:
\begin{itemize}
%%%%%%%%%%%%%%%%%%%%%%
\item [{\it i)}] 
We fix the initial time at $N_{\rm ref}$ and an initial small value of $\epsilon_{\rm I}$. As long as the number of e-folds falls within the interval $N\in [N_{\rm ref},\,N_{\rm I}]$, the ansatz forces $\eta_{\rm I}$ to remain constant and negative; 
the solution of eq.\,\eqref{eq:MainEqEta}, predicting a scaling of the form $\epsilon(N) \propto e^{-2\eta N}$,
give rise to an exponential variation of $\epsilon$ during this phase. 
However, as the value of $\eta_{\rm I}$ is taken to be small 
(with the aim of reproducing the conventional slow-roll dynamics),
the evolution of $\epsilon$ is tamed. 
%%%%%%%%%%%%%%%%%%%%%%
\item [{\it ii)}]
Within the subsequent interval $N\in [N_{\rm I},\,N_{\rm II}]$ 
we impose $\eta_{\rm II} > (3+\epsilon)/2 \simeq 3/2$.  
A negative value of $\eta$ is associated to a period of negative friction, 
and the Hubble parameter $\epsilon$ is forced to decrease abruptly  
down to values $O(\ll \epsilon_{\rm I})$.
This phase realist the Ultra-Slow Roll (USR)
evolution typically advocated to generate enhanced spectra at small scales, within single field models of inflation. 
%%%%%%%%%%%%%%%%%%%%%%
\item [{\it iii)}]
 Subsequently, when the number of e-folds falls within  $N\in [N_{\rm II},\,N_{\rm III}]$, impose $\eta_{\rm III} = 0$. 
This forces $\epsilon$ to remain constant at the tiny value reached at the end of the negative friction phase.
%%%%%%%%%%%%%%%%%%%%%%
\item [{\it iv)}]
 The final phase is characterized by $\eta_{\rm IV} < 0$, which is a necessary requirement to bring $\epsilon$ back to $O(1)$ values
 and cause the end of inflation.  
\end{itemize}
The sharpness of the transition between each phase is
controlled by the parameters $\delta N_{\rm I,II,III}$.
In the limit of vanishing  $\delta N \to 0$, 
one obtains step transitions which are, however, unphysical. 
The expectations about the time evolution of $\epsilon$ 
qualitatively described above are confirmed by solving numerically eq.\,(\ref{eq:HubbleParameters}), 
adopting the parametrisation of $\eta(N)$ and using the initial condition $\epsilon(N_{\rm ref}) = \epsilon_{\rm I}$ 
imposed at the initial reference time $N_{\rm ref}$.
The solution is shown in fig.\,\ref{fig:PlotEpsilonEta}
assuming the parameters reported in table\,\ref{eq:ModelTab}.
As we will see in the following, the free parameters entering in the ansatz \eqref{eq:MainEqEta} will have a clear and direct connection to physical observables (as highlighted in table\,\ref{eq:ModelTab} below) and can be adjusted to devise consistent inflationary dynamics producing interesting late time signatures. We will come back to this point later on.

\subsection{Model parameters}

In the following, we shall discuss in detail the free parameters entering in eq.\,(\ref{eq:MainEqEta}), whose interpretation is summarised in table~\ref{eq:ModelTab}. 
\begin{itemize}
\item[$\circ$] 
The values of the parameters $\epsilon_{\rm I}$, $\eta_{\rm I}$ and $V_{\rm ref}$ are fixed by requiring consistency with large scale CMB observations.
In our model, this constraint is simply fulfilled by the dynamics of the first phase before $N_\text{ref}$ extending backwards up to CMB scales. 
We define $k_{\rm ref} = 0.05$ Mpc$^{-1}$ as the scale that exits the Hubble horizon at time $N_{\rm ref}$, that is the CMB pivot scale \cite{Planck:2018jri}, and use the slow roll relations 
\begin{equation}
\eta_{\rm I} = (n_s - 1 + 4\epsilon_{\rm I})/2
   ~~~~~~~~~ \text{and} ~~~~~~~~~
\epsilon_{\rm I} = r/16,
\end{equation}
linking the Hubble parameters at the pivot scale with the scalar spectral index $n_s$ and the tensor-to-scalar ratio $r$.  
In other words, we anchor the initial values $\eta_{\rm I}$ and $\epsilon_{\rm I}$ to CMB observables.  
Moreover, the amplitude of the power spectrum at the pivot scale, $A_s$, is related (via $H^2$) to $V_{\rm ref}$ by means of the 
Friedmann equation. We find 
\begin{equation}
    V_{\rm ref} = 24\pi^2\epsilon_{\rm I}(1 - \epsilon_{\rm I}/3)A_s.
\end{equation}
We fix $n_s$, $A_s$ and $r$ consistently with observations at CMB scales (with the value of $r$ within reach in next-generation CMB measurements). 
This, in turn, will 
directly nail down the fundamental parameters 
$\epsilon_{\rm I}$, $\eta_{\rm I}$ and $V_{\rm ref}$ of our phenomenological model \cite{Lidsey:1995np}. 
For definiteness, we take $n_s = 0.965$, $r = 0.005$ and 
$A_s = 2.1 \times 10^{-9}$ \cite{Planck:2018jri}. 
% Consequently to this procedure, our model is, by construction, in agreement with CMB observables. 

{\renewcommand{\arraystretch}{1.4}
\begin{table*}[!t!]%[htp]
\renewcommand{\arraystretch}{1.55}
\begin{center}
% \begin{adjustbox}{min width=1\textwidth}
\begin{tabular}{||c||c|c|c||c|c||}
\hline\hline
\textbf{Model parameter} &
% \textbf{{\color{verdes}{Value}}}
\textbf{{\color{verdes}{Model (1)}}} &
% \textbf{{\color{blue}{Value}}}
\textbf{{\color{blue}{Model (2)}}}  &
% \textbf{{\color{cornellred}{Value}}} 
\textbf{{\color{cornellred}{Model (3)}}} &
\textbf{Spectral feature} &
\textbf{Phenomenology}
\\
\hline 
$\boldsymbol{\epsilon_{\rm I}}$ &
\multicolumn{3}{c||}{$3.125\times10^{-4}$} &
\multirow{2}{*}{tilt of $P_{\mathcal{R}}$ at CMB scales}  &
\multirow{2}{*}{$n_s$, $r$}  
\\
\cline{1-4} 
$\boldsymbol{\eta_{\rm I}}$ &
\multicolumn{3}{c||}{$-1.68\times 10^{-2}$} & &
\\
\hline
$\boldsymbol{V_{\rm ref}}$ &
\multicolumn{3}{c||}{$1.55\times 10^{-10}$} &
amplitude of $P_{\mathcal{R}}$ at CMB scales &
$A_s$
\\
\hline 
$\boldsymbol{N_{\rm ref}}$ &
\multicolumn{3}{c||}{$0$} &
pivot scale  $k_{\star}$ &
CMB
\\
\hline 
\hline 
 $\boldsymbol{N_{\rm I}}$ &
 \multicolumn{3}{c||}{$15.5$} &
 large-scale edge of the plateau $k_{\rm min}$ &
 solar-mass PBHs
 \\
\hline 
$\boldsymbol{\delta N_{\rm II}}$ &
$0.50$ &
$0.46$ &
$0.60$ &
bump at $k_{\rm min}$ &
peak of solar-mass PBHs 
 \\
 \hline 
 $\boldsymbol{\eta_{\rm II}}$ & 
 2.709 &
 2.710 &
 2.735 &
 \multirow{2}{*}{height of the plateau} & 
 \multirow{2}{*}{PBH$=$DM: $f_{\rm PBH} \overset{!}{=} 1$}
 \\
 \cline{1-4} 
 $\boldsymbol{\Delta N_{\rm USR}}$ &
 \multicolumn{3}{c||}{$2.9$} &
 &
 \\
 \hline 
 $\boldsymbol{\eta_{\rm III}}$ &
 \multicolumn{3}{c||}{$0$} & 
plateau & 
multi-scales $\Omega_{\rm GW}(f)$ signal  
\\
\hline 
$\boldsymbol{\Delta N_{\rm plateau}}$ &
\multicolumn{3}{c||}{$16.6$} &
small-scale edge of the plateau $k_{\rm max}$ &
asteroid-mass PBHs
\\
\hline 
$\boldsymbol{\delta N_{\rm III}}$ &
$0.50$ &
$0.50$ &
$1.31$ &
bump at $k_{\rm max}$ &
peak of asteroid-mass PBHs 
\\
\hline
\hline
$\boldsymbol{N_{\rm IV}}$ &
\multicolumn{3}{c||}{$55$} &
\multirow{2}{*}{drop-off} &
\multirow{2}{*}{end of inflation}
\\
\cline{1-4} 
$\boldsymbol{\eta_{\rm IV}}$ &
-0.554 &
-0.554 &
-0.560 &&
\\ 
\hline\hline
\end{tabular}
% \end{adjustbox}
% \vspace{-0.35cm}
\caption{{
Free parameters of our model together with their numerical benchmark values. 
We define $\Delta N_{\rm USR}\equiv N_{\rm II} - N_{\rm I}$ and  
$\Delta N_{\rm plateau}\equiv N_{\rm III} - N_{\rm II}$.
Dimensionful quantities are written in units of the reduced Planck mass. 
$\delta N_{\rm I} = 0.50$ is kept fixed.
}}\label{eq:ModelTab}
\end{center}
\end{table*}
}

    \item[$\circ$] 
    The value of $N_{\rm I}$ sets the beginning of the USR phase and controls the comoving wavenumber at which the power spectrum 
    of curvature perturbations starts increasing with respect to its slow-roll value. 
    In order to reproduce the results of ref.\,\cite{DeLuca:2020agl}, we need an early growth of the power spectrum 
 at scales set by the value of $k_{\rm min}$. 
 As a rule of thumb, we estimate the corresponding value of $N_{\rm I}$ by means of the logarithmic scaling 
\begin{align}
N_{\rm I} \simeq \log\left(
\frac{k_{\rm min}}{k_{\star}}
\right)\,.
\end{align}
The above estimate represents a first guess for $N_{\rm I}$ 
around which we tune its final value by the accurate solving of the MS equation.
    \item[$\circ$] The values of $\eta_{\rm II}$ and $\Delta N_{\rm USR}\equiv N_{\rm II} - N_{\rm I}$ 
control the  height of the plateau in the power spectrum. 
These values are tuned in order to 
get the right abundance of dark matter in the form of PBHs. 
   \item[$\circ$] We set  $\eta_{\rm III} = 0$ in order to generate a plateau in the power spectrum.
\item[$\circ$] The $e$-fold interval $\Delta N_{\rm plateau}\equiv N_{\rm III} - N_{\rm II}$ controls 
 the broadness of the plateau.
  In order to reproduce the results of ref.\,\cite{DeLuca:2020agl} we need a broad plateau that covers 
 approximately the range of comoving wavenumbers 
 $k_{\rm max}/k_{\rm min} \approx 10^{9}$.
 As a rule of thumb, we estimate the corresponding value of $\Delta N_{\rm plateau}$ by means of the logarithmic scaling
  \begin{align}
\Delta N_{\rm plateau}\equiv N_{\rm III} - N_{\rm II} \simeq \log\left(
\frac{k_{\rm max}}{k_{\rm min}}
\right)\,.
\end{align}
The above estimate represents a first guess for $\Delta N_{\rm plateau}$
around which we tune its final value by the accurate solving of the MS equation.
 \item[$\circ$] We fix $N_{\rm IV} = 55$ in order to get a long enough inflationary phase to solve the horizon and flatness problems.
 Consequently, the value of $\eta_{\rm IV}$ is tuned in order to get $\epsilon = 1$ at $N_{\rm IV} = 55$. 
\item[$\circ$] The parameters $\delta N_{\rm I,II,III}$ control the 
 sharpness of the transitions in the evolution of $\eta(N)$ at $e$-fold times, respectively, $N_{\rm I}$, $N_{\rm II}$ and $N_{\rm III}$. 
 The limit $\delta N \to 0$ corresponds to a step transition.
In short, these parameters control the bump-like features that are present in the power spectrum at the two edges of the plateau region (see the detailed discussion in sec.~\,\ref{sec:feat}). 
These parameters, therefore, play a very special role in our analysis. 
This is because the computation of the PBH abundance is exponentially sensitive to the shape of the power spectrum, and small variations are capable of producing very sizable effects. 
\end{itemize}

\subsection{Curvature perturbations}

Once the background evolution is specified, one can compute the spectrum of gauge-invariant comoving curvature perturbation generated during inflation and transferred to the radiation fluid after reheating. 
As long as the slow-roll approximation is valid, this can be computed as
\begin{equation}
P_{\mathcal{R}}(k) = \frac{H^2}{8\pi^2 \epsilon}\,,\label{eq:SlowRollPS}
\end{equation} 
where the Hubble parameters are evaluated at horizon crossing of modes $k$. 
To get an expectation of what spectrum of curvature perturbations 
would result from eq.~\eqref{eq:MainEqEta}, one could  na\"{\i}vely 
reverse the evolution of $\epsilon$ shown in fig.\,\ref{fig:PlotEpsilonEta},
that clearly features an exponential growth 
followed by a plateau region inherited from the term $1/\epsilon$.

In order to confirm this intuition beyond the slow-roll approximation, we compute $P_{\mathcal{R}}(k)$ by solving the Mukhanov-Sasaki (MS) equation \cite{Sasaki:1986hm,Mukhanov:1988jd}
\begin{align}\label{eq:M-S}
\frac{d^2 u_k}{dN^2} &+ (1-\epsilon)\frac{du_k}{dN} + 
\left[
\frac{k^2}{(aH)^2} + (1+\epsilon-\eta)(\eta - 2) - \frac{d}{dN}(\epsilon - \eta)
\right]u_k = 0\,,
\end{align}
which was shown to describe the properties of perturbations at the linear level even with the inclusion of quantum diffusion effects \cite{Ballesteros:2020sre}; it should be noted, however, that stochastic effects may become relevant, beyond the linear order, during the USR phase~\cite{Pattison:2017mbe,Biagetti:2018pjj,Ezquiaga:2019ftu,Pattison:2021oen,Figueroa:2021zah}. 
% We, therefore, solve in Fourier space the MS equation (with sub-horizon Bunch-Davies initial conditions) and compute the power spectrum $P_{\mathcal{R}}(k)$ of the gauge-invariant comoving curvature perturbation, 
We remark that we do not include in our analysis any non-linear effects related to the dynamics of curvature perturbations (e.g. \cite{Namjoo:2012aa,Chen:2013eea,Cai:2018dkf,Passaglia:2018ixg,Biagetti:2021eep}). 
Non-gaussian effects, for a given mode $k$, are mostly controlled by the value of $\eta$ after the mode settles to its final conserved value\,\cite{Atal:2018neu,Atal:2019cdz,Taoso:2021uvl}. Modes that contribute to the plateau become constant during phase {\it iii)} with $\eta_{\rm III} = 0$ and, therefore, should have negligible non-gaussianity. 
Modes that contribute to the right-side edge and the subsequent fall-off of the power spectrum,  settle to their final constant value during phase {\it iv)} with $\eta_{\rm IV}$ non-zero and negative. However, in all realizations of our model 
we consider in this work (see table\,\ref{eq:ModelTab}) 
the actual value of $|\eta_{\rm IV}|$ is small, and we do not expect large corrections\,\cite{Young:2022phe} (reabsorbable by a small re-tuning of $\eta_{\rm II}$). 
Furthermore, we model the transitions at the beginning and end of the USR phase in a smooth way, and this has the effect of further suppress local non-gaussianity\,\cite{Cai:2018dkf,Passaglia:2018ixg}. 
Finally, assessing the impact of non-linear stochastic effects on our model deserves a separate analysis beyond the scope of this work.

We solve the MS equation  with sub-horizon Bunch-Davies initial conditions at $N \ll N_k$, where $N_k$ indicates the horizon crossing time for the mode $k$, that is the time at which we have $k = a(N_k)H(N_k)$. This is implemented as 
\begin{equation}
u_k(N \ll N_k) = \frac{1}{\sqrt{2 k}}, 
\qquad \text{and} \qquad
\frac{du_k (N \ll N_k)}{dN} = -  \frac{k}{\sqrt{2} a(N)H(N)}, 
\end{equation}
where, without loss of generality, we choose the phase of $u_k$ such that it is real initially.
We then compute the power spectrum $P_{\mathcal{R}}(k)$ of the gauge-invariant comoving curvature perturbation 
$\mathcal{R}$ given by 
\begin{align}\label{eq:PS}
P_{\mathcal{R}}(k) = \frac{k^3}{2\pi^2}\left|\frac{u_k(N)}{z(N)}\right|^2_{N > N_{\rm F}(k)}\,,~~~~~~~~~
{\rm with\,\,\,}\mathcal{R}_k(N) = -\frac{u_k(N)}{z(N)}\,,~~~~{\rm and\,\,\,}z(N) = a(N)\frac{d\phi(N)}{dN}\,.
\end{align}
In eqs.\,(\ref{eq:M-S},\,\ref{eq:PS}) $\mathcal{R}_k(N)$ and $u_k(N)$ are time-dependent Fourier mode 
corresponding to a fixed comoving wavenumber $k\equiv |\vec{k}|$. 
The power spectrum $P_{\mathcal{R}}(k)$ does not depend on time because the meaning of 
eq.\,(\ref{eq:PS}) is that $P_{\mathcal{R}}(k)$ must be evaluated 
after the time $N_{\rm F}(k)$ at which the mode $|u_k(N)/z(N)|$ freezes to the constant value that is conserved until its 
horizon re-entry. 
We then have
\begin{align}
N_{\rm F}(k) \equiv {\rm max}\{N_k,N_{\rm II}\}\,.
\end{align}
Modes that exit the horizon {\it before} the time $N_{\rm II}$ (that is modes such that $N_k < N_{\rm II}$) are 
not conserved  (even though super-horizon) because they experience afterward the negative friction phase. Consequently, 
for these modes their contribution to  eq.\,(\ref{eq:PS}) must be evaluated at any time $N > N_{\rm II} > N_k$ after the negative 
friction phase ends. Contrariwise, modes 
that exit the horizon {\it after} the time $N_{\rm II}$ (that is modes such that $N_k > N_{\rm II}$) freeze 
to their constant value after they become super-horizon. Consequently, 
as customary, the contribution of these modes to  eq.\,(\ref{eq:PS}) must be evaluated at any time $N > N_k > N_{\rm II}$.

It is sometime useful to rewrite the MS equation in the form
\begin{align}
\frac{d^2\mathcal{R}_k}{dN^2} + (3+\epsilon - 2\eta)\frac{d\mathcal{R}_k}{dN} + \frac{k^2}{(aH)^2} \mathcal{R}_k = 0\,.
\end{align}
Assuming $\epsilon \approx 0$, constant $\eta$ and constant $H$, this equation admits the solution 
\begin{align}
\mathcal{R}_k(N) \propto e^{-\left(\frac{3}{2}-\eta\right)N}
\left[
c_1\,J_{\frac{3}{2}-\eta}\left(\frac{k}{k_{\star}}e^{-N}\right)
\Gamma\left(
\frac{5}{2}-\eta
\right) + 
c_2\,J_{-\frac{3}{2}+\eta}\left(\frac{k}{k_{\star}}e^{-N}\right)
\Gamma\left(
-\frac{1}{2}+\eta
\right)
\right]\,,\label{eq:AnalRk}
\end{align}
where $J_{\alpha}(x)$ are Bessel functions of the first kind and $\Gamma(x)$ is the Euler gamma function. 
We are interested in the sub-Hubble regime, meaning that in the argument of the Bessel function $ke^{-N}/k_{\star} \gg 1$. 
In this limit the asymptotic behavior of the Bessel function is controlled, at the first order, by the scaling 
$J_{\alpha}(x) \sim 1/\sqrt{x}$.

\begin{figure}[h]
\leavevmode
\centering
\includegraphics[width=0.65\textwidth]{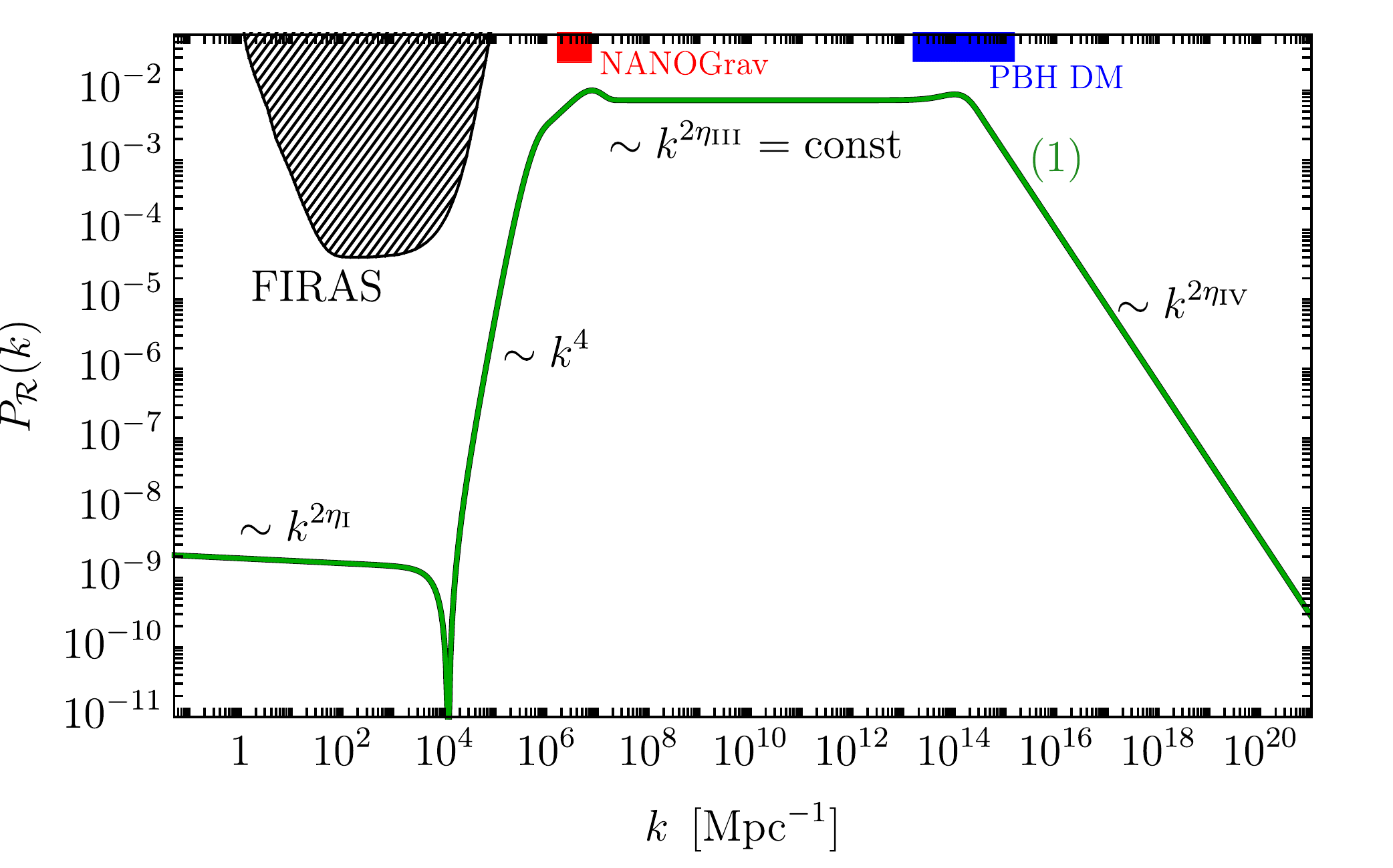}  
\caption{\label{fig:PS}
Power spectrum (solid green) corresponding to the model 
described in the second column of table\,\ref{eq:ModelTab}. 
The region shaded in gray corresponds to the region constrained by CMB spectral distortions\,\cite{Fixsen:1996nj}. 
To guide the eye, we indicate in red the frequency range 
$2.5\times 10^{-9} < f\,[{\rm Hz}] < 1.2\times 10^{-8}$  characterising the NANOGrav signal and in blue 
the mass range 
$10^{-16} < M_{\rm PBH}\,[M_{\odot}] < 10^{-12}$ in which PBHs may comprise the totality of DM.
}
\end{figure}

We show the numerical result of this procedure in fig.\,\ref{fig:PS}. 
% When slow roll is applicable, we find the scaling $P_{\mathcal{R}} \sim 1/\epsilon \sim e^{2\eta N} \sim k^{2\eta}$ where in the last step we used the horizon crossing condition $k=aH$. 
During the transition from the initial slow-roll phase to the plateau, we note that our model gives the steepest growth $\sim k^4$\,\,\cite{Byrnes:2018txb}.
The numerical solution of the MS equation in fig.\,\ref{fig:PS} 
shows that the USR dynamics encoded in eq.\,(\ref{eq:MainEqEta}) correctly  gives a plateau 
in the power spectrum of curvature perturbations that is compatible with the result of ref.\,\cite{DeLuca:2020agl}. 
The numerical values of the parameters used in fig.\,\ref{fig:PS} are summarized in table\,\ref{eq:ModelTab} (second row).

Two aspects of our approach are truly remarkable. 
First, {\it all} free parameters entering in 
eq.\,(\ref{eq:MainEqEta}) have a neat and simple relation to a physical observable; this is summarized in the last two columns of 
 table\,\ref{eq:ModelTab}, and discussed in full detail in the following sections. 
 This is contrary to what usually happens if one takes the conventional route of starting from the potential and then studying the dynamics. 
 The free parameters entering the scalar potential usually give very little intuition about the physics of PBH formation. 
Second, our analysis is not just a mere  rewording of what done in  ref.\,\cite{DeLuca:2020agl}; 
on the contrary, our approach discloses a much richer phenomenology that we shall now discuss.
Furthermore, it will allow us to derive the inflationary potential that realise such scenario.

\subsection{On the formation of a raised plateau in the power spectrum}

\begin{figure}[!h!]
\begin{center}
$$
\includegraphics[width=.495\textwidth]{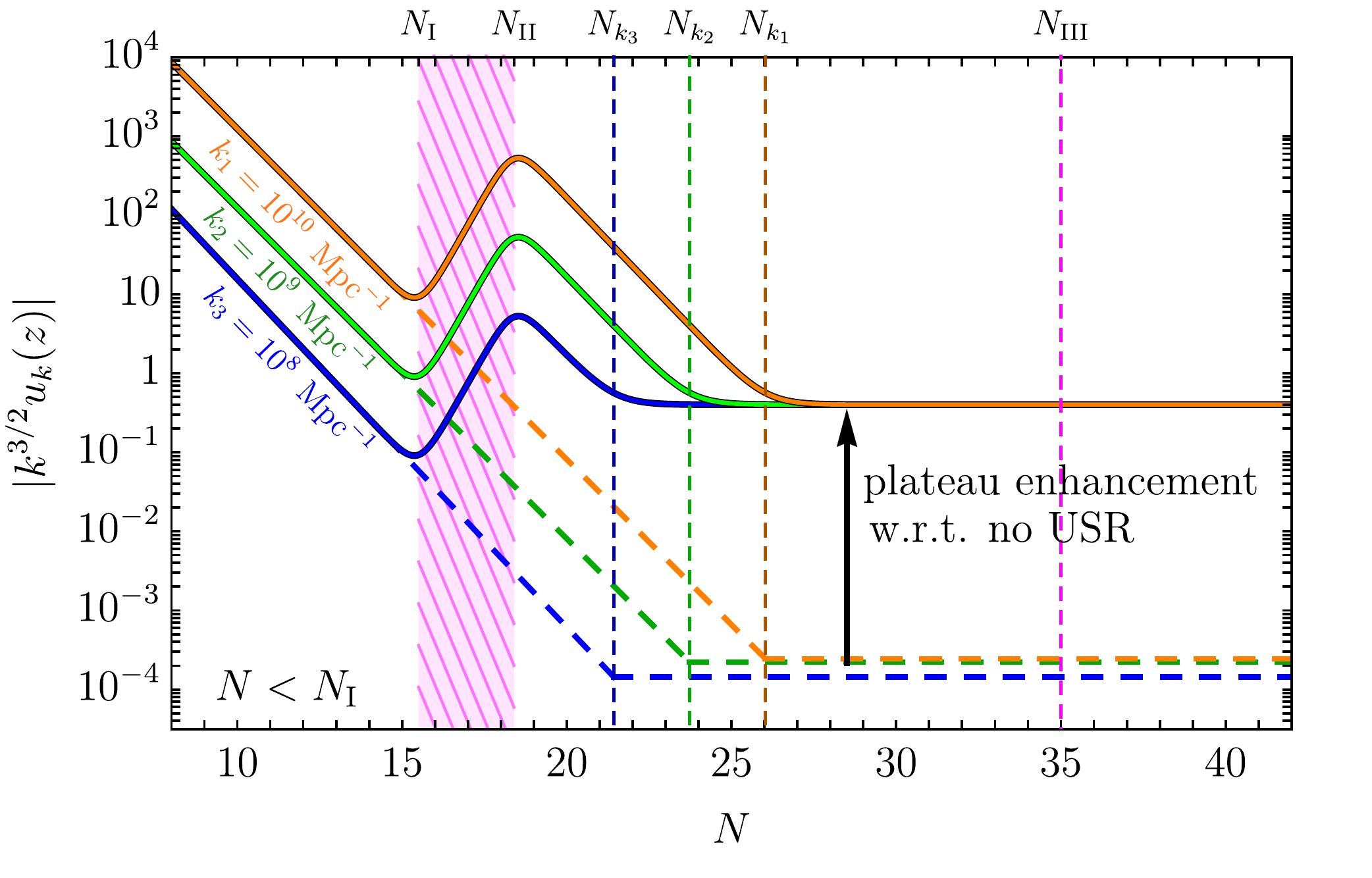}~\includegraphics[width=.495\textwidth]{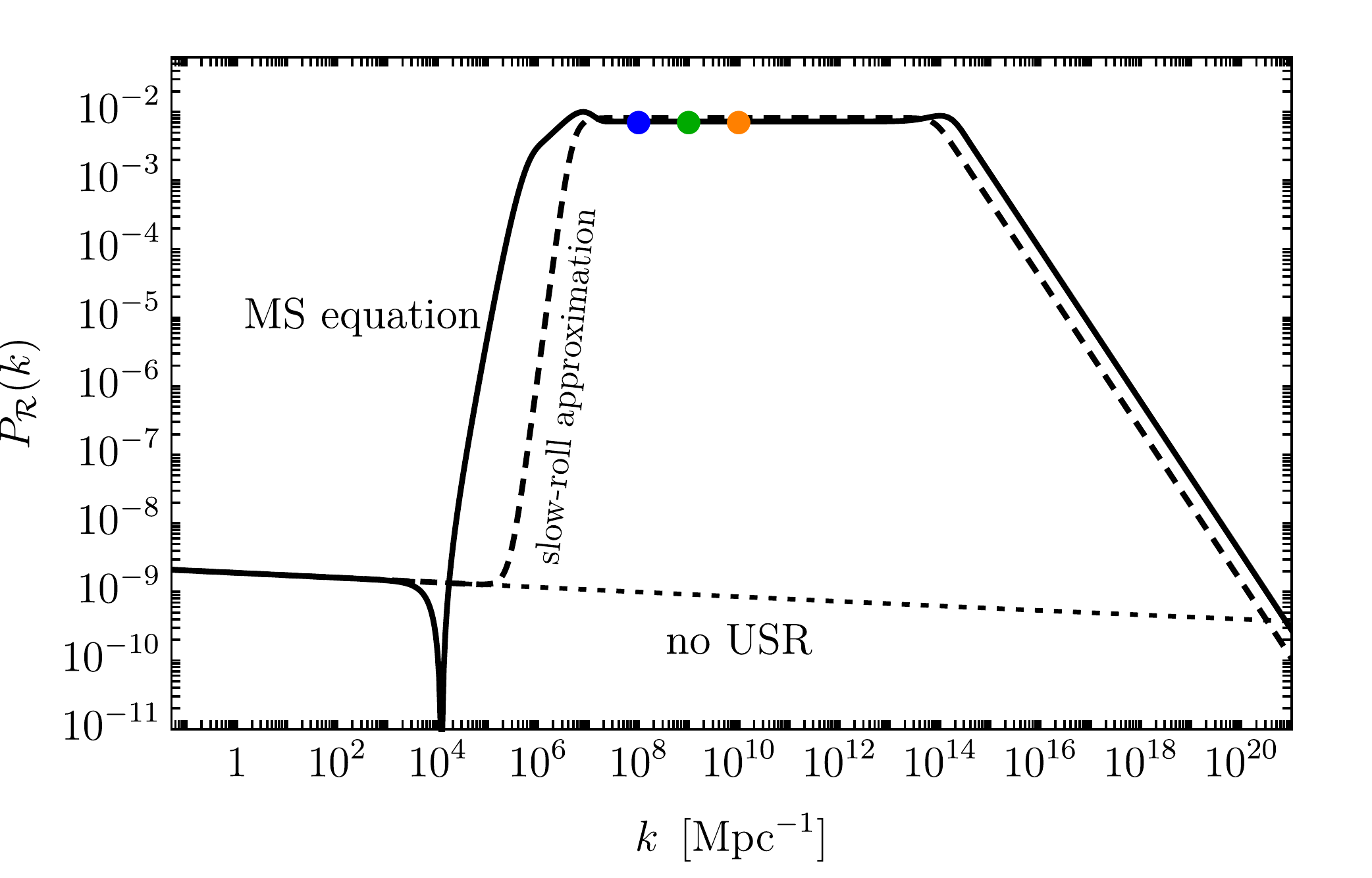}$$
\caption{
Time evolution (in terms of the $e$-fold number) 
of $k^{3/2}|u_k(N)/z(N)|$ for three different modes with $k_{1,2,3} = 10^{10,9,8}$ Mpc$^{-1}$ 
obtained by numerically solving the MS equation (solid lines). The dashed lines 
correspond to the absence of the USR phase. 
The vertical lines labeled with $N_k$ mark the $e$-fold time of horizon crossing for the mode with comoving wavenumber $k$.
As shown in the right panel (colored dots, one for each $k_{1,2,3}$), these modes  
contribute to the plateau of the power spectrum. 
In this plot we also show the power spectrum corresponding to the absence of the USR phase (dotted) and the one computed by means of the slow-roll approximation (dashed).
 }\label{fig:EvoModes}  
\end{center}
\end{figure}

The modes that form the plateau are those that exit the horizon during the phase 
$N_{\rm II} < N < N_{\rm III}$ with $\eta = 0$.  
We show in fig.\,\ref{fig:EvoModes} the time evolution of three representative modes of this kind 
for which $N_{\rm II} < N_k < N_{\rm III}$.
We shall analyze the dynamics in three subsequent steps, and arrive at a simple analytical understanding of the plateau's formation.

\begin{itemize}
\item[$\circ$] $N < N_{\rm I}$.
The modes are sub-Hubble ($k\gg aH$). 
The modulus of the function $k^{3/2}|u_k(N)/z(N)|$ exponentially decays while its real and imaginary parts oscillate. For different $k$, the modes decay equally fast (see
 left panel of fig.\,\ref{fig:EvoModes}). 
 Using eq.\,(\ref{eq:AnalRk}) and neglecting $\eta$ since $|\eta_{\rm I}| \ll 1$, we simply have 
 $k^{3/2}|u_k(N)/z(N)| \sim e^{-N}$. This time-dependence is confirmed numerically in fig.\,\ref{fig:EvoModes}.
 
 The difference in normalization -- the function $k^{3/2}|u_k(N)/z(N)|$ is bigger for larger $k$, see
 left panel of fig.\,\ref{fig:EvoModes} -- can be traced back to the Bunch-Davies initial condition. 
Deep in the sub-Hubble regime, we have
\begin{align}\label{eq:BD}
k^{3/2}\left|\frac{u_k(N)}{z(N)}\right| \sim k^{3/2}\times \frac{1}{\sqrt{k}} = k\,.
\end{align} 
Since the subsequent time evolution is universal, we conclude 
that the difference between two modes with comoving wavenumbers $k_1$ and $k_2 < k_1$ is simply given by  
$k_1/k_2$ as a consequence of eq.\,(\ref{eq:BD}). 
This is confirmed numerically if we compare the modes with $k_{1,2,3}$ (that differ between each other by one 
order-of-magnitude) displayed in the 
left panel of fig.\,\ref{fig:EvoModes}. 

\item[$\circ$] $N_{\rm I} < N < N_{\rm II}$.

The modes enter in the negative-friction phase, and they are now exponentially enhanced.
The key point is that modes with different $k$ experience, during this phase, the same amount of 
exponential growth. The latter is fixed by the value of $\eta_{\rm II}$ and the duration of the negative-friction phase 
$\Delta N_{\rm USR} = N_{\rm II} - N_{\rm I}$\,\cite{Ballesteros:2020qam}. 
This is again a consequence of eq.\,(\ref{eq:AnalRk}); since $\eta_{\rm II} > 3/2$,  
the factor $e^{-\left(3/2-\eta\right)N}$ gives an exponential growth that is bigger for longer 
$\Delta N_{\rm USR}$.
Consequently, at the end of the negative-friction phase modes with different $k_1$ and $k_2 < k_1$ will 
still differ between each other by the factor $k_1/k_2$. 
This is confirmed numerically if we compare at $N=N_{\rm II}$ the modes with $k_{1,2,3}$ in the 
left panel of fig.\,\ref{fig:EvoModes}.

\item[$\circ$] $N_{\rm II} < N < N_{\rm III}$.

The modes exit from the negative friction phase. 
The function $k^{3/2}|u_k(N)/z(N)|$ decays exponentially fast in the sub-Hubble regime until the time 
$N = N_k$ at which the mode crosses the Hubble horizon and settles to its final constant value. 
During this phase the time-dependence is again given by
\begin{align}
k^{3/2}\left|\frac{u_k(N)}{z(N)}\right| \sim e^{-N}\,.\label{eq:FinalScaling}
\end{align}
This follows from the time-dependence of eq.\,(\ref{eq:AnalRk}) with $\eta_{\rm III} = 0$.
The key point is that now the value of $N_k$ is larger for increasing $k$ since we have $N_k = \log(k/k_{\star})$.   
This means that modes with different $k_1$ and $k_2 < k_1$ will experience, before horizon crossing, a different amount of 
exponential suppression: the mode with $k_1 > k_2$ will exit the horizon after the $k_2$ mode.  
Consequently, the $k_1$ mode will get, compared to the $k_2$ mode, an extra suppression given by the factor
\begin{align}
e^{-\log(k_1/k_2)} = \frac{k_2}{k_1}\,.
\end{align} 
This extra suppression will precisely cancel the initial enhancement 
of the $k_1$ mode compared to the $k_2$ mode, as discussed below eq.\,(\ref{eq:BD}), so that they eventually settle 
precisely on the same value. 
This compensating mechanism produces the plateau. 
It should be stressed that this exact compensation is possible because we set $\eta_{\rm III} = 0$ 
(otherwise the scaling in eq.\,(\ref{eq:FinalScaling}) would have been different).
In the 
left panel of fig.\,\ref{fig:EvoModes} the time evolution of the three modes with $k_{1,2,3}$ 
clearly shows how the initial mismatch  during $N<N_{\rm I}$ gets precisely reabsorbed during the phase 
$N_{\rm II} < N < N_{\rm III}$ with $\eta = 0$.

\end{itemize}

In conclusion, the formation of the plateau follows from the same mechanism that originates a scale-invariant 
power spectrum in the slow-roll limit when both $\epsilon \to 0$ and $|\eta|\to 0$. 
Modes $k^{3/2}|u_k(N)/z(N)|$ with larger $k$ starts from larger values 
in the Bunch-Davies vacuum 
but exponentially decay for longer time before horizon crossing. 
In our model the presence of negative friction 
introduces an intermediate phase of exponential growth which however affects all modes in the same way: the net effect 
is that  of an exponential enhancement of the plateau value compared to the case in which the negative friction phase was absent.
This is evident from the evolution of the modes shown in the left panel of fig.\,\ref{fig:EvoModes}.
All in all, the mechanism that generates the plateau in our model is not fundamentally different compared to what discussed in ref.\,\cite{Leach:2001zf} (often dubbed Wands duality, see ref.\,\cite{Wands:1998yp}). 
However, the discussion presented here in terms of the evolution of individual modes gives a particularly limpid interpretation of the mechanism.

{As a final remark, we reiterate the importance of solving numerically the MS equation for the computation of the power spectrum. In the right panel of fig.\,\ref{fig:EvoModes} we show the comparison with the slow-roll approximation in eq.\,(\ref{eq:SlowRollPS}). 
The slow-roll approximation captures well the overall features of the power spectrum but it misses the right modelling of the transition regions at the two edges of the plateau. These two parts of the power spectrum, as we shall discuss next, are of crucial importance for the phenomenology of PBHs.}

\subsection{Features at the edges of the plateau}\label{sec:feat}

\begin{figure}[t]
\begin{center}
$$\includegraphics[width=.495\textwidth]{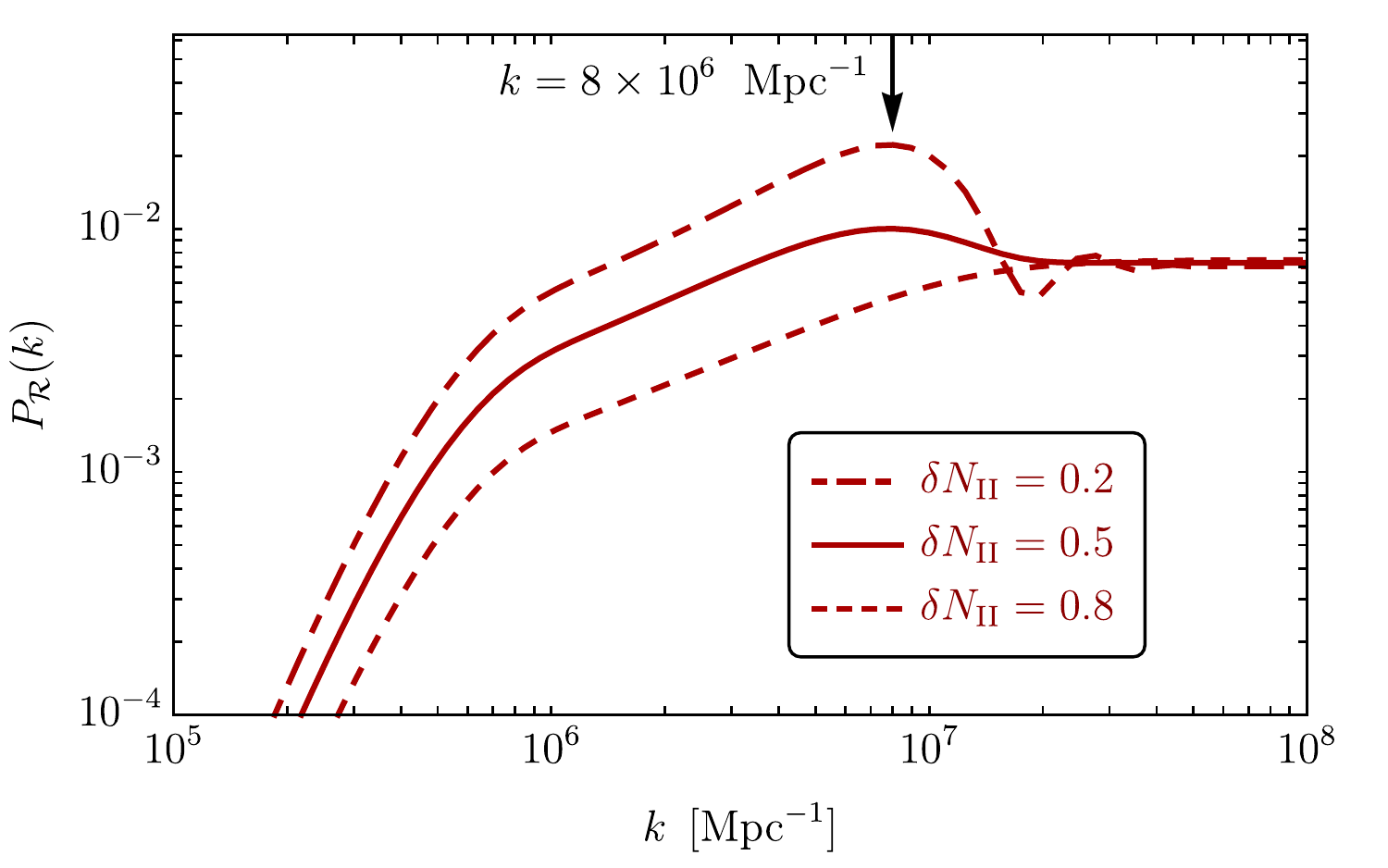}~\includegraphics[width=.495\textwidth]{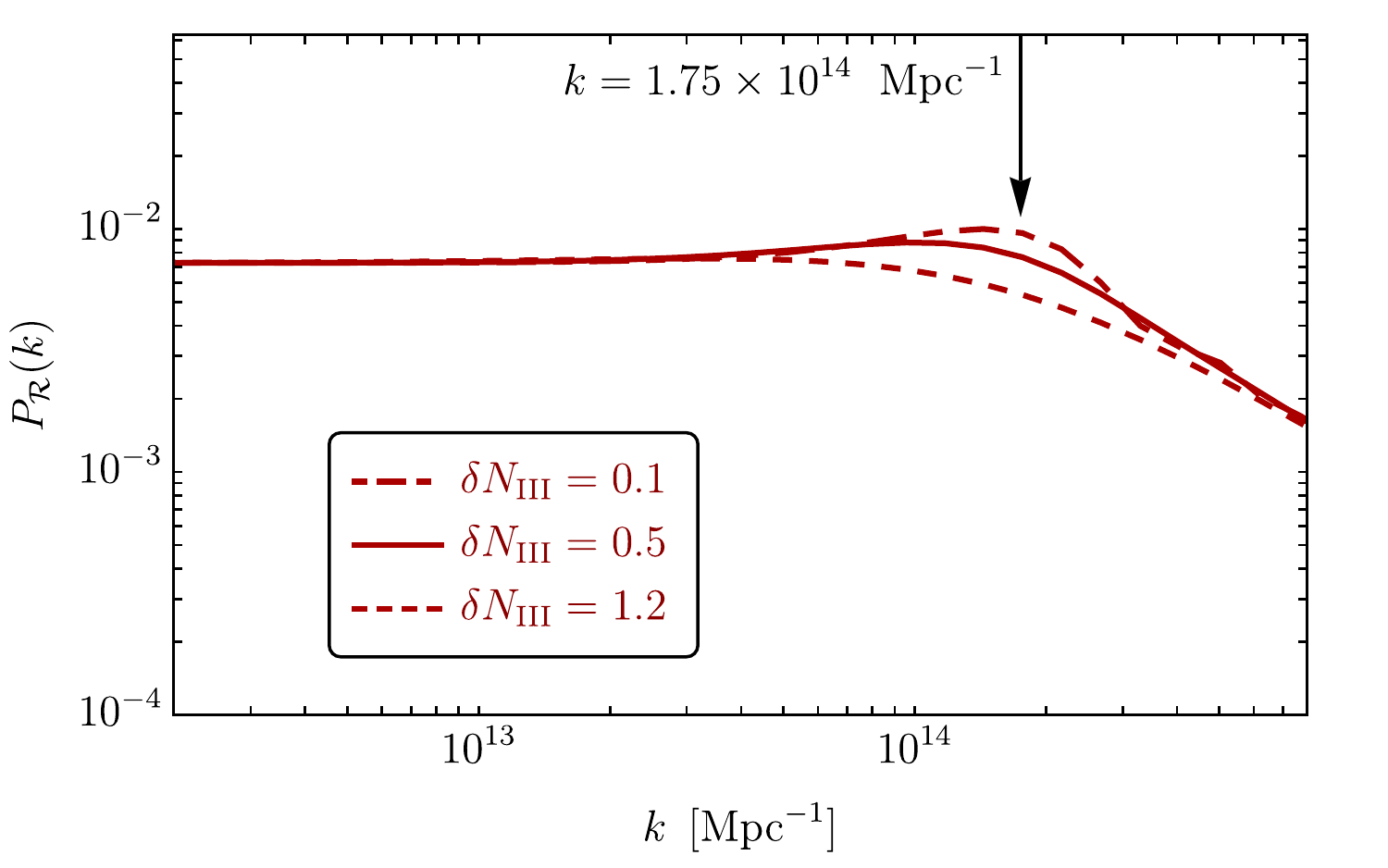}$$
\caption{
Power spectrum for different choices of $\delta N_{\rm II}$ (left panel) and $\delta N_{\rm III}$ (right panel). All other parameters are fixed according to the first column in table\,\ref{eq:ModelTab}. 
We zoom in the transition regions at the two edges of the plateau.
 }\label{fig:DeltaN} 
\end{center}
\end{figure}

Let us discuss here the role of $\delta N_{\rm I,II,III}$ previously anticipated. 
First, we take $\delta N_{\rm I} = \delta N_{\rm III} = 0.50$ fixed, and
consider a variation of $\delta N_{\rm II}$ with respect to the value $\delta N_{\rm II} = 0.50$ (that is the one used in the benchmark model corresponding to the first column in table\,\ref{eq:ModelTab}). We show our result in the left panel of fig.\,\ref{fig:DeltaN}. We note that $\delta N_{\rm II}$ controls the shape of the power spectrum at the left-side edge of the plateau. In particular, a sharper transition (smaller $\delta N_{\rm II}$) results in the formation of a bump-like feature at $k_{\rm min}$; on the contrary, a wider transition (larger $\delta N_{\rm II}$) smooths out the bump.

Second, we take $\delta N_{\rm I} = \delta N_{\rm II} = 0.50$ fixed, and
consider a variation of $\delta N_{\rm III}$ with respect to the value $\delta N_{\rm III} = 0.50$. 
We show our result in the right panel of fig.\,\ref{fig:DeltaN}. We note that $\delta N_{\rm III}$ controls the shape of the power spectrum at the right-side edge of the plateau. In particular, a sharper transition (smaller $\delta N_{\rm III}$) results in the formation of a bump-like feature at $k_{\rm max}$; on the contrary, a wider transition (larger $\delta N_{\rm III}$) smooths out the bump.

\begin{figure}[!t!]
\begin{center}
$$\includegraphics[width=.495\textwidth]{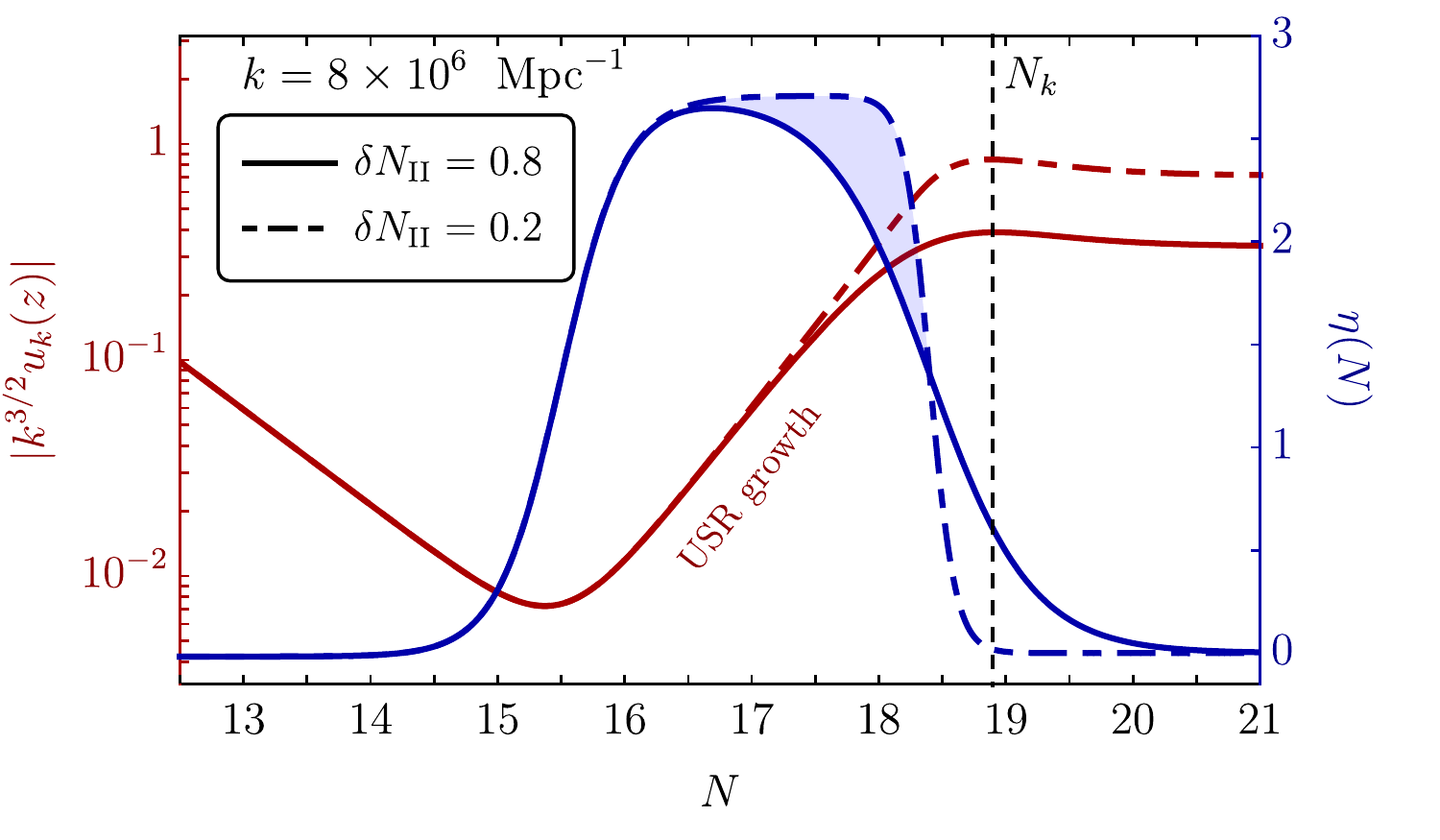}
~\includegraphics[width=.495\textwidth]{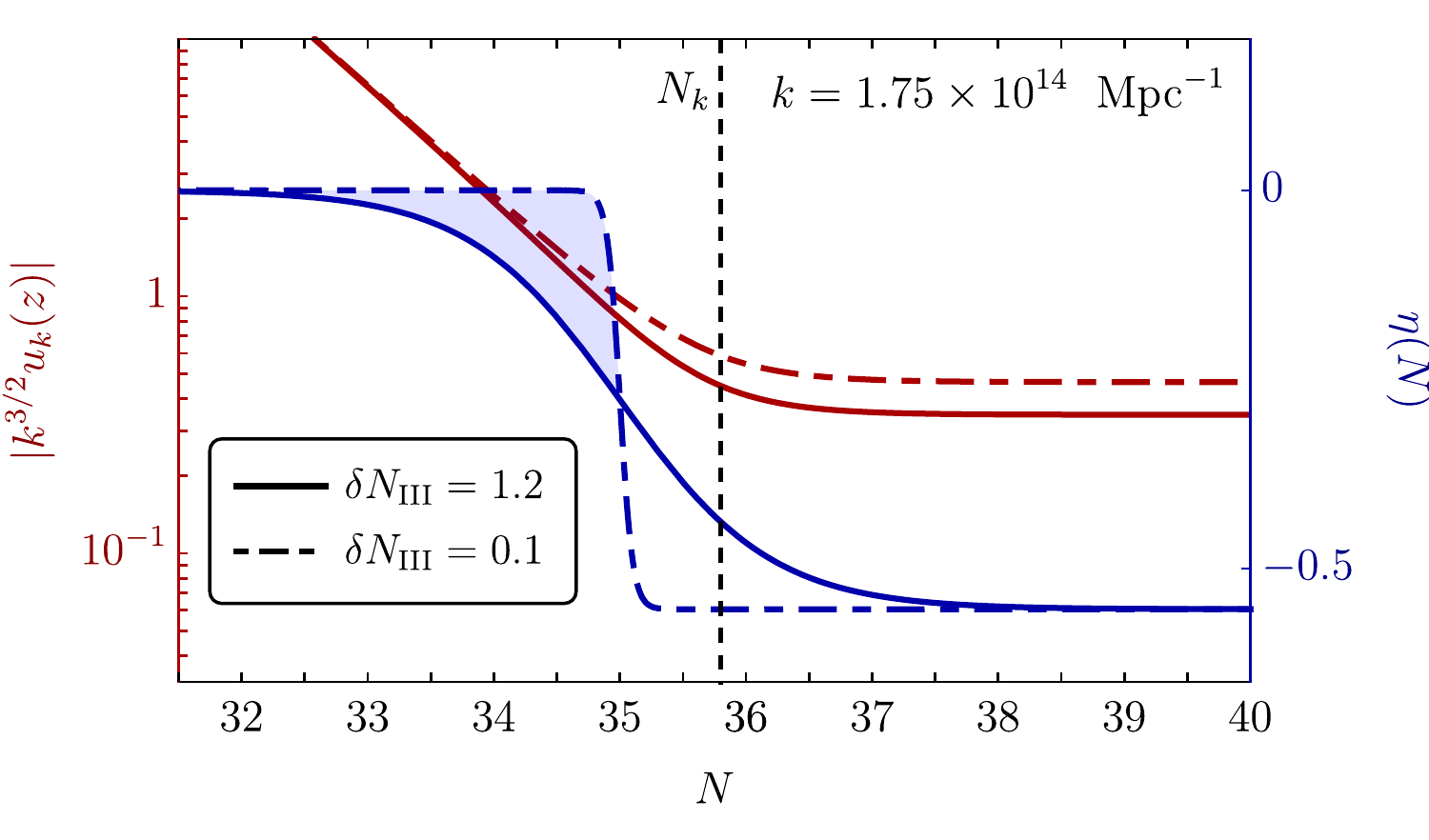}$$
\caption{
Left panel. We plot the time evolution of the quantity 
$k^{3/2}\left|u_k/z\right|$ for $k=8\times 10^{6}$ Mpc$^{-1}$ for two different values of $\delta N_{\rm II}$ (left-side $y$-axis, lines in red). All other free parameters 
are fixed to the values collected in the second column of table\,\ref{eq:ModelTab}. 
We superimpose the time evolution of the Hubble parameter $\eta$ for the same two values of $\delta N_{\rm II}$ (right-side $y$-axis, lines in blue). The vertical dashed line marks the horizon crossing time while    (from table\,\ref{eq:ModelTab}) we have $N_{\rm II} = N_{\rm I} + \Delta N_{\rm USR} = 17.9$. 
In the region shaded in blue 
we have that $\eta(N)$ for $\delta N_{\rm II} = 0.2$ is larger than  
$\eta(N)$ for $\delta N_{\rm II} = 0.8$;
this region highlights the difference between the two choices of $\delta N_{\rm II}$ in terms of the evolution of $\eta(N)$: the USR phase lasts for a slightly longer time if we consider smaller $\delta N_{\rm II}$ (that is a sharper transition at $N_{\rm II}$). 
Right panel. Same as in the left panel but for  
$k=1.75\times 10^{14}$ Mpc$^{-1}$ and for two different values of $\delta N_{\rm III}$ (with all other free parameters 
kept fixed according to the second column of table\,\ref{eq:ModelTab}).
In the region shaded in blue we have that $\eta(N)$ for $\delta N_{\rm III} = 0.1$ is larger than  
$\eta(N)$ for $\delta N_{\rm III} = 1.2$; 
this region highlights the difference between the two choices of $\delta N_{\rm III}$ in terms of the evolution of $\eta(N)$: 
the mode experiences, before horizon crossing, a stronger exponential suppression for increasing 
$\delta N_{\rm III}$, as discussed in eq.\,(\ref{eq:CorrectedEvo}).
 }\label{fig:DeltaNBump} 
\end{center}
\end{figure}
Let us give a closer look at the last point.
As discussed in the main text, the dependence on the parameters 
$\delta N_{\rm II}$ and $\delta N_{\rm III}$ is an important result from a phenomenological point of view since the bumps at the left- and right-side edges of the plateau directly control the abundance of, respectively, solar- and asteroid-mass PBHs. 
It is, therefore, natural to ask what is the physical origin of the effect that we described in fig.\,\ref{fig:DeltaN}. 
To answer this question, it is instructive to consider the dynamics of individual modes.

\subsubsection{Variation of $\delta N_{\rm II}$}\label{sec:DeltaNII}

    We focus on the left panel of fig.\,\ref{fig:DeltaN}, and---for definiteness---consider the evolution of the mode 
    with $k=8\times 10^{6}$ Mpc$^{-1}$ (black arrow). The contribution of this mode to the power spectrum, as shown in the left panel of fig.\,\ref{fig:DeltaN}, is enhanced (suppressed) for a sharper (smoother) transition at $N=N_{\rm II}$. 
    We show the time evolution of this mode, both for 
    $\delta N_{\rm II} = 0.2$ and $\delta N_{\rm II} = 0.8$, 
    in the left panel of fig.\,\ref{fig:DeltaNBump} (left-side of the plot, lines in red). We superimpose the time evolution of the Hubble parameter $\eta(N)$ (right-side of the plot, lines in blue). 
    We note that this mode (as well as the other modes that form the left-side edge of the plateau) crosses the Hubble horizon right after the end of the USR phase. 
    As explained in the previous section, during the USR phase the mode 
    gets exponentially enhanced. The key point is that 
    the amount of USR depends on the sharpness of the transition 
    at $N=N_{\rm II}$. As evident in the left panel of fig.\,\ref{fig:DeltaN}, 
    a very sharp transition (like in the case with $\delta N_{\rm II} = 0.2$) gives to the same mode more time to exponentially grow. 
    This is highlighted by the region shaded in blue in the left panel of fig.\,\ref{fig:DeltaN}. 
    In the case with $\delta N_{\rm II} = 0.2$ the mode has more time to grow before horizon crossing and, if compared with the evolution of the same mode but in the case of a smoother transition ($\delta N_{\rm II} = 0.8$), it settles to a higher final value. 
    This is the reason why the bump at the left-side edge of the plateau stands out more and more as one takes decreasing values of 
    $\delta N_{\rm II}$.
    
    Before proceeding, there is one more point 
    that is worth discussing. 
    As evident from the left panel of fig.\,\ref{fig:DeltaNBump}, the bump only concerns modes that cross the horizon right after the transition time $N_{\rm II}$. Plateau modes, that is modes that cross the horizon deeper during the $\eta_{\rm III} = 0$ phase, are not sensitive on the specific value of $\delta N_{\rm II}$. The reason is illustrated  in the left panel of fig.\,\ref{fig:DeltaNBump2}. In this figure we plot the dynamics of one of the modes that contribute to the plateau. For definiteness, we take $k=10^8$ Mpc$^{-1}$. 
    This mode crosses the horizon at time $N_k > N_{\rm II}$ when the value of $\eta$, for both choices $\delta N_{\rm II} = 0.2$ and $\delta N_{\rm II} = 0.8$, eventually settled to the value 
    $\eta_{\rm III} = 0$. After the end of the USR phase and before crossing the horizon at time $N_k$, the mode exponentially decays according to the scaling 
        \begin{align}
k^{3/2}\left|\frac{u_k(N)}{z(N)}\right| \sim 
e^{-(1 - \eta_{\rm III})N}\,.\label{eq:TransitionScaling}
    \end{align}
What happens is that if we take the case of a smooth transition 
the value of $\eta_{\rm III}$ is not exactly equal to zero after $N>N_{\rm II}$ but, since the $\tanh$ function has a sizable width, 
it transits through a phase in which $\eta_{\rm III} > 0$. Consequently, the mode has a slower exponential decay compared to the case of a sharp transition in which we have, from eq.\,(\ref{eq:TransitionScaling}), 
the scaling $e^{-N}$ immediately after $N_{\rm II}$. 
Because of the symmetry of the $\tanh$ function, 
the slower exponential decay for $N>N_{\rm II}$ precisely compensate the exponential growth for $N < N_{\rm II}$ so that, independently on $\delta N_{\rm II}$, the final value of the mode after its horizon crossing will be the same. This compensation 
is evident in the numerical result displayed in the left panel of fig.\,\ref{fig:DeltaNBump2}. 
Importantly, this compensation works only for modes that exit the horizon at times $N_k$ after that the transition from $\eta_{\rm II}$ to $\eta_{\rm III}$ is completed (so that they can experience while sub-horizon both sides of the $\tanh$ transition at $N_{\rm II}$).

\subsubsection{Variation of $\delta N_{\rm III}$}

    We focus on the right panel of fig.\,\ref{fig:DeltaN}, and---for definiteness---consider the evolution of the mode 
    with $k=1.75\times 10^{14}$ Mpc$^{-1}$ (black arrow). The contribution of this mode to the power spectrum, as shown in the right panel of fig.\,\ref{fig:DeltaN}, is enhanced (suppressed) for a sharper (smoother) transition at $N=N_{\rm III}$. 
    We show the time evolution of this mode, both for 
    $\delta N_{\rm III} = 0.1$ and $\delta N_{\rm III} = 1.2$, 
    in the right panel of fig.\,\ref{fig:DeltaNBump} (left-side of the plot, lines in red). We superimpose the time evolution of the Hubble parameter $\eta(N)$ (right-side of the plot, lines in blue). 
    We note that this mode (as well as the other modes that form the right-side edge of the plateau) crosses the Hubble horizon right after the transition at time $N_{\rm III}$. The key point is the following. As discusses in the previous section, during its
    sub-Hubble evolution at times $N<N_k$, the mode evolves as
    \begin{align}
k^{3/2}\left|\frac{u_k(N)}{z(N)}\right| \sim 
e^{-(1 - \eta_{\rm III})N}\,,
    \end{align}
which is the same time-dependence discussed in eq.\,(\ref{eq:FinalScaling}) but with  $\eta_{\rm III}$ explicitly written. 
If we consider the case of a very smooth transition, from the $\eta(N)$ evolution displayed in right panel of fig.\,\ref{fig:DeltaNBump} we see that the mode experiences a non-zero value of $\eta_{\rm III} < 0$ already before the transition time at $N_{\rm III}$ while in the case of a sharper transition stays closer to $\eta_{\rm III} = 0$ for longer time. This is highlighted by the region shaded in blue in right panel of fig.\,\ref{fig:DeltaNBump}. 
Consequently, in the case $\delta N_{\rm III} = 1.2$ (smoother transition) the mode, before horizon crossing and for $N<N_{\rm III}$, experiences a short phase
during which it evolves as
    \begin{align}
k^{3/2}\left|\frac{u_k(N)}{z(N)}\right| \sim 
e^{-(1 + |\eta_{\rm III}|)N}\,,\label{eq:CorrectedEvo}
    \end{align}
    with $\eta_{\rm III} < 0$ non-zero and negative because of 
    the effect of the transition region. 
    The mode, therefore, undergoes a phase of exponential suppression that is slightly faster compared with the case of a sharper transition (for which $\eta_{\rm III}$ remains closer to zero until the actual transition at $N=N_{\rm III}$). 
    This is evident in right panel of fig.\,\ref{fig:DeltaNBump}: 
    in the case with $\delta N_{\rm III} = 1.2$ the mode is more suppressed and, if compared with the evolution of the same mode but in the case of a smoother transition ($\delta N_{\rm III} = 0.1$), it settles to a lower final value. 
   This is the reason why the bump at the right-side edge of the plateau becomes smoother and smoother as we increase the value of  
    $\delta N_{\rm III}$.   
    
\begin{figure}[t]
\begin{center}
$$\includegraphics[width=.5\textwidth]{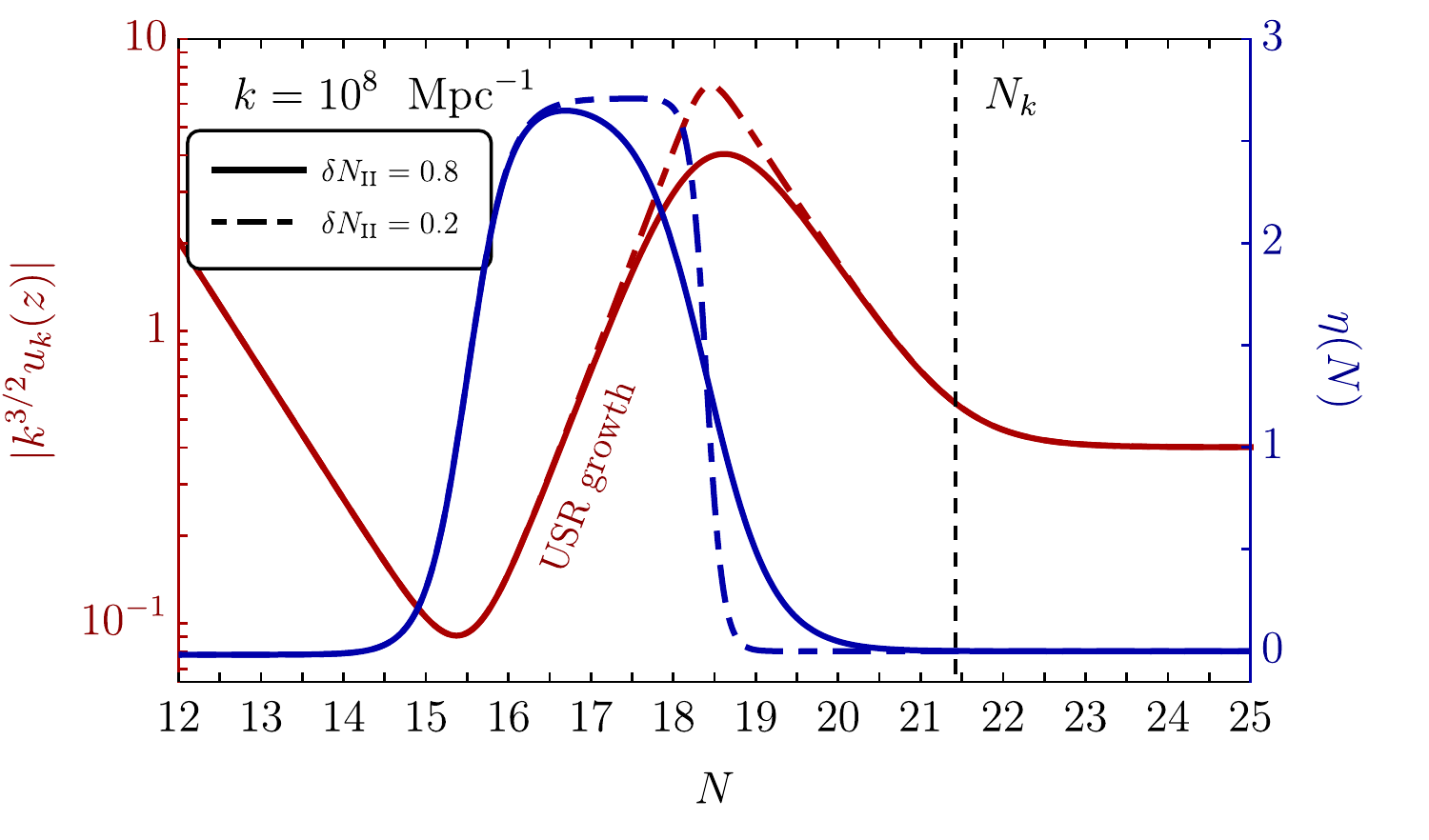}
~~\includegraphics[width=.5\textwidth]{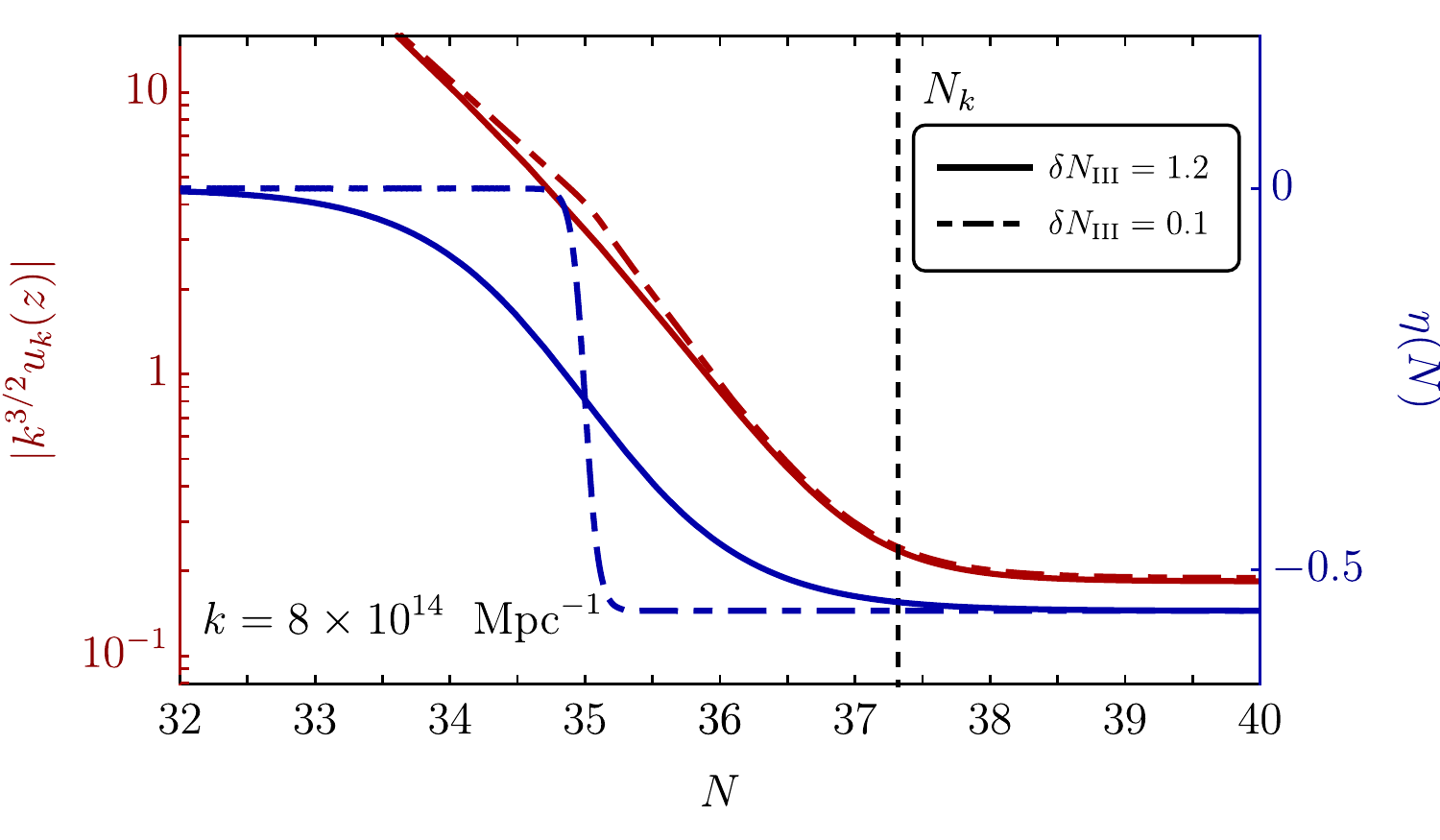}$$
\caption{
Left panel.
Same as in the left panel of fig.\,\ref{fig:DeltaNBump} but for the time evolution of the plateau mode with $k=10^8$ Mpc$^{-1}$. Right panel. 
Same as in the right panel of fig.\,\ref{fig:DeltaNBump} but for the time evolution of the mode with $k=2\times 10^{14}$ Mpc$^{-1}$.
 }\label{fig:DeltaNBump2} 
\end{center}
\end{figure}

    We note that this effect is again limited to those modes that exit the horizon right after the transition at 
    $N = N_{\rm III}$. 
    Modes that exit the horizon before the transition 
    time $N_{\rm III}$ (like the plateau modes) are already super-horizon, and, therefore, conserved, at time $N_{\rm III}$; modes that cross the Hubble horizon well after the transition time $N_{\rm III}$ (more specifically, after that $\eta$ completed the transition from $\eta_{\rm III}$ to $\eta_{\rm IV}$) experience a compensating effect that is completely analogue to the one discussed before at time $N_{\rm II}$. This is shown in the  right panel 
    of fig.\,\ref{fig:DeltaNBump2} for the mode with $k=2\times 10^{14}$ Mpc$^{-1}$. 
    Consider the smooth transition with $\delta N_{\rm III} = 0.8$ (solid line). The faster exponential decrease right before $N = N_{\rm III}$ is compensated by a slower exponential decrease right after the transition so that the two effects compensate at horizon crossing. 

\begin{figure}[t]
\begin{center}
$$\includegraphics[width=.495\textwidth]{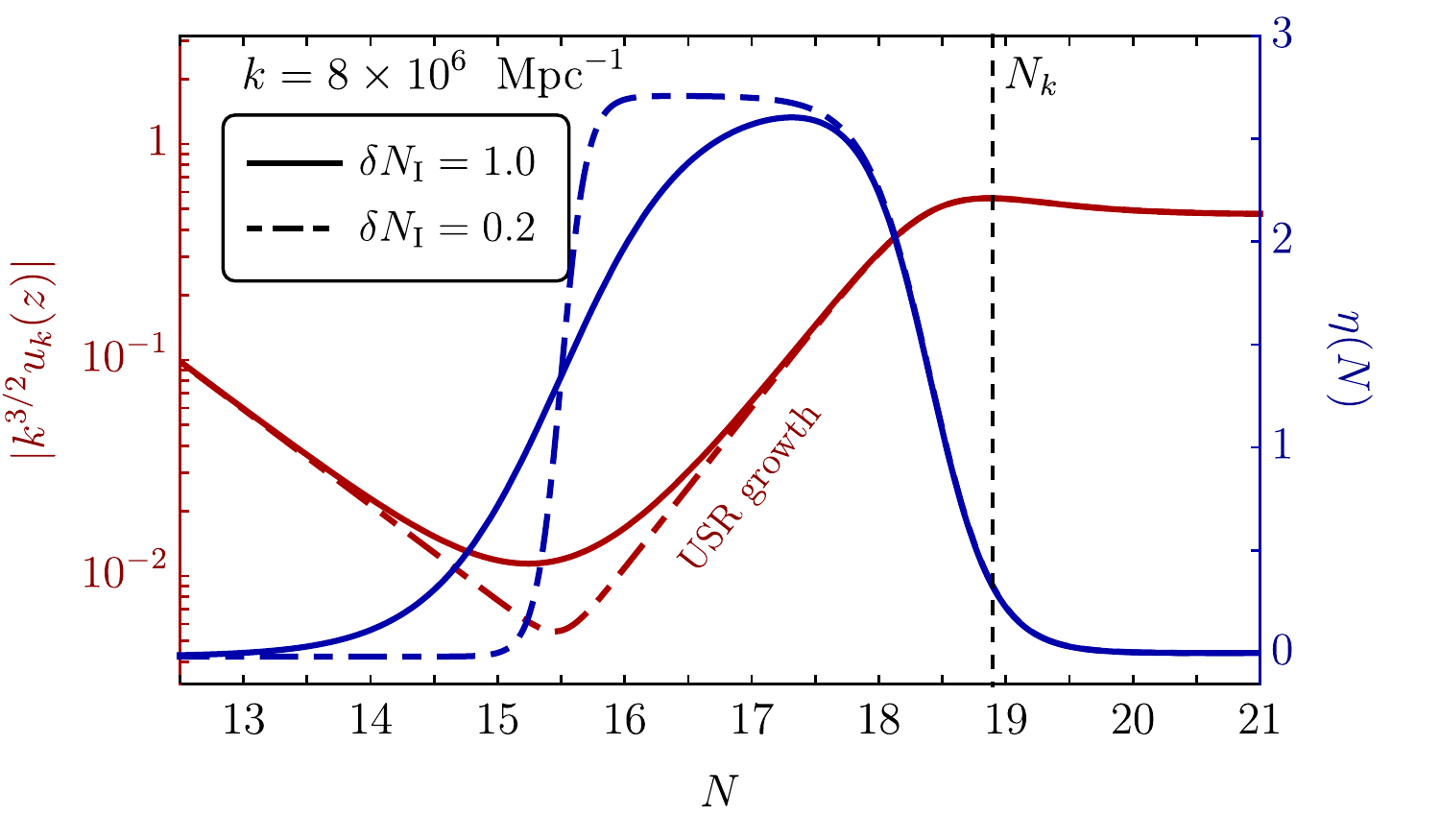}
~\includegraphics[width=.495\textwidth]{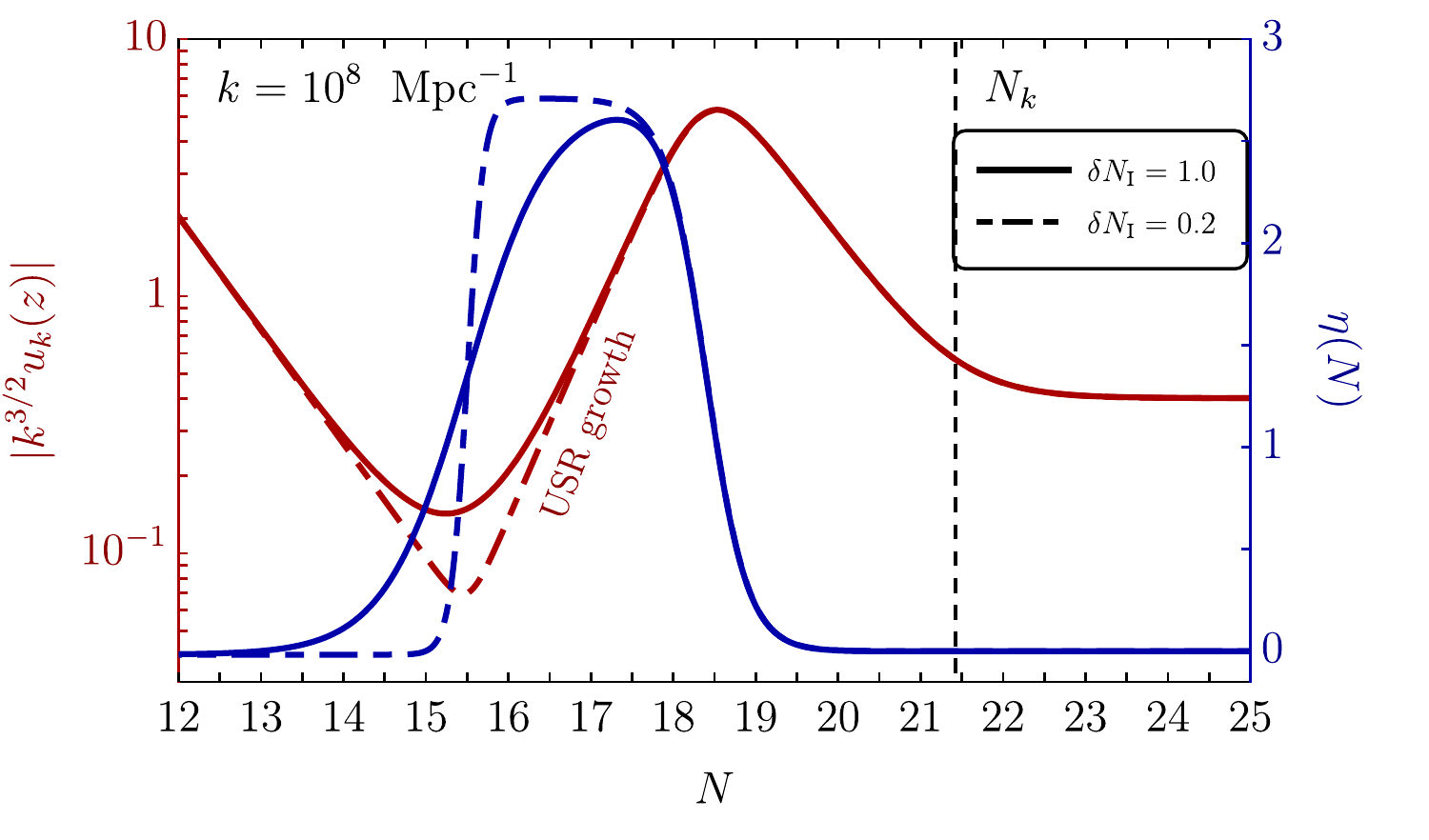}$$
\caption{
Same as in fig.\,\ref{fig:DeltaNBump} but now with $\delta N_{\rm II} = \delta N_{\rm III} = 0.5$ fixed while we vary   
$\delta N_{\rm I} = 0.2$ (dot-dashed lines) and
$\delta N_{\rm I} = 1$ (solid lines). 
Left panel. Time evolution of the mode with $k= 8\times 10^6$ Mpc$^{-1}$ (one of the modes that contribute to the bump-like feature at the left-side edge of the plateau). 
Right panel. Time evolution of one of the plateau modes with  $k=10^8$ Mpc$^{-1}$.
 }\label{fig:DeltaFirstTransition} 
\end{center}
\end{figure}

\subsubsection{Variation of $\delta N_{\rm I}$}

Before concluding this section, we quickly comment about the dependence on the parameter $\delta N_{\rm I}$ which controls the sharpness of the first transition at $N=N_{\rm I}$.  
 The point we want to make is that changing this parameter does not alter neither the bump-like feature at the left-side edge of the plateau  nor the subsequent plateau. 
 The reason is summarized in fig.\,\ref{fig:DeltaFirstTransition}. 
 We fix $\delta N_{\rm II} = \delta N_{\rm III} = 0.5$ and consider two cases with
 $\delta N_{\rm I} = 0.2$ and 
 $\delta N_{\rm I} = 1$. 
 In the left panel, we plot the time evolution of the mode with $k=2\times 10^6$ Mpc$^{-1}$. This mode crosses the Hubble horizon right after the end of the USR phase, and contributes to the bump-like feature at the left-edge of the plateau. 
 Changing $\delta N_{\rm I}$ does not 
 alter the final value of the mode because we observe, in the case of a smoother transition ($\delta N_{\rm I} = 1$, solid lines), a compensation between a slower exponential decay right before $N_{\rm I}$ and the subsequent faster exponential growth right after $N_{\rm I}$. 
 In the right panel, we plot the time evolution of the plateau mode with $k= 10^8$ Mpc$^{-1}$. In analogy to the previous discussion, the compensation 
 right before and after $N_{\rm I}$ cancels out any difference between the final conserved value of the modes if 
 $\delta N_{\rm I} = 0.2$ and 
 $\delta N_{\rm I} = 1$ are compared. 
 Motivated by this analysis, 
 in the explicit realizations of our model (see table\,\ref{eq:ModelTab}) we fix $\delta N_{\rm I} = 0.5$.

In conclusion, we showed how the features at both ends of the plateau of curvature perturbations are directly controlled by $\delta N_{\rm II}$ and $\delta N_{\rm III}$ and can be simply interpreted in terms of the dynamics of the perturbation modes.

%%%%%%%%%%%%%%%%%%%%%%%%%%%%%%%%%%%%%%%%%%%%%%%%
\section{The abundance of PBHs}
\label{sec:fPBH}
%\label{sec:results}
%%%%%%%%%%%%%%%%%%%%%%%%%%%%%%%%%%%%%%%%%%%%%%%%

In this section we review how one can compute the abundance of PBHs. Here we adopt the formalism developed in ref.\,\cite{inprepQCD} 
that include the dependence on the equation of state, which deviates from perfect radiation around the QCD epoch when PBHs of around the solar mass are formed. 
We assume that the universe was dominated by relativistic particles at energies higher than what currently included in the standard model of particle physics, leading to a perfect radiation fluid dominating the universe above the electro-weak scale. 
We mention, however, that a different equation of state may be possible, implying modifications of the PBH formation \cite{Khlopov:1980mg,Green:1997pr,Musco:2012au,Harada:2016mhb,Carr:2017edp,Carr:2018nkm,Escriva:2020tak,deJong:2021bbo,DeLuca:2021pls} and induced SGWB \cite{Inomata:2019ivs,Inomata:2019zqy,Domenech:2019quo,Domenech:2020kqm,Hook:2020phx} discussed in the next section.

After matter-radiation equality the dark matter fraction consisting of
PBHs can be expressed as
\begin{align}\label{eq:fPBH}
\Omega_{\rm PBH} =
\int d \log M_H 
% \frac{1}{\Omega_{\rm CDM}}
\left(
\frac{M_{\rm eq}}{M_H}
\right)^{1/2}\beta(M_H)
\,,~~~~~~~~
f_{\rm PBH}(M_{\rm PBH}) = 
\frac{1}{\Omega_{\rm CDM}}\frac{d\Omega_{\rm PBH}}{d\log M_{\rm PBH}}\,,
\end{align}
where 
 $M_H$ is the horizon mass at the time of horizon re-entry, 
 $M_{\rm eq} \simeq 3 \times 10^{17}$ $M_{\odot}$ the horizon mass at matter-radiation equality,
and $\Omega_{\rm CDM}$ is the cold dark matter density of the Universe 
($\Omega_{\rm CDM} \simeq 0.12\,h^{-2}$, with $h = 0.674$ for the Hubble parameter). 
The approximate relation between the horizon mass $M_H$ and comoving wavenumber $k_H$ is given by 
\begin{align}\label{eq:MHk}
M_H \simeq  17 \times \left(\frac{g_*}{10.75}\right)^{-1/6} \left(
\frac{k_H}{10^6\,{\rm Mpc}^{-1}}
\right)^{-2}\,M_{\odot}\,,
\end{align}
where $g_*$ is the number of degrees of freedom of relativistic particles
with $g_* = 106.75$ deep in the radiation epoch. We include the temperature dependence of $g_*$ following ref.\,\cite{Saikawa:2018rcs}.\footnote{We convert the temperature dependence into a functional dependence on the horizon mass $M_H$ by means of the relation 
$M_H\simeq 1.5\times 10^5 (g_*/10.75)^{-1/2}(T/\textrm{MeV})^{-2}\,M_{\odot}$.}
 The mass of the resulting PBH is given by\,\cite{Young:2019yug}
 \begin{align}\label{eq:MainNL}
    M_{\rm PBH} = \mathcal{K} M_H\left[
    \left(\delta_{\rm L} - \frac{1}{4\Phi}\delta_{\rm L}^2\right) - \delta_c 
    \right]^{\gamma}\,.
  \end{align}
 Eq.\,(\ref{eq:MainNL}) automatically takes into account the non-linear relation 
between curvature perturbations and the density  contrast field $\delta$  \cite{DeLuca:2019qsy,Young:2019yug}.  
More concretely, $\delta_{\rm L}$ represents the linear Gaussian component of the  density  contrast field while 
$\delta_c$ is the threshold value for gravitational collapse that refers to the full density  contrast field.  
  In full generality, $\mathcal{K}(M_H)$, $\gamma(M_H)$, $\Phi(M_H)$ and $\delta_c(M_H)$ are functions of the horizon mass.
During the radiation epoch,
  $\mathcal{K}$ typically takes a value between 3 and 5 for perturbations produced by a nearly scale-invariant spectrum \cite{Germani:2018jgr,Escriva:2021pmf}, $\gamma \simeq 0.36$, $\delta_c \simeq 0.56$ and  $\Phi = 2/3$ for a radiation fluid with equation of state parameter $\omega = p/\rho = 1/3$. In our analysis, we include the full $M_H$ dependence of the above quantities following refs.\,\cite{inprepQCD,Muscoinprep}.
  This is an important point since the equation of state parameter $\omega$ reduces by around 30\% and the critical threshold $\delta_c$ decreases by around 10\% 
during the QCD phase
transition \cite{Muscoinprep}.
This leads to a boost in the PBH mass distribution by at least two orders of
magnitude compared to a Universe in which the equation of state parameter remains constantly equal to that of radiation,
$\omega = 1/3$.

The expression for $\beta$ in eq.\,(\ref{eq:fPBH}) accounts for the fraction of each Hubble volume which collapses
to form a PBH. 
Assuming threshold statistics, we have
\begin{align}
\beta(M_H) & = 
\int_{\delta_c}^{\infty}\frac{M_{\rm PBH}}{M_H}P(\delta)d\delta = 
\mathcal{K}
\int_{\delta_{\rm L}^{\rm min}}^{\delta_{\rm L}^{\rm max}}
\left(\delta_{\rm L} - \frac{1}{4\Phi}\delta_{\rm L}^2 - \delta_c\right)^{\gamma}
P_{\rm G}(\delta_{\rm L})d\delta_{\rm L}\,,\label{eq:GaussianTerm1}\\
P_{\rm G}(\delta_{\rm L}) & = \frac{1}{\sqrt{2\pi}\sigma(R_H)}e^{-\delta_{\rm L}^2/2\sigma(R_H)^2}\,,
\label{eq:GaussianTerm}
\end{align}
where in eq.\,(\ref{eq:GaussianTerm1}) we changed variable from the full density contrast to its linear (hence gaussian) component. 
The characteristic size of perturbations is identified by the scale $r_m$ where the maximum of the mass excess (or compaction function) is found \cite{Musco:2018rwt} and it is larger then the inverse of the comoving spectral number $k$. For nearly scale invariant spectra, the two are related by the condition $r_m k \equiv \kappa \simeq 4.49$ \cite{Musco:2020jjb}.
The peak of the compaction function sets the corresponding horizon crossing $r_m =1/aH \equiv R_H $, where $R_H$ is the comoving Hubble radius; 
its relation with $M_H$ can be read from eq.\,(\ref{eq:MHk}) 
at the time of horizon re-entry $k_H = 1/r_m$.
The extrema of integrations in eq.\,(\ref{eq:GaussianTerm1}) are 
\begin{align}
\delta_{\rm L}^{\rm min} = 2\Phi\left(
1 - \sqrt{1-\frac{\delta_c}{\Phi}}\,
\right)\,,~~~~~~~~~\delta_{\rm L}^{\rm max} = 2\Phi\,.
\end{align}
The variance that enters in eq.\,(\ref{eq:GaussianTerm}) refers to the linear component of the density contrast and 
can be computed by integrating the power spectrum of curvature perturbations 
\begin{align}\label{eq:Sigma}
\sigma^2(R_H) = \frac{4}{9}\Phi^2
\int_{0}^{\infty}(k R_H)^4W^2(k R_H)T^2(k R_H) P_{\mathcal{R}}(k)
d \log k\,.
\end{align}
In eq.\,(\ref{eq:Sigma})
 we include  the Fourier transform of the top-hat window
function in real space $W(k R_H)$ (used to smooth the field over a finite volume) and the linear transfer function $T(k R_H)$ (which describes the damping of
perturbations on sub-horizon scales). We use\footnote{It should the noted that 
the transfer function in eq.\,(\ref{eq:WeT}) is strictly valid in a radiation-dominated phase.}
\begin{align}\label{eq:WeT}
W(y)  = 3\bigg[
\frac{\sin(y) - y\cos(y)}{y^3}\bigg] \,,
~~~~~~~~T(y)  = 3\bigg[
\frac{\sin(y/\sqrt{3}) - (y/\sqrt{3})\cos(y/\sqrt{3})}{(y/\sqrt{3})^3}\bigg]\,.
\end{align}
Following ref.\,\cite{Byrnes:2018clq}, we make another change of variables  
from $\delta_{\rm L}$ to $M_{\rm PBH}$ by inverting eq.\,(\ref{eq:MainNL}). 
We arrive at the final formula\footnote{It should be noted that ref.\,\cite{DeLuca:2020agl} computes the 
abundance of PBHs in the gaussian approximation, and includes the effect of non-linearities by means of a final rescaling 
of the amplitude of the power spectrum by a factor of $2$ (following the prescription given in refs.\,\cite{DeLuca:2019qsy,Young:2019yug}).}
\begin{align}\label{eq:massfunctionintegral}
&f_{\rm PBH}(M_{\rm PBH}) = \frac{1}{\Omega_{\rm CDM}}\int_{M_H^{\rm min}}^{\infty}
\left(
\frac{M_{\rm eq}}{M_H}
\right)^{1/2} 
\frac{
e^{-\frac{8}{9\sigma(R_H)^2}\left[
1 - 
\sqrt{\Lambda
% 1 - \frac{3}{2}\delta_c(M_H) - \frac{3}{2}\left(\frac{M_{\rm PBH}}{\mathcal{K}M_H}\right)^{1/\gamma}
}\,
\right]
^2}
}{
\sqrt{2\pi}\sigma(R_H)
\Lambda
%\left[ 1-\frac{3}{2}\delta_c(M_H) - \frac{3}{2}\left(\frac{M_{\rm PBH}}{\mathcal{K}M_H}\right)^{1/\gamma} \right]
^{1/2}}\left(
\frac{M_{\rm PBH}}{\gamma M_H}
\right)\left(
\frac{M_{\rm PBH}}{\mathcal{K}M_H}
\right)^{1/\gamma}d\log M_H
\,,
\end{align}
where we conveniently defined
\begin{equation}
    \Lambda \equiv 1 - \frac{\delta_c}{\Phi} - \frac{1}{\Phi}\left(\frac{M_{\rm PBH}}{\mathcal{K}M_H}\right)^{1/\gamma}
\end{equation}
in which the right-hand side can be integrated numerically to give $f(M_{\rm PBH})$ for each value of the PBH mass.
The  lower limit of integration follows from the condition $\Lambda > 0$
 (notice this must be the case due to the term $\sqrt{\Lambda}$ appearing in eq.~\eqref{eq:massfunctionintegral}). 
As far as the numerical value of $\delta_c$ is concerned, it takes the value of $\delta_c = 0.56$ in a radiation-dominated universe in the case of a 
broad power spectrum of curvature perturbations and including the non-linear 
relation between curvature perturbations and the density contrast field.\footnote{Ref.\,\cite{DeLuca:2020agl} 
takes the value $\delta_c = 0.51$ which is the value that corresponds to the gaussian approximation, 
see ref.\,\cite{Germani:2018jgr}.}
It is additionally reduced and modulated when 
$M_H \approx M_\odot$ when the collapse takes place across the QCD epoch \cite{Muscoinprep}.

\begin{figure}[t]
\leavevmode
\centering
\includegraphics[width=0.65\textwidth]{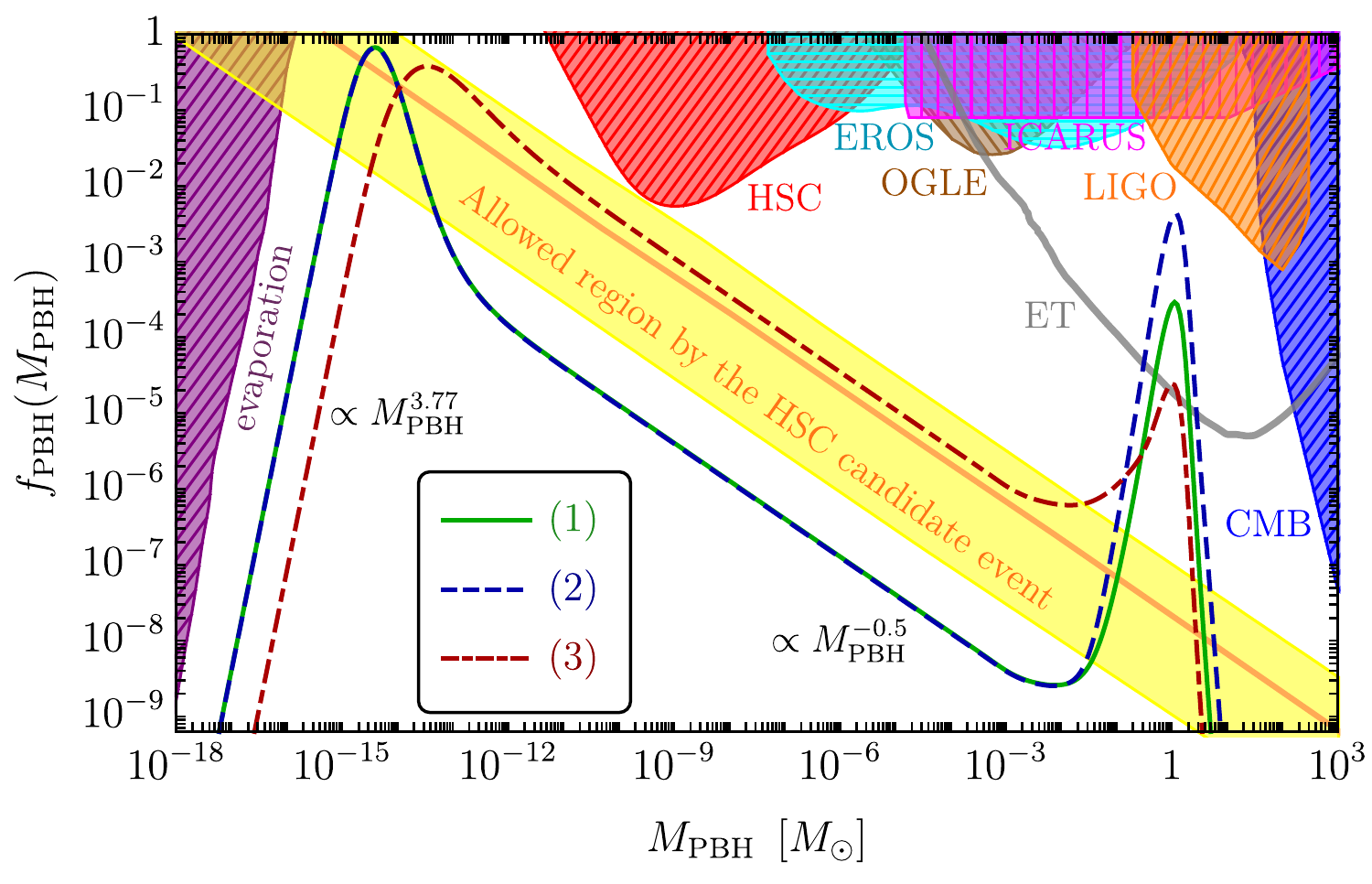}  
\caption{\label{fig:PBHAbundanceFull}
Fraction of DM in the form of PBHs with mass $M_{\rm PBH}$. 
We plot the most stringent constraints (meshed regions, cf. the SM) and 
the mass function resulting from three benchmark realizations of our model 
(labeled as 
{\color{verdes}{$(1)$}}, 
{\color{blue}{$(2)$}}, 
{\color{cornellred}{$(3)$}}, see table\,\ref{eq:ModelTab}). 
The yellow band 
corresponds to the allowed region for a PBH mass function $\propto M_{\rm PBH}^{-0.5}$ consistent with the Subaru Hyper Suprime-Cam (HSC)
microlensing candidate event\,\cite{Niikura:2017zjd} (see also \cite{Sugiyama:2020roc}). 
The gray line indicates the minimum PBH abundance 
required to have at least one PBH merger event per year at the
Einstein Telescope (ET) experiment, see ref.\,\cite{DeLuca:2021hde}.
%assuming Poisson statistics, neglecting PBH %clustering and accretion, and for a %monochromatic PBH mass distribution
}
\end{figure}

The parameters of the dynamics are chosen in such a way that the integral 
\begin{align}
f_{\rm PBH} \equiv \frac{\Omega_{\rm PBH}}{\Omega_{\rm CDM}} = \int f_{\rm PBH}(M_{\rm PBH}) d\log M_{\rm PBH} \approx 1\,,
\end{align}
which means that we get $\approx 100\%$ of DM in the form of PBHs.
In fig.\,\ref{fig:PBHAbundanceFull} we show the following constraints (see ref.\,\cite{Green:2020jor} for a review and\,\href{github.com/bradkav/PBHbounds}{\faGithub/bradkav/PBHbounds}).
Envelope of evaporation constraints (see also \cite{Saha:2021pqf,Laha:2019ssq,Ray:2021mxu}): EDGES\,\cite{Mittal:2021egv}, 
CMB\,\cite{Clark:2016nst}, INTEGRAL\,\cite{Laha:2020ivk,Berteaud:2022tws}, 511 keV\,\cite{DeRocco:2019fjq}, Voyager\,\cite{Boudaud:2018hqb}, 
EGRB\,\cite{Carr:2009jm};
microlensing constraints from the Hyper-Supreme Cam (HSC), ref.\,\cite{Niikura:2017zjd}; 
microlensing constraints from EROS, ref.\,\cite{EROS-2:2006ryy}; 
microlensing constraints from OGLE, ref.\,\cite{Niikura:2019kqi}; 
Icarus microlensing event, ref.\,\cite{Oguri:2017ock}; 
constraints from modification of the CMB spectrum due to accreting PBHs, ref.\,\cite{Serpico:2020ehh};
direct constraints on PBH-PBH mergers with LIGO, refs.\,\cite{LIGOScientific:2019kan,Kavanagh:2018ggo} (see also \cite{Wong:2020yig,Hutsi:2020sol,DeLuca:2021wjr,Franciolini:2021tla}).

Recently, it was suggested that observations of Sun-like stars in dwarf galaxies may constrain the PBH abundance  to be below $f_{\rm PBH} \lesssim 0.3$ in part of the asteroidal mass window \cite{Esser:2022owk}, i.e. for masses $M_{\rm PBH}\lesssim 10^{-12} M_\odot$.
Similar constraints were set in the past by studying neutron stars and white dwarfs in DM-rich environments like dwarf galaxies \cite{Capela:2013yf,Capela:2012jz,Capela:2014ita} (but see ref.\,\cite{Montero-Camacho:2019jte} for a discussion on their validity),
for which no direct observations exist to date.
On the contrary, ref.\,\cite{Esser:2022owk} focuses on main sequence stars. The newly derived bound, however, requires {\it assuming} a maximum allowed fraction $\xi$ of disrupted stars that can be compatible with current observations, given the lack of precise modelling of the initial number. 
${\cal O}(1)$ differences on $\xi$ may completely relax this bound \cite{Esser:2022owk}. 
Therefore, we decided not to report it in fig.~\,\ref{fig:PBHAbundanceFull}.
We conclude by pointing out that, even taking at face value the bound that follows from the assumptions made in ref.\,\cite{Esser:2022owk}, it would still be possible to tune the asteroidal mass peak in fig.~\ref{fig:PBHAbundanceFull} in order to evade the constraint with small modifications of the parameters reported in Tab.~\ref{eq:ModelTab}.

The resulting mass distribution is shown in fig.\,\ref{fig:PBHAbundanceFull} for three benchmark realizations of our model, all of which are chosen to reproduce all the DM in the form of PBHs ($f_{\rm PBH} \overset{!}{=}  1$). 
The values of the parameters are collected in table\,\ref{eq:ModelTab}.

The first noticeable feature of the resulting mass distribution
$f_{\rm PBH}(M_{\rm PBH})$
is the characteristic scaling $\propto M_{\rm PBH}^{-1/2}$ for masses produced by modes in the enhanced plateau (see fig.\,\ref{fig:PBHAbundanceFull}). 
This is because nearly scale invariant power spectra induce the formation of PBHs of various masses with equal probability $\beta(M_H)$ \cite{MoradinezhadDizgah:2019wjf,DeLuca:2020ioi}, but smaller PBHs form earlier and their abundance is redshifted compared to heavier ones. This naturally induces a more prominent contribution to the DM from the light portion of the mass spectrum. 
Additionally, the smoothness of the transition between phases II$\to$III and III$\to$IV, controlled by the parameters $\delta N_{\rm II}$ and $\delta N_{\rm III}$ respectively, determines the spectral features at the sides of the enhanced plateau, which are magnified in the PBH abundance due to its exponential dependence on ${\cal P}_{\cal R}(k)$.
 In particular, as already discussed in the previous section, a sharper transition produces a more prominent oscillatory feature (see e.g. \cite{Dalianis:2021iig,Cole:2022xqc,Karam:2022nym}), whose main peak greatly enhances the relative PBH abundance at the corresponding mass.

 Focusing on the first transition, which corresponds to the formation of heavier (solar mass) PBHs, $\delta N_{\rm II}$ allows to boost $f_{\rm PBH}$ 
 (on top of the softening of the QCD equation of state, 
 whose only impact on the enhancement of the mass function can be measured by looking at case 
 \textbf{{\color{cornellred}{(3)}}}
 in fig.~\ref{fig:PBHAbundanceFull}), to a much larger value which may potentially produce observable PBH mergers at current and future ground-based GW experiments \cite{Clesse:2020ghq,Franciolini:2021tla,DeLuca:2021hde,Pujolas:2021yaw,Ng:2022agi,Martinelli:2022elq,inprepQCD}. 
 On the other hand, a smaller $\delta N_{\rm III}$ would induce a peak at small masses.
 In the absence of a bump at asteroidal masses (i.e. case {\color{cornellred}{$(3)$}}),  $f_{\rm PBH}(M_{\rm PBH})$ would be compatible with  the HSC detection of a candidate lens \cite{Niikura:2017zjd} (indicated in fig.\,\ref{fig:PBHAbundanceFull} with a yellow band).  On the other hand, for fixed abundance $f_{\rm PBH} = 1$, a more pronounced peak (i.e. {\color{verdes}{$(1)$}} and {\color{blue}{$(2)$}}) would decrease the amplitude of the whole tail $\propto M_{\rm PBH}^{-1/2}$,  potentially evading future HSC constraints \cite{Sugiyama:2020roc}.

\color{black}

\subsection{On the maximum mass of PBHs in USR scenarios} 

We now discuss the maximum mass of PBHs that can be generated within our model.
This is a delicate issue which is mostly related to the shape of the power spectrum at the left-side edge of the plateau. 
To make this point more clear, we start from some preliminary considerations. 

\begin{figure}[!t!]
\begin{center}
$$\includegraphics[width=.495\textwidth]{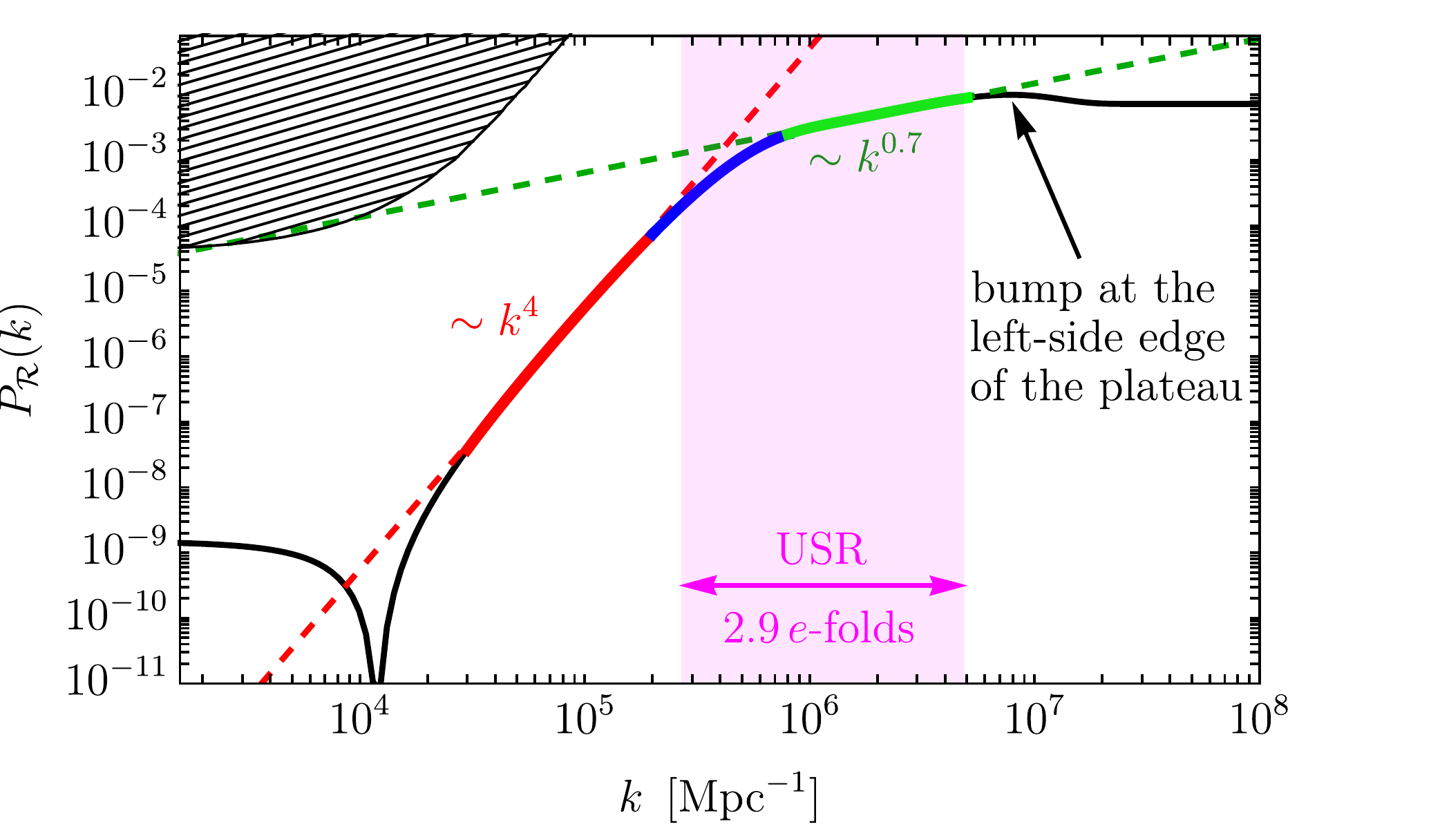}
~\includegraphics[width=.495\textwidth]{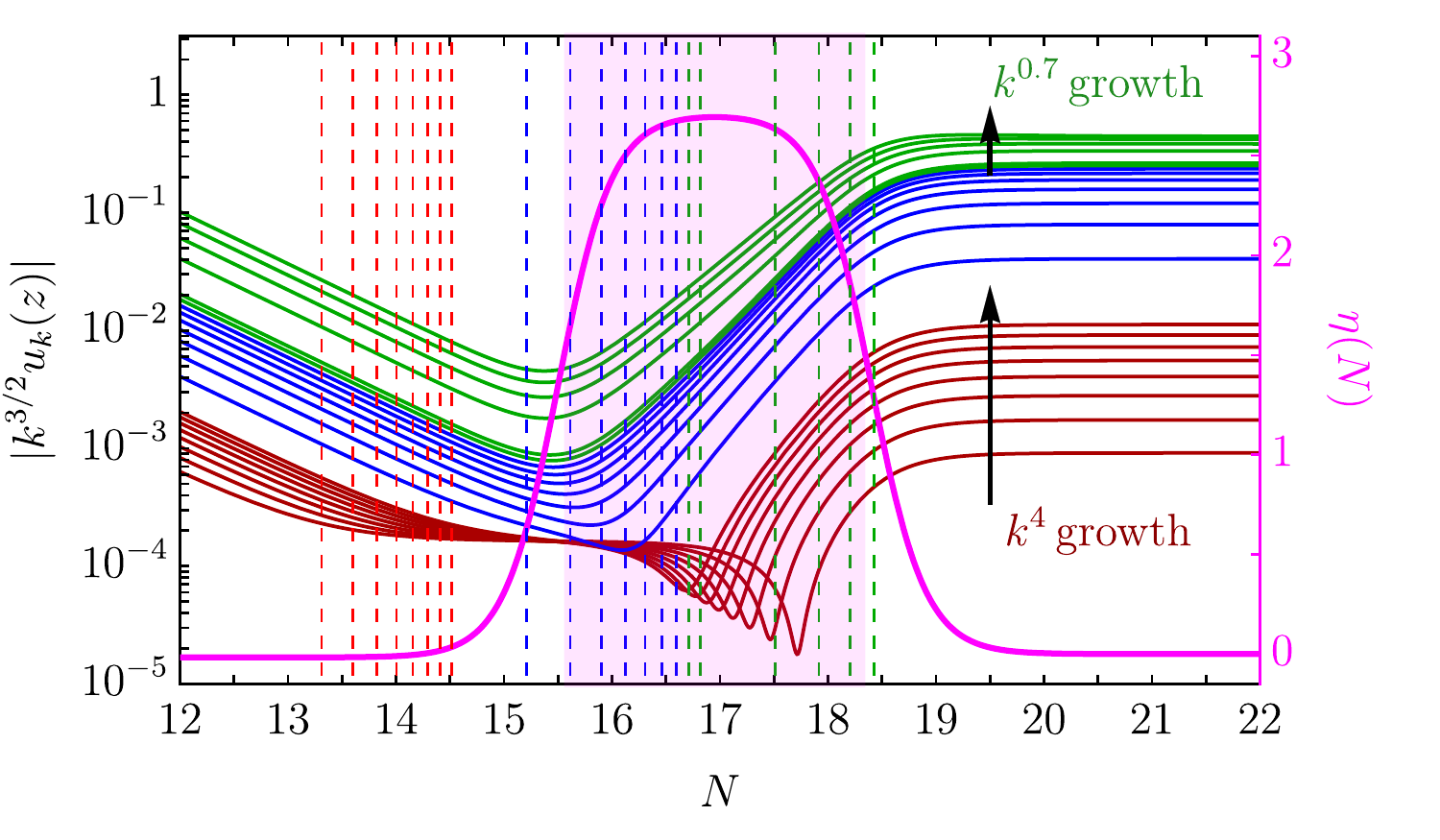}$$
\caption{
Left panel. Zoom in on the left-side edge of the plateau of the power spectrum in fig.\,\ref{fig:PS}. We highlight in magenta the values of $k$ such that the horizon crossing condition 
$k=a(N_k)H(N_k)$ is solved for $N_{\rm I} \leqslant N_k \leqslant N_{\rm II}$. 
Right panel. Dynamical evolution of the perturbation modes with $k$ in the red, blue and green part of power spectrum (cf. the use of different colors in the left panel: the red (green) part of the power spectrum corresponds to the $\sim k^{4}$ ($\sim k^{0.7}$) growth while the blue part lies in between). The vertical lines mark the horizon crossing time, and make clear that for the blue and green modes we have 
$N_{\rm I} \leqslant N_k \leqslant N_{\rm II}$.
 }\label{fig:Anatomy} 
\end{center}
\end{figure}

In the left panel of fig.\,\ref{fig:Anatomy} we zoom in on this part of the power spectrum. 
For definiteness, we consider the model dubbed ${\color{verdes}{(1)}}$ in table\,\ref{eq:ModelTab}.
Some comments are in order. 
First, the region shaded in magenta represents the interval of comoving wavenumber $k$ such that the horizon crossing condition $k=a(N_k)H(N_k)$ falls inside the time interval $N_{\rm I} \leqslant N_k \leqslant N_{\rm II}$. 
Second, we highlight in red the part of the power spectrum that features 
the power-law growth $P_{\mathcal{R}}(k)\sim k^4$. 
We note that this part of the power spectrum lies immediately before the region shaded in magenta; this suggests that the modes that contribute to the  
 $\sim k^4$ growth are those for which the horizon crossing condition happens right  before the beginning of the USR phase. 
 On the other hand, as already discussed at length in the previous section, the bump-like feature (highlighted with a black arrow in the left panel of fig.\,\ref{fig:Anatomy}) lies immediately after the magenta region, consistently with the fact that the modes that contribute to the bump at small $k$ are those for which the horizon crossing condition takes place right after the USR phase. 
 We now focus on the transition region that connects the $\sim k^4$ growth to the bump at the left-side edge of the plateau. 
 This part of the power spectrum is formed by those modes for which horizon crossing takes place {\it during} the USR phase.
 In this region the slope of the power spectrum gradually decreases from 
 $\sim k^4$ to $\sim k^{0.7}$ (regions highlighted first in blue, then in green in the left panel of fig.\,\ref{fig:Anatomy}). 
 In order to make more explicit the interplay between the horizon crossing condition and the USR phase, in the right panel of fig.\,\ref{fig:Anatomy} we show the time evolution of the 
 individual modes that contribute to the  red, blue and green part of the power spectrum; we superimpose the time evolution of $\eta$, and the vertical lines mark the horizon crossing time for each mode.
  This plot confirms what already realized  before at the level of the power spectrum: the red (blue and green) modes cross the horizon right before (during) the USR phase.

Bearing in mind the above discussion, we now come back to the issue of the maximum PBH mass.  
In fig.\,\ref{fig:PBHAbundanceFull}, 
the solar-mass bump in the distribution $f_{\rm PBH}(M_{\rm PBH})$ is generated, at the level of the power spectrum, by the bump at the left-side edge of the plateau.
Consequently, the rule of thumb is very simple: if we move the bump in 
$P_{\mathcal{R}}(k)$ towards smaller $k$ we will get heavier PBHs since the solar-mass peak will shift toward increasing values of $M_{\rm PBH}$. 
In our model, we point out two ways to accomplish  this change.
\begin{itemize}
\item[{\it i)}] 
    The simplest option is to anticipate the beginning of the USR phase. 
\begin{figure}[!t!]
\begin{center}
$$\includegraphics[width=.495\textwidth]{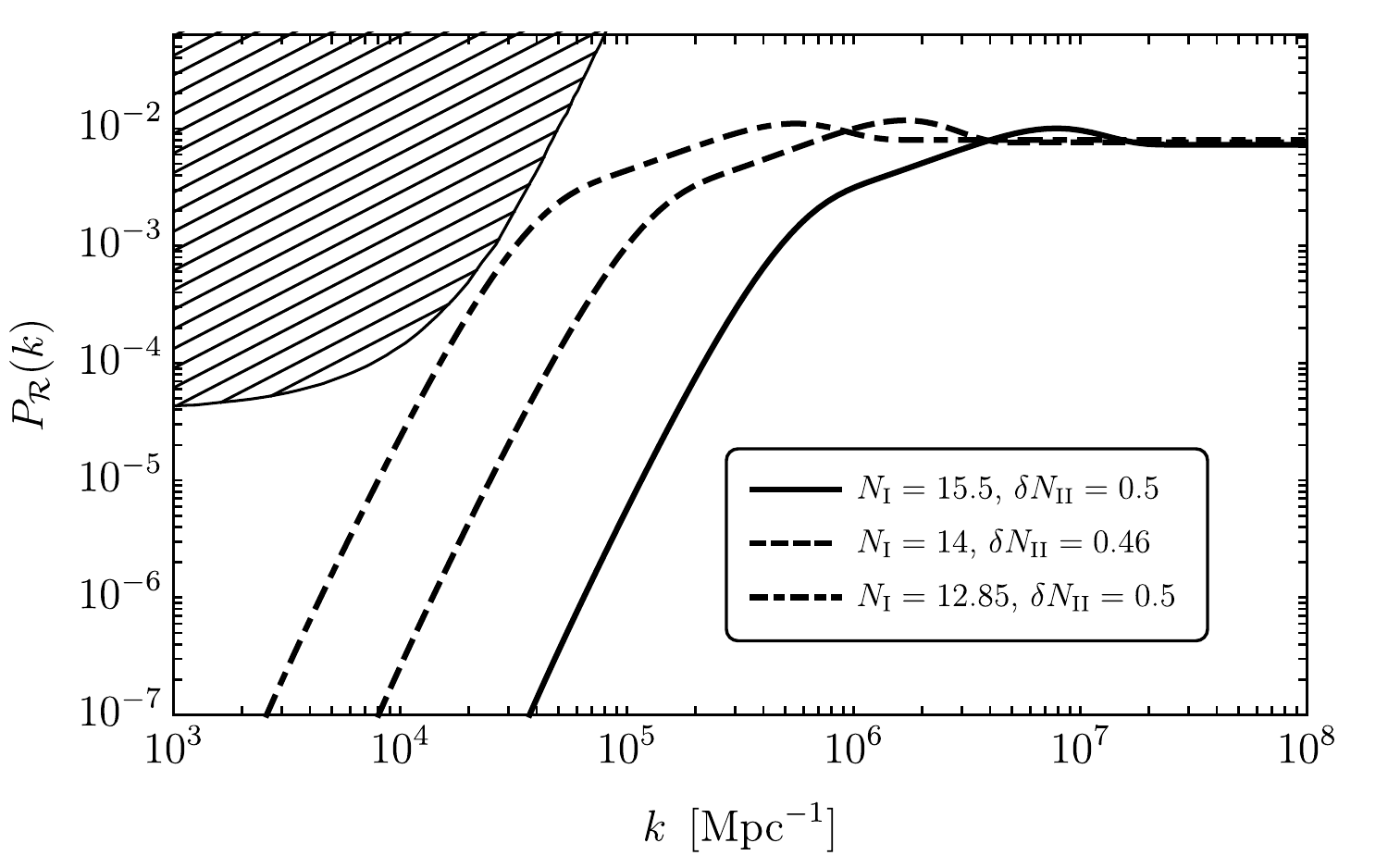}
~\includegraphics[width=.495\textwidth]{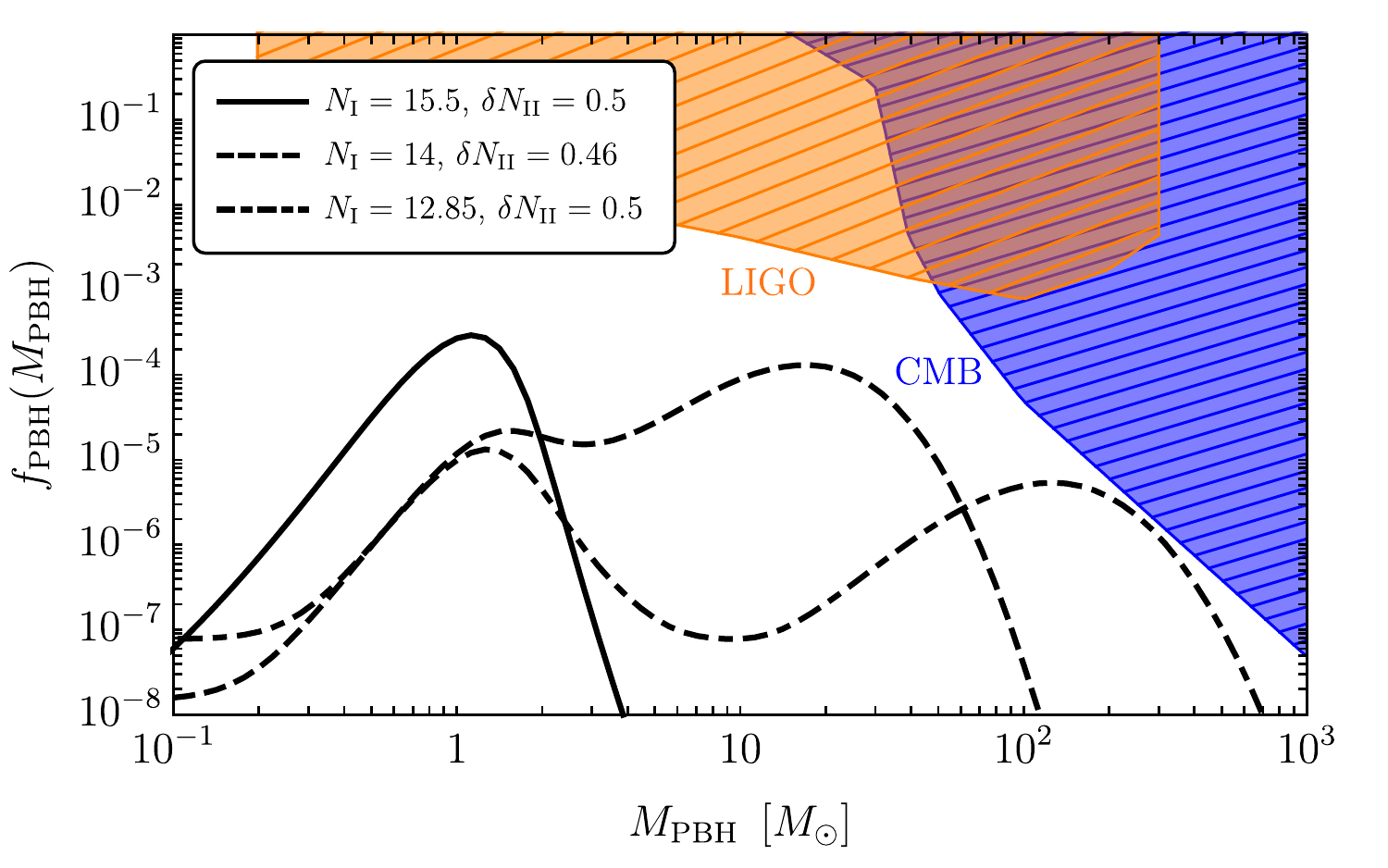}$$
\caption{
Left panel. Zoom in on the left-side edge of the plateau of the power spectrum in fig.\,\ref{fig:PS}; we consider two variations of model {\color{verdes}{$(1)$}} that have different values of $N_{\rm I}$ and $\delta N_{\rm II}$ (cf. the plot legend for details).  
Right panel. Fraction of DM in the form of PBHs with mass $M_{\rm PBH}$. 
We zoom in on the solar-mass range and show the abundance corresponding to the three models discussed in the left panel.
 }\label{fig:MaxMass} 
\end{center}
\end{figure}
    Technically, this means taking smaller values of $N_{\rm I}$. 
    This has the net effect of a shape-invariant shift of the left-side edge of the power spectrum towards smaller $k$. 
    From the left panel of fig.\,\ref{fig:Anatomy}, we see that this shift is possible until the 
    power spectrum (more precisely, the part of it in between the red and blue region) clashes with the FIRAS bound. 
    %We find that this happens 
    %for $N_{\rm I} \simeq 12.85$ (we %remind that the benchmark value for %our models in table\,%\ref{eq:ModelTab} is $N_{\rm I} %\simeq 15.5$). 
    {We illustrate our findings in fig.\,\ref{fig:MaxMass}. We consider model {\color{verdes}{$(1)$}} (solid black lines in fig.\,\ref{fig:MaxMass}) and take increasingly smaller values of $N_{\textrm{I}}$. As expected, the left-side edge of the plateau shifts rigidly towards  smaller $k$ (left panel in fig.\,\ref{fig:MaxMass}). 
    For definiteness, we focus on two specific modifications of model {\color{verdes}{$(1)$}}. 
    First, consider the dashed black lines in fig.\,\ref{fig:MaxMass} that correspond to $N_{\rm{I}} = 14$. In the right panel of fig.\,\ref{fig:MaxMass}, we show the corresponding mass function $f_{\textrm{PBH}}(M_{\textrm{PBH}})$. 
    The latter exhibits a characteristic double-peak shape. This is because we are now separating the peak due to the softer QCD equation of state (that sits at around $M_{\textrm{PBH}} = 1\,M_{\odot}$) from the peak that is due to the bump at the left-side edge of the power spectrum (that now shifted towards smaller $k$ thus larger $M_{\textrm{PBH}}$). The height  of the latter, as explained in section\,\ref{sec:DeltaNII}, is controlled by $\delta N_{\rm II}$, and the model that corresponds to the dashed black lines in fig.\,\ref{fig:MaxMass} has 
    $\delta N_{\rm II} = 0.46$ thus slightly smaller than the benchmark value $\delta N_{\rm II} = 0.5$; this is because the second peak is no longer boosted by the QCD phase transition (which, as discussed, takes place at around $M_{\textrm{PBH}} = 1\,M_{\odot}$), and we compensate this effect with a smaller $\delta N_{\rm II}$. 
    In this configuration the model produces a sizable abundance of PBHs with a mass function peaked at around $M_{\rm PBH} \simeq 20\,M_{\odot}$. 
    As shown in the right panel of fig.\,\ref{fig:MaxMass}, the upper bound on the abundance of these PBHs is given by LIGO data. 
    } 
    This is an interesting point since it shows that a rigid shift of the USR dynamics presented in the main text may generate a sufficiently abundant population of PBHs within the so-called lower mass gap,  that is in the range $\approx [2.2\divisionsymbol 6]\,M_\odot$ (see e.g. \cite{LIGOScientific:2021psn,Farah:2021qom}), {if we just take  a value of $N_{\rm I}$ slightly larger than the one discussed above} and in the so-called upper mass gap, that is above $\approx 50\,M_\odot$, {if we just take  a value of $N_{\rm I}$ slightly smaller than the one discussed above. We refer to \cite{inprepQCD} for a comprehensive discussion about the role that these PBHs may have in the gravitational-wave merger events detectable by the LVKC.} 
    
    {We now consider a second, much  smaller value for $N_{\rm I}$; the dot-dashed black lines in fig.\,\ref{fig:MaxMass} correspond to $N_{\rm{I}} = 12.85$. From the  plot of the  power spectrum in the left panel of fig.\,\ref{fig:MaxMass} we see that this value of $N_{\rm I}$ almost saturates the region allowed by the FIRAS bound. The corresponding mass distribution of PBHs is shown in the right panel of fig.\,\ref{fig:MaxMass}. The second peak is now very close to the CMB bound, and corresponds to PBHs with mass $M_{\rm PBH} \simeq 10^2\,M_{\odot}$ or larger. However, we remark that in this case the upper bound on the abundance of these PBHs is given by the CMB constraint.
    In the model that corresponds to the dot-dashed black lines in fig.\,\ref{fig:MaxMass} we take $\delta N_{\rm II} = 0.5$; if we take smaller values of $\delta N_{\rm II}$ the second peak at large PBH mass will be enhanced, in conflict with the CMB bound.}
    
%If we just insist on changing $N_{\rm %I}$, the discussion above has shown that %the FIRAS bound provides an obstruction. %In our model there exists, however, a %second possibility.
   
   \item[{\it ii)}] 
   Consider again the power spectrum in the left panel of fig.\,\ref{fig:Anatomy}.   The idea is to alter the slope of the blue and green region such to connect more directly the red growth $\sim k^4$ to the bump.
   Thanks to our preliminary discussion, we have the right intuition about how to achieve this goal: we just need to 
   shorten the duration of the USR phase and reach the first peak earlier.
  As a simple consequence, the interval of modes for which the horizon crossing condition takes place within the USR phase will be reduced. 
\begin{figure}[!t!]
\begin{center}
$$\includegraphics[width=.33\textwidth]{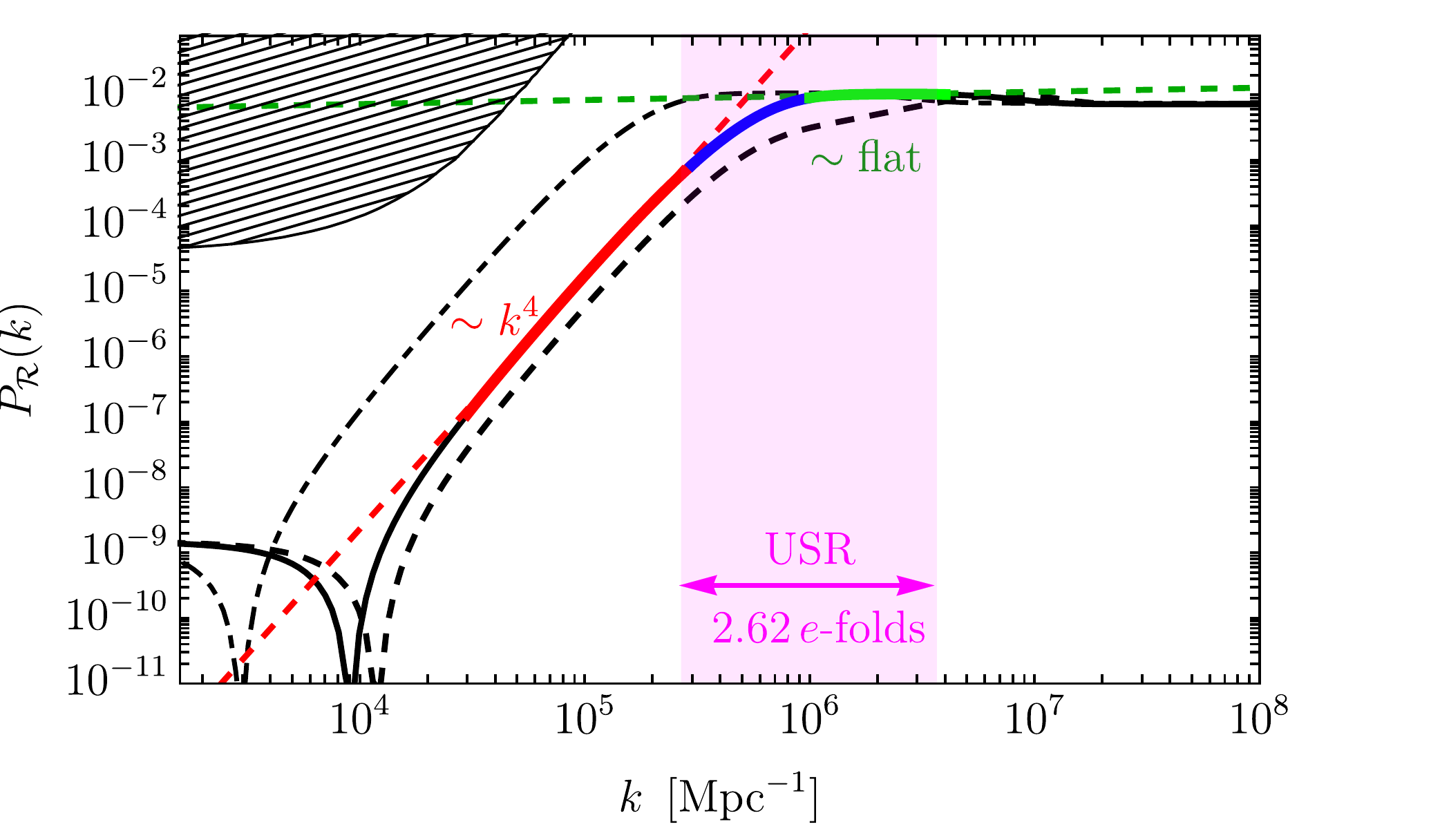}~\includegraphics[width=.33\textwidth]{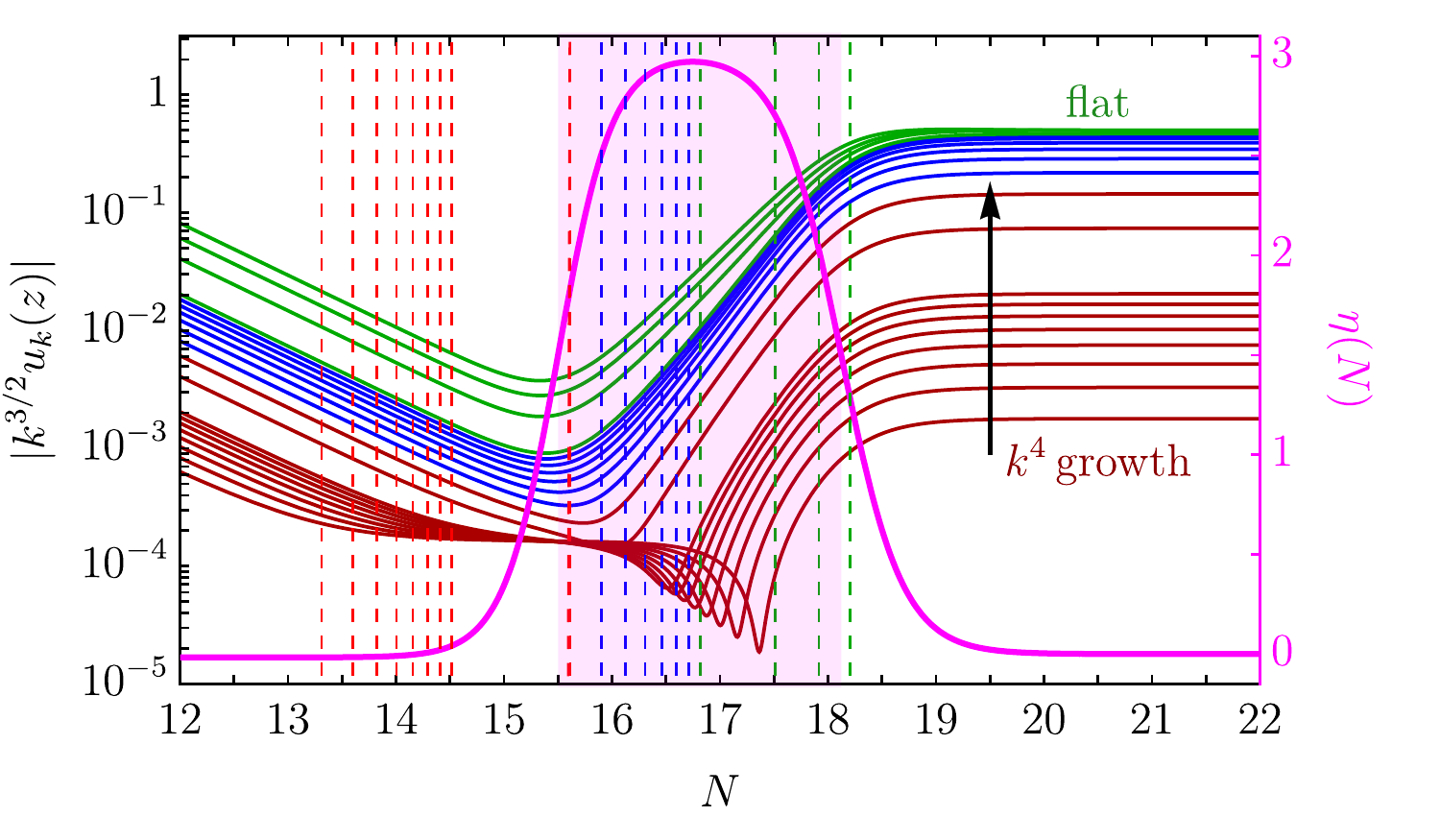}~\includegraphics[width=.33\textwidth]{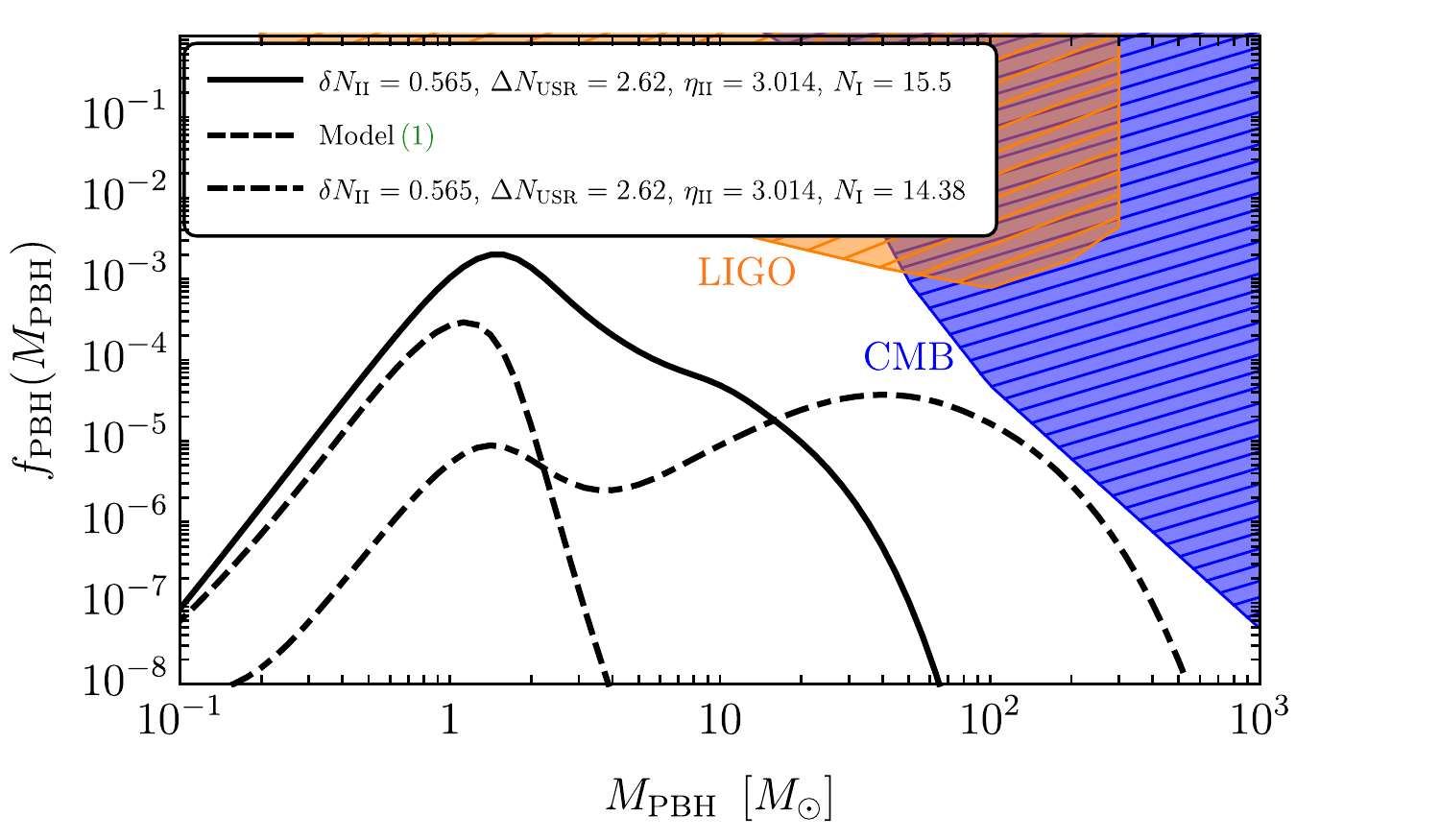}$$
\caption{
Left panel. The dashed black line is the power spectrum shown in the left panel of fig.\,\ref{fig:Anatomy}; the solid black line is the power spectrum that is obtained taking $\Delta N_{\rm USR} = 2.62$ (instead of the benchmark value $\Delta N_{\rm USR} = 2.9$) and $\eta_{\rm II} \simeq 3$ (instead of the benchmark value $\eta_{\rm II} \simeq 2.7$); the dot-dashed black line is the power spectrum obtained for the same parameters given above but with an anticipated USR phase ($N_{\rm I} = 14.38$ instead of $N_{\rm I} = 15.5$).
The red part of the power spectrum corresponds to the $\sim k^{4}$ growth while the green part is approximately  flat; the blue part lies in between. 
Central panel. Dynamical evolution of the perturbation modes with $k$ in the red, blue and green part of power spectrum discussed in the left panel. The vertical lines mark the horizon crossing time, and make clear that the for blue and green modes we have 
$N_{\rm I} \leqslant N_k \leqslant N_{\rm II}$. 
Right panel. The dashed black line corresponds to the PBH mass distribution in model {\color{verdes}{$(1)$}}. The solid black line is obtained taking a shorter URS phase with $\Delta N_{\rm USR} = 2.6$, $\eta_{\rm II} = 3.014$ and  
$\delta N_{\rm II} = 0.565$ (but with the same $N_{\rm I} = 15.5$ as in model {\color{verdes}{$(1)$}}). 
The dot-dashed black line corresponds to the same model that gives  the solid black line but with an anticipated USR phase, $N_{\rm I} = 14.38$.
 }\label{fig:Anatomy2} 
\end{center}
\end{figure}  
In the left panel of fig.\,\ref{fig:Anatomy2} we show the power spectrum that we get if we modify model 
{\color{verdes}{(1)}} by taking a  
shorter USR phase. We consider 
$\Delta N_{\rm USR} = 2.62$ instead of the benchmark value $\Delta N_{\rm USR} = 2.9$.

Importantly, it should be noted that, 
in order to maintain the same height of the plateau in the power spectrum, decreasing the value of $\Delta N_{\rm USR}$ should be compensated by a larger value of $\eta_{\rm II}$. This simply follows from the exponential growth in front of eq.\,(\ref{eq:AnalRk}). 
In the left panel of fig.\,\ref{fig:Anatomy2}, in fact, we are forced to consider 
 $\eta_{\rm II} \simeq 3$ (instead of the benchmark value $\eta_{\rm II} \simeq 2.7$). 
 This simple fact has a very profound implication. Since during phase I we have 
 $\eta_{\rm I} \simeq 0$, 
 a Wands duality \cite{Wands:1998yp} is established 
between phase I and phase II: 
 phases with $\eta$ and $3-\eta$ (that is, in our case, $\eta_{\rm I} \simeq 0$ and 
 $\eta_{\rm II} \simeq 3 = 3-\eta_{\rm I}$) are dual in the sense that they give rise the same spectral slope. This means that we expect a flattening of the power spectrum during the USR phase.
 
 The numerical analysis shown in the left panel of fig.\,\ref{fig:Anatomy2} fully confirms our intuition. 
 For completeness, in the central panel of fig.\,\ref{fig:Anatomy2} we show the time evolution of red, blue and green modes together with their horizon crossing time (vertical lines). 
 As expected, at the left-side edge of the plateau the power spectrum now has, as a consequence of the duality, a flat region (instead of just a bump-like feature) that is quickly connected to the $\sim k^4$ growth. From a phenomenological viewpoint,   this simple modification has a far-reaching implication since it means that it will be possible to  generate, at the level of the distribution $f_{\rm PBH}(M_{\rm PBH})$, not just a peak (as in the case of the bump-like feature) but a broader mass distribution
 in the LVKC detectable mass range, and extending towards heavier PBHs well within the upper mass gap. {We illustrate this point in 
 the right panel of fig.\,\ref{fig:Anatomy2} in which we show the PBH mass distribution of the benchmark model {\color{verdes}{$(1)$}} (black dashed line) compared with the one obtained for $\Delta N_{\rm USR} = 2.62$, $\eta_{\rm II} = 3.014$ and $\delta N_{\rm II} = 0.565$.}
 
 Furthermore, from the left panel of fig.\,\ref{fig:Anatomy2}, we also see 
 that, in principle, we have enough room to combine {\it i)} and 
 {\it ii)} and push the power spectrum towards the FIRAS bound by taking smaller values of $N_{\rm I}$.
 Interestingly, we find that if we combine 
 {\it i)} and 
 {\it ii)}
 it is not possible to saturate the FIRAS bound (the minimum allowed value of $N_{\rm I}$ turns out to be about $14.4$). 
 The reason is that, since we now have a broader distribution in $f_{\rm PBH}(M_{\rm PBH})$ in the solar-mass range instead of a narrow peak, before saturating the FIRAS bound we would clash with the CMB constraint on accreting PBHs, shown in blue in 
 fig.\,\ref{fig:PBHAbundanceFull}. 
 {The black dot-dashed line in the right panel of fig.\,\ref{fig:Anatomy2} corresponds to the same model that gives the solid black line discussed before but with an anticipated USR phase ($N_{\rm I} = 14.38$ instead of $N_{\rm I} = 15.5$). The PBH mass distribution saturates the CMB bound even though the left-side edge of the plateau in the power spectrum is far from the FIRAS bound (cf. the dot-dashed black line in the left panel).}
 \end{itemize}
In conclusion, the USR dynamics studied in this paper 
may easily accommodate a population of solar-mass PBHs with a (potentially broad) mass distribution that extends up to the constraint provided by the modification of the CMB spectrum due to PBH accretion.  
A more quantitative and detailed discussion will be presented in 
ref.\,\cite{inprepQCD}.

As a final remark, we would like to stress that the above discussion shows very clearly the power of our approach.  
Starting from a well-defined physical question (what is the maximum mass of PBHs) we were able, in very few steps, to pinpoint a neat connection with the underlying dynamics that made extremely  clear the correct way to get to the desired answer. 
%is to anticipate the beginning of the

 \section{The scalar-induced GW signal}\label{sec:GW}
Next, we compute the second-order 
gravitational-wave signal sourced by scalar perturbations\,\cite{Tomita:1975kj,Matarrese:1993zf,Acquaviva:2002ud,Mollerach:2003nq,Ananda:2006af,Baumann:2007zm} (see ref.\,\cite{Domenech:2021ztg} for a recent review). 
The current energy density of 
gravitational-waves as function of their frequency $f$  is given by 
\begin{align}\label{eq:NGGW}
\Omega_{\rm GW}(f) = \frac{c_g\Omega_r}{36}\int_{0}^{\frac{1}{\sqrt{3}}}dt\int_{\frac{1}{\sqrt{3}}}^{\infty}ds
\left[
\frac{(t^2-1/3)(s^2-1/3)}{t^2 - s^2}
\right]^{2}\left[\mathcal{I}_c(t,s)^2 + \mathcal{I}_s(t,s)^2\right]
P_{\mathcal{R}}\left[\frac{k\sqrt{3}}{2}(s+t)\right]
P_{\mathcal{R}}\left[\frac{k\sqrt{3}}{2}(s-t)\right]\,,
\end{align}
where $\Omega_r$ is the current energy density of radiation 
and $\mathcal{I}_c$ and $\mathcal{I}_s$ are two functions that  can  be  computed  analytically (see, for instance, refs.\,\cite{Espinosa:2018eve,Kohri:2018awv}). 
The parameter $c_g$ defined as 
\begin{equation}
c_g	\equiv \frac{g_*(M_H)}{g_{*}^0}
\left( \frac{g_{*S}^0}{g_{*S} (M_H)}\right) ^{4/3}
\end{equation}  accounts for the change of the effective degrees of freedom of the thermal radiation $g_*$ and $g_{*S}$ (where the superscript $^0$ indicates the values today) during the evolution (assuming Standard Model physics), and it is of order $c_g=0.4$ for modes related to the formation of asteroid-mass PBHs.
The frequency $f$ is related to the comoving wavenumber $k$ by the relation 
\begin{align}
k \simeq 6.47\times 10^{14} \left(
\frac{f}{{\rm Hz}}
\right)\,\,{\rm Mpc}^{-1}\,,
\end{align}
so  that the two sides of the flat power spectrum in fig.\,\ref{fig:PS} correspond to 
$f = O(0.1)$ Hz (for $k_{\rm max} \simeq 10^{14}$ Mpc$^{-1}$) and 
$f = O(10^{-9})$ Hz (for $k_{\rm min} \simeq 10^{-9}k_{\rm max}$). 
These frequencies are related to the formation of PBHs with asteroidal \cite{Bartolo:2018rku,Bartolo:2018evs,Balaji:2022rsy} and solar masses \cite{Vaskonen:2020lbd}, respectively (see fig.\,\ref{fig:PS}).

\begin{figure}[b]
\leavevmode
\centering
\includegraphics[width=0.65\textwidth]{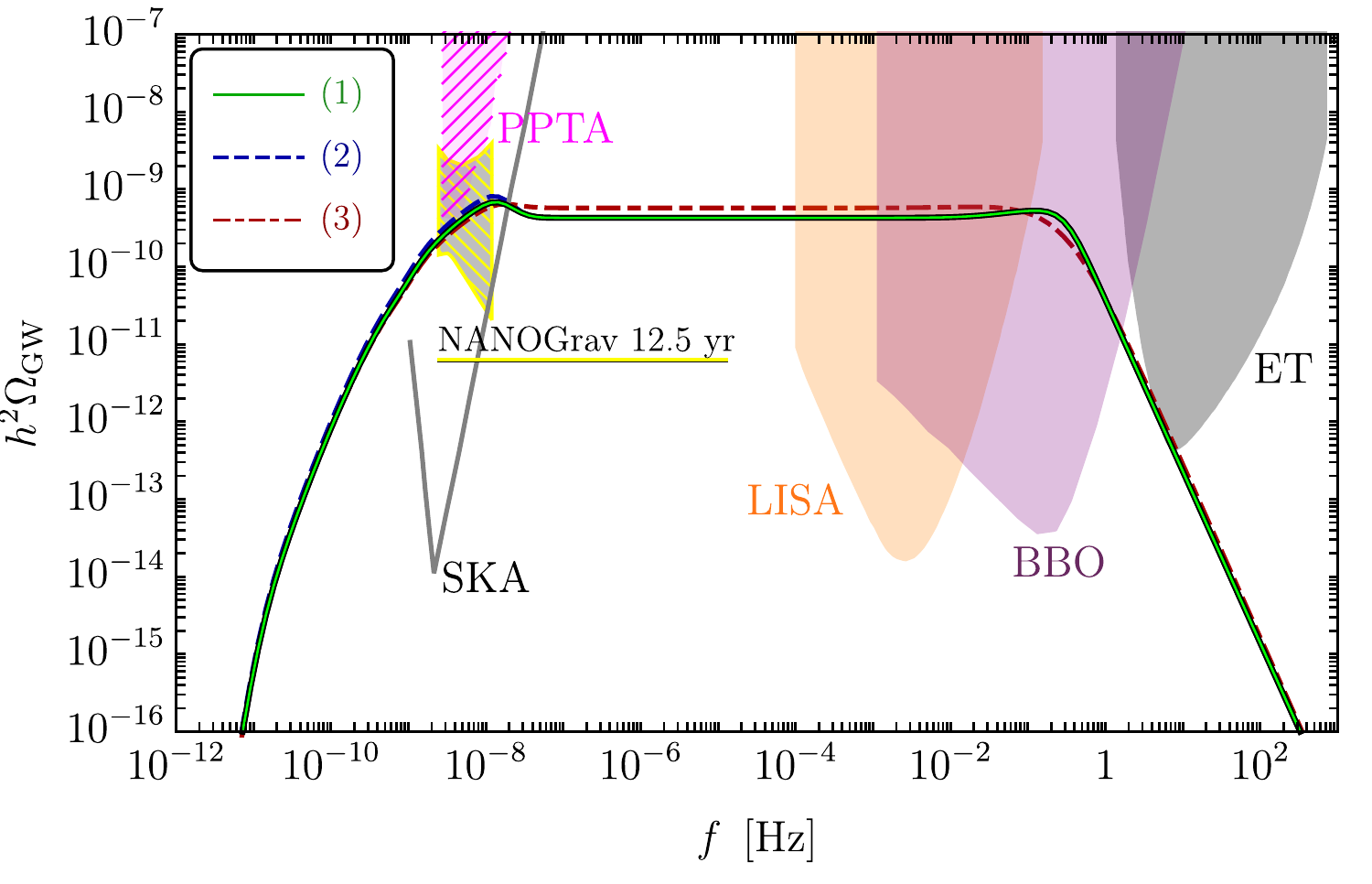}  
\caption{\label{fig:GWSignal}
Fraction of the energy density in GWs relative to the critical energy density of the Universe as
function of the frequency. We show the power-law integrated sensitivity curves \cite{Thrane:2013oya} of future ground- and space-based GW experiments (as derived in app. C of ref.\,\cite{Bavera:2021wmw}) as well as previous Parkes Pulsar Timing Array (PPTA) constraint\,\cite{Shannon:2015ect}, NANOGrav putative band\,\cite{NANOGrav:2020bcs} and SKA projected sensitivity\,\cite{Janssen:2014dka}. 
We plot the signals predicted by our model 
in the three realizations proposed in table\,\ref{eq:ModelTab}.
}
\end{figure}

A robust prediction of this scenario, as highlighted in ref.\,\cite{DeLuca:2020agl}, is the generation of a nearly scale invariant SGWB (shown in fig.~\ref{fig:GWSignal}) crossing both  PTA experiments and LISA. Due the quadratic dependence of the SGWB amplitude to the spectrum amplitude, 
one finds much milder features mirroring the large enhancements observed in the PBH mass distribution. However, it is interesting to notice that $\delta N_{\rm II}$ would potentially modify the spectral tilt within the PTA frequency range, ranging from flat to slightly red in scenarios {\color{cornellred}{$(3)$}}
and {\color{verdes}{$(1)$}}, respectively.
In all cases, such spectrum has a frequency dependence in the PTA range which is different from the one emitted by massive BH binaries  (e.g. \cite{Middleton:2020asl}).

{Before proceeding, 
let us comment on a number of approximations that we have done in the computation of the scalar-induced GW signal. 
First, we remark that eq.\,(\ref{eq:NGGW}) and the value $c_g=0.4$ are strictly valid only during the radiation epoch with $g_* = O(100)$. In principle, one should modify 
eq.\,(\ref{eq:NGGW}) to include the effect of  the QCD quark-hadron phase transition (along the lines of what is done in ref.\,\cite{Abe:2020sqb}). This is particularly relevant for the comparison with experimental data in the low-frequency region of PTA and NANOGrav.  
Future confirmation of the NANOGrav signal with 
additional spectral information will make this computation extremely relevant for an appropriate comparison between theoretical predictions and data. We leave this analysis for future work. 
Second, as done throughout this work, we neglect possible primordial non-Gaussian corrections in the computation of the scalar-induced GW signal. We refer to refs.\,\cite{Yuan:2020iwf,Adshead:2021hnm,Abe:2022xur,Chang:2022nzu} for a discussion about the impact of these effects.
}

\color{black}

\section{The reconstructed potential}\label{sec:potential}

From $\eta(N)$ and $\epsilon(N)$, that capture the inflationary dynamics and connects it to the various late time observable, we can reconstruct the 
scalar potential $V(\phi)$.
This is the final aim of the reverse engineer approach and one of the main results of our paper.

\subsubsection{From dynamics to the inflationary potential}
Once the Hubble parameters are known, one can compute the inflationary potential by means of 
\begin{align}
V(N) & = V(N_{\rm ref})\exp\left\{
   -2\int_{N_{\rm ref}}^{N}dN^{\prime}\left[\frac{\epsilon(3-\eta)}{3-\epsilon}\right]
   \right\}\,,
   ~~~~~~~~~
\phi(N)  = \phi(N_{\rm ref}) \pm \int_{N_{\rm ref}}^N dN^{\prime}\sqrt{2\epsilon}\,,   \label{eq:recPot1}
% \label{eq:recPot2}
\end{align}
where in the second equation we consider the minus sign having in mind a large-field model in which the field value decreases as inflation proceeds. 
Combining $V(N)$ and $\phi(N)$, we reconstruct the 
profile $V(\phi)$ of the inflationary potential in field space\,\cite{Byrnes:2018txb}. 
We will discuss further details of the reconstruction procedure and
the interpretation of the potential reminder of this section.
We mention here that eq.\,\eqref{eq:recPot1} shows the convenience of modelling the inflationary dynamics directly at the level of $\eta$ instead of $V(\phi)$. 
This is because the Hubble parameters enters at the exponent of the definition of $V(N)$, 
and thus allow for a much finer control on power spectral features 
when performing the reverse engineering procedure.

Using the reconstructed potential $V(\phi)$, 
one can also solve the inflaton equation of motion 
\begin{align}\label{eq:EoM}
\frac{d^2\phi}{dN^2} + \left[3 - \frac{1}{2}\left(\frac{d\phi}{dN}\right)^2\right]
\left[\frac{d\phi}{dN} + \frac{d\log V(\phi)}{d\phi}
\right] = 0\,,
\end{align}
and, in turn, compute the time evolution of the Hubble parameters in eq.\,(\ref{eq:HubbleParameters}) 
and the Hubble rate by means of the relations
\begin{align}
\epsilon  = \frac{1}{2}\left(\frac{d\phi}{dN}\right)^2\,,~~~~~~~~~~~\eta = 
3 - \frac{V^{\prime}(\phi)[- 6 + (d\phi/dN)^2]}{2V(\phi)(d\phi/dN)}
\,,~~~~~~~~~~~
(3-\epsilon)H^2 = V(\phi)\,.
\end{align}
As far as the Hubble parameters are concerned, these
 equations are nothing but a rewriting of eq.\,(\ref{eq:HubbleParameters}) in terms of the classical 
 field dynamics while the last equation is the Friedmann equation. 
 These equations are valid under the assumption that the energy density of the 
 inflating Universe is given entirely by the scalar field $\phi$.
As a consistency check, we correctly find the same functional dependence illustrated in 
fig.\,\ref{fig:PlotEpsilonEta} (but now obtained as an output instead of an input).

\begin{figure}[!htb!]
%\boxed{ \eta_{\rm I} = -0.2\,,~~~\eta_{\rm II} = 3.5\,,~~~\eta_{\rm III} = 0\,,~~~\eta_{\rm IV} = -1.5\,,~~~
%N_{\rm II}-N_{\rm I} = 2.5\,,~~~
%N_{\rm III}-N_{\rm II} = 7.5\,,~~~\delta N = 0.13\,,~~~\epsilon_{\rm I} = 10^{-3}}\vspace{-0.3cm}
\begin{center}
$$\includegraphics[width=.495\textwidth]{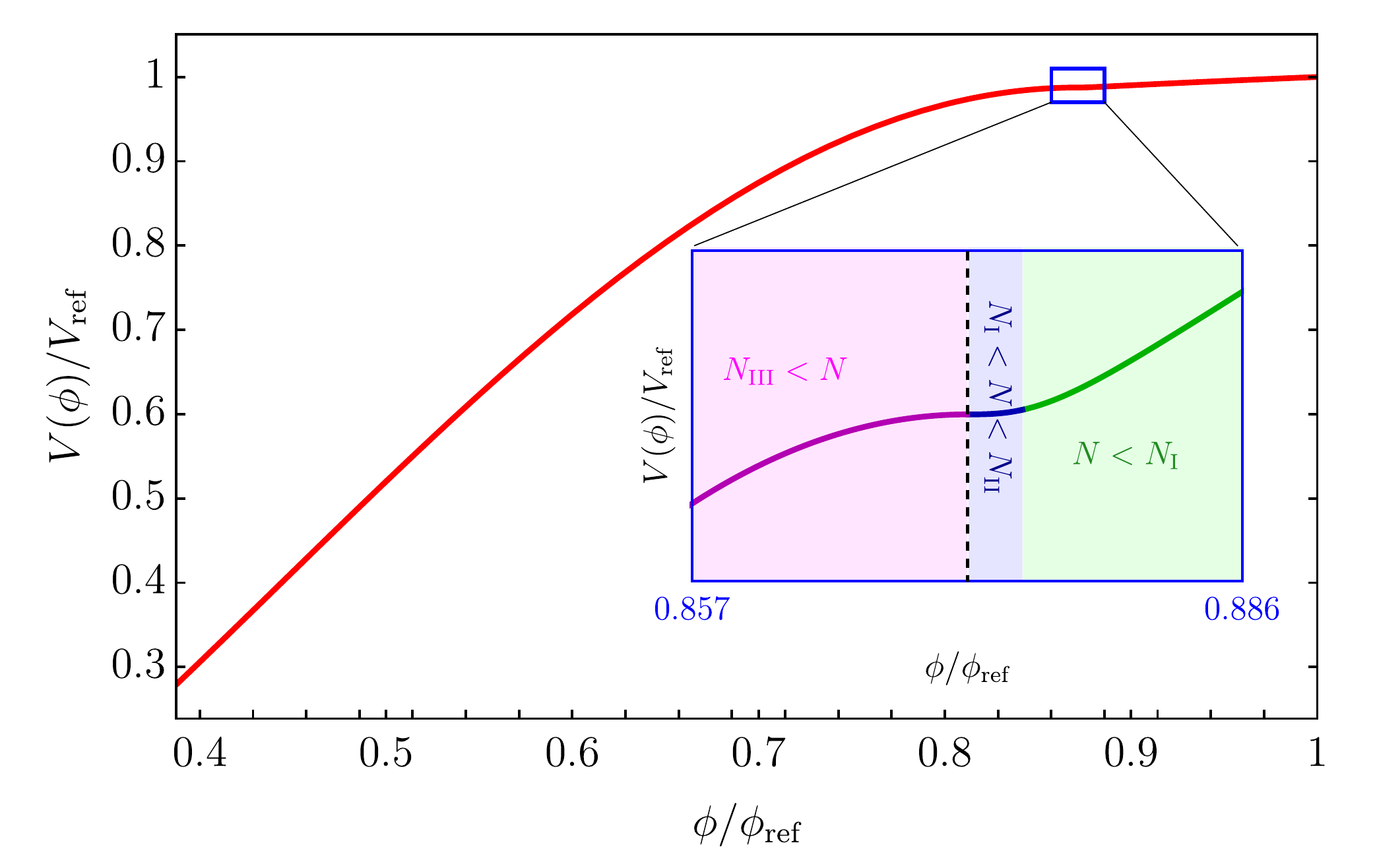}~\includegraphics[width=.495\textwidth]{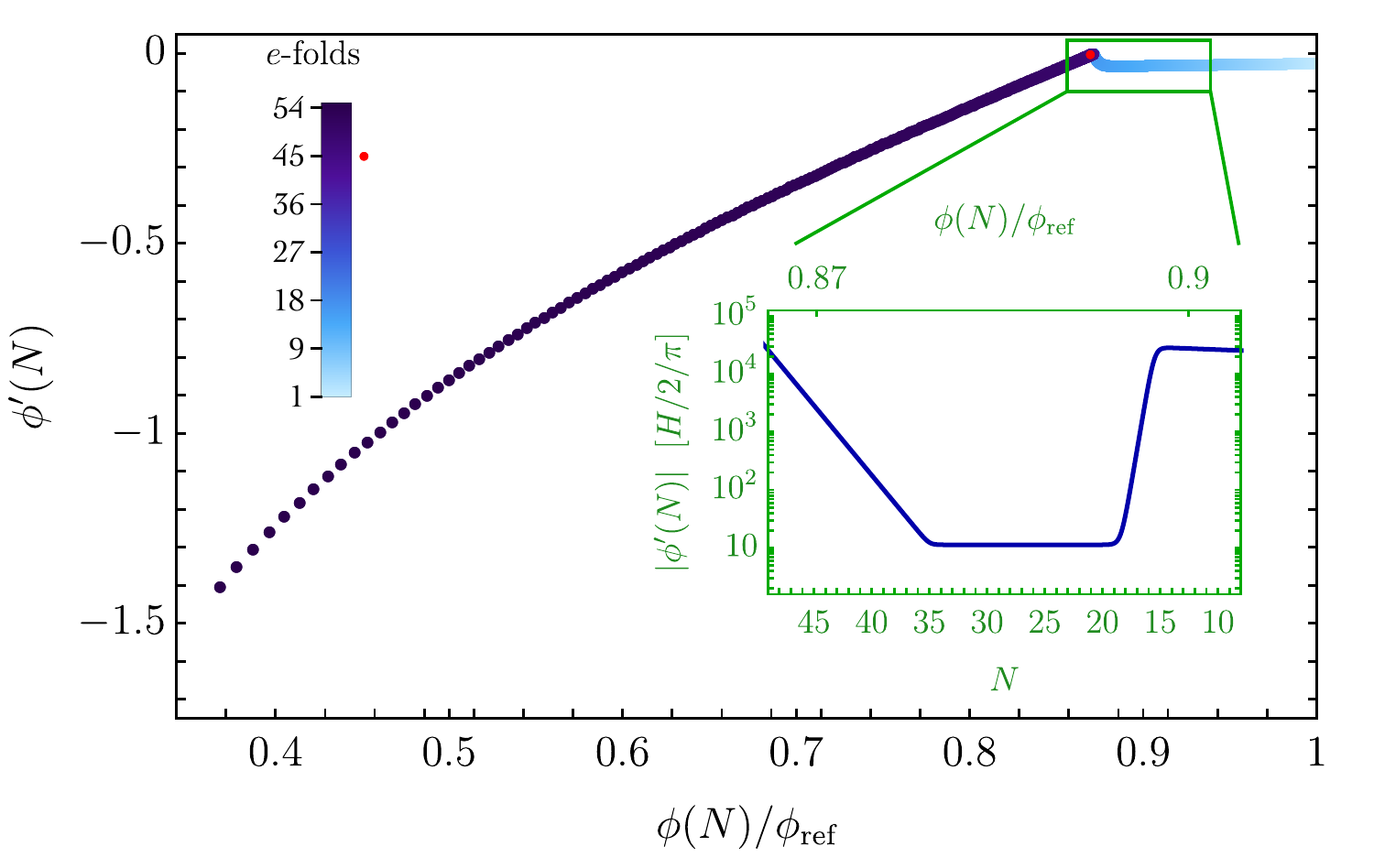}$$
\caption{ 
Left panel. Reconstructed potential for the model in the second column of table\,\ref{eq:ModelTab}. 
In the inset plot, we zoom in the transition region, and 
we indicate with different colors the various stages of the classical inflaton dynamics.  
It should be noted that during the time interval $N_{\rm II} < N < N_{\rm III}$ (shown as dashed vertical line) the inflaton field has 
almost zero velocity and it remains approximatively stuck in field space (with this part of the dynamics that lasts for about 
 $\Delta N_{\rm plateau}  \simeq 18$ $e$-folds) before entering the last stage that ends inflation.  
Right panel. Phase-space classical dynamics of the inflaton field. 
Different gradations of blue correspond, according to the inset legend, to increasing $e$-fold time $N$ starting 
from $N_{\rm ref} = 0$ (lighter) to the end of inflation $N_{\rm IV} = 55$ (darker). 
The red dot corresponds to $N = 45$. 
On the $y$-axis, the field has units of reduced Planck mass and we use the notation $\phi^{\prime} = d\phi/dN$. 
In the inset plot, we plot the modulus of the velocity in units of $H/2\pi$ as function of $N$ (with the corresponding field values on the top $x$-axis), and we focus on the USR phase and the subsequent phase during which $\eta_{\rm III} = 0$. 
We note that, in units of $H/2\pi$, we have 
$|\phi^{\prime}(N)| \gg 1$ during the whole dynamics with $|\phi^{\prime}(N)| = O(10)$ during the phase with $\eta_{\rm II} = 0$.
 }\label{fig:Benchmark2}  
\end{center}
\end{figure}

\begin{figure}[!h!]
\begin{center}
$$
\includegraphics[width=.495\textwidth]{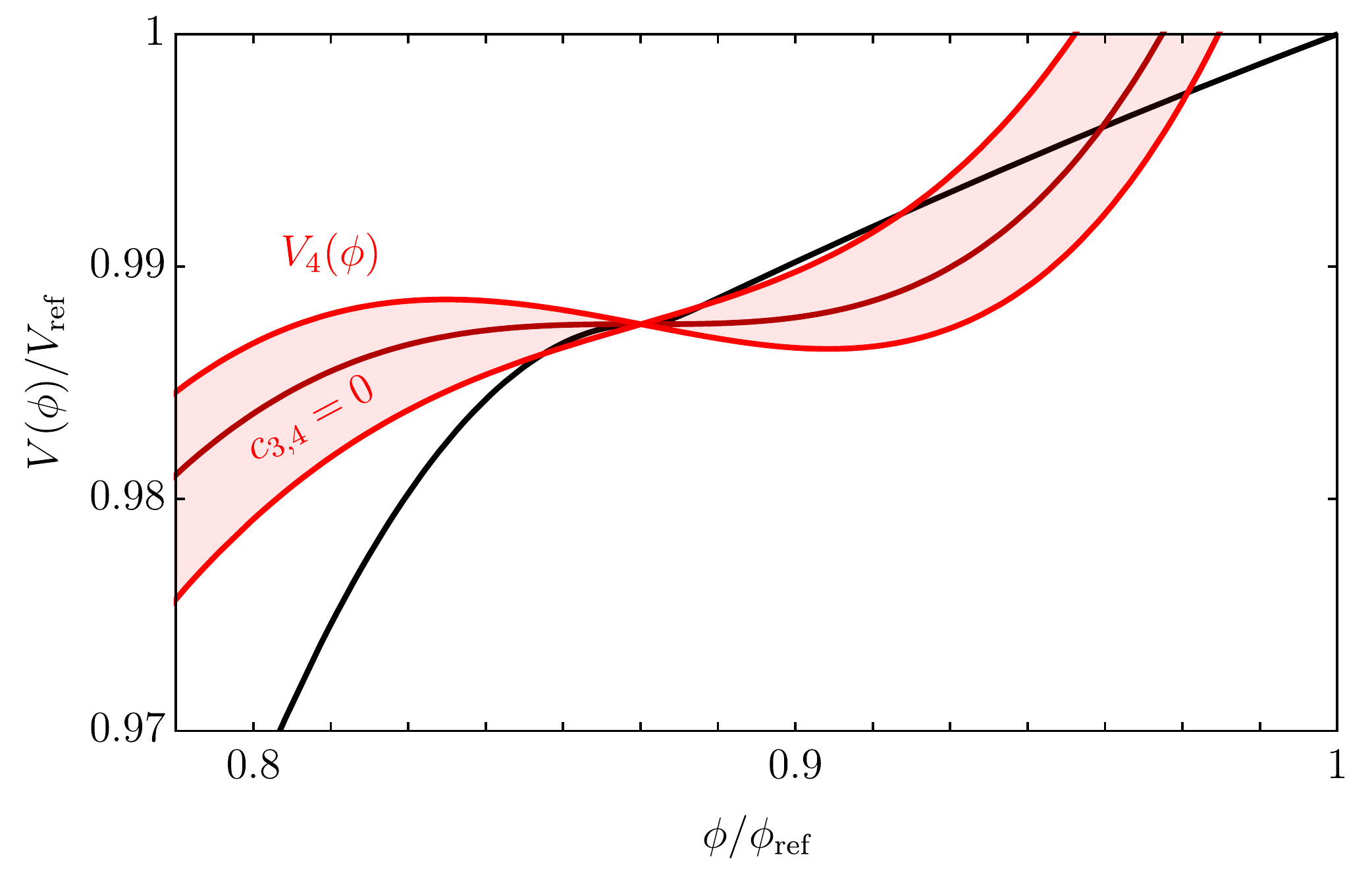}~\includegraphics[width=.495\textwidth]{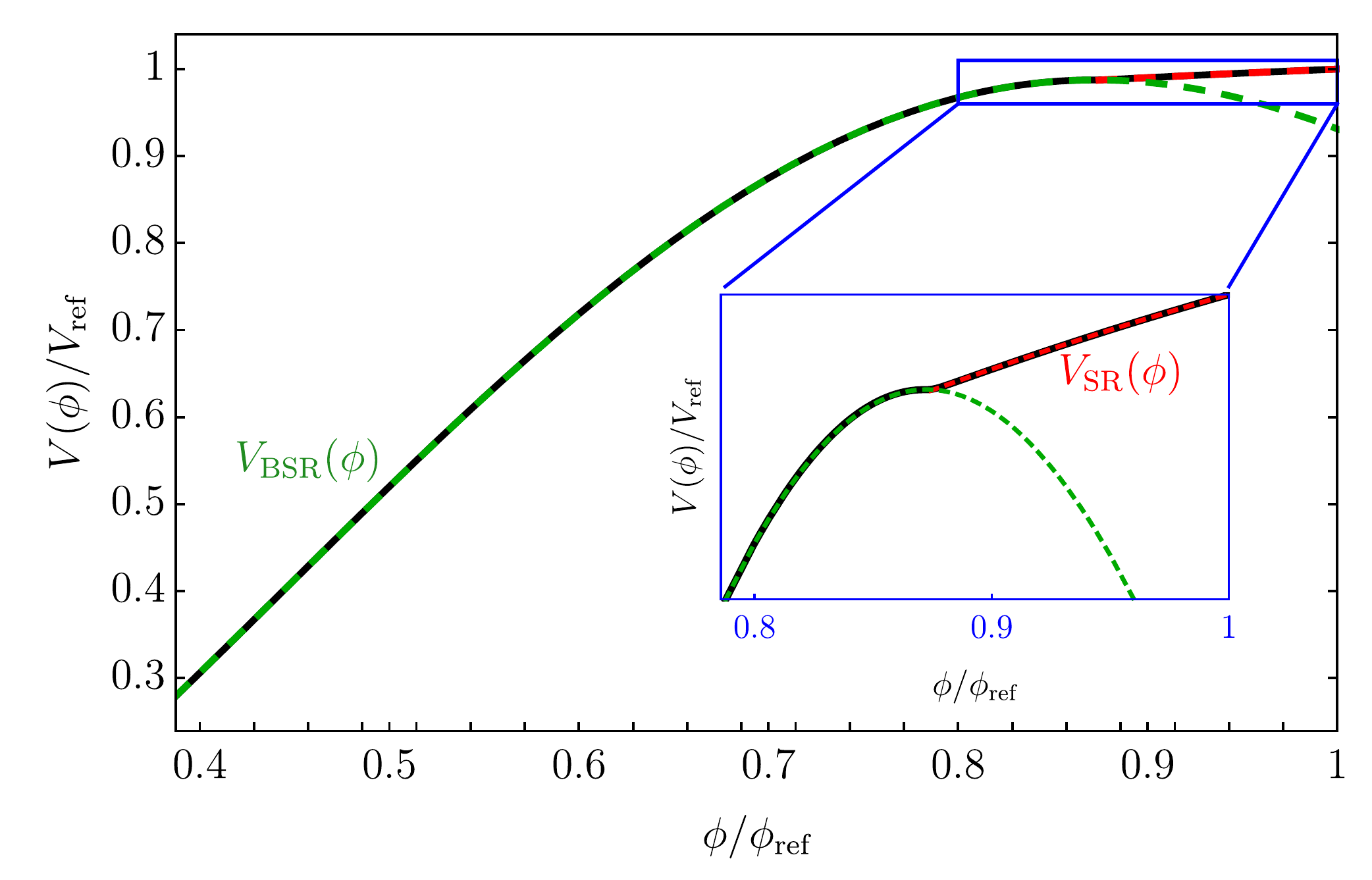}$$
\caption{
Left panel. We compare the reconstructed potential (solid black line) 
with a benchmark potential featuring an approximate stationary inflection point (anchored to the field value at which 
$V^{\prime}(\phi_0) = V^{\prime\prime}(\phi_0) = 0$ with $\phi_0$ chosen to be the field value at $N = N_{\rm II}$). The latter is given by eq.\,(\ref{eq:Vn}) with $n=4$ and $c_{3,4} = 0$ (solid red line). 
The region shaded in red is obtained for non-zero $c_{3,4}$ at the percent level. 
Right panel. We model the reconstructed potential (solid black line) with a combination 
of the potential in eq.\,(\ref{eq:VSR}) (dashed red, label $V_{\rm SR}$) and  
 eq.\,(\ref{eq:VBSR}) (dashed green, label $V_{\rm BSR}$). 
 For the potential $V_{\rm SR}(\phi)$ we use the values of the parameters given in table\,\ref{eq:ModelTab} (since this is 
 the actual analytical solution of the system in eq.\,(\ref{eq:recPot1}) during the initial slow-roll phase).
 In the case of $V_{\rm BSR}(\phi)$, we take 
 $V_0 \simeq 0.988$, $\lambda \simeq 0.714$, $c_4 \simeq -0.748$ and $\phi_0 \simeq 3.054$; 
 we extract these values from a fit of the reconstructed potential considering the region $\phi < \phi_{\rm I} = \phi(N_{\rm I})$. 
{In the reconstructed potential we take $\phi_{\textrm{ref}} = 3.5$.} 
 }\label{fig:PotentialFit}  
\end{center}
\end{figure}

The presence of an USR phase is typically associated with an (approximate) stationary inflection point in the potential of the inflaton. 
This is what we obtain by following the reconstruction procedure.
In the left panel of fig.\,\ref{fig:Benchmark2} we show the reconstructed potential that corresponds to  model {\color{verdes}{$(1)$}} (cf. table\,\ref{eq:ModelTab}). 
 In the right panel of the same figure, we plot the 
inflationary trajectory in the phase space of the inflaton field. 
At first sight, the reconstructed
potential is characterized by a flattish region at large field values 
(where we fit the CMB observables) followed by a steeper decrease that ends inflation. 
However, a closer look (see the inset plot in the left panel of fig.\,\ref{fig:Benchmark2}) reveals 
the presence of a transition region which plays a crucial role for the manifestation of the 
USR dynamics. 
During the time-interval $N_{\rm I}< N < N_{\rm II}$ the field breaks its slow-rolling and spends the next $e$-fold 
time-interval $N_{\rm II}< N < N_{\rm III}$ almost stuck in field space 
retaining just the right amount of inertia to cross the transition region and ends inflation.

{It is important to stress that the part of the dynamics that corresponds to the formation of the plateau in the power spectrum, that is the  $e$-fold time interval $N_{\rm II} < N < N_{\rm III}$ (cf. the schematic evolution in fig.\,\ref{fig:PlotEpsilonEta}), is hidden within a tiny region in field space (to the point of being just a vertical line in the left panel of fig.\,\ref{fig:Benchmark2}). 
The reverse engineering approach proposes in this paper, therefore, seems to be the right language to capture and describe such a finely-tuned part of the inflationary dynamics.}

\subsection{Interpretation within single-field models}
Consider for instance the following potential
\begin{align}
V_n(\phi) = 
\frac{V_0}{(n-2)^2}\left\{
%(n-1)n(c_2-1)\left(
%\frac{\phi}{\phi_0} 
%\right)^2 
\left[
-4c_4 (n-1) + n(n-1+c_3)
\right]\left(
\frac{\phi}{\phi_0} 
\right)^2
+
n(1-c_3)\left(
\frac{\phi}{\phi_0} 
\right)^{2n-2}
-4(n-1)(1-c_4)\left(
\frac{\phi}{\phi_0} 
\right)^n
\right\}\,.\label{eq:Vn}
\end{align}
This potential (of the type $\phi^2+\phi^3+\phi^4$ for $n=3$ and $\phi^2+\phi^4+\phi^6$ for $n=4$), by construction, features at $\phi=\phi_0$ 
a stationary inflection point (i.e. $V^{\prime}(\phi_0) = V^{\prime\prime}(\phi_0) = 0$) if $c_{3,4} = 0$.
Values $c_{3,4} \neq 0$ parametrize deviations from this exact configuration (approximate stationary infection point). 
By construction, $V_0 \equiv V_n(\phi_0)$. 

For illustration, we compare in the left panel of fig.\,\ref{fig:PotentialFit} the functional form 
given by $V_4(\phi)$ with the reconstructed potential. 
We take $\phi_0 = \phi_{\rm II} \equiv \phi(N_{\rm II})$ (that is the field value at which for the reconstructed potential we have
$V^{\prime}(\phi_0) \approx V^{\prime\prime}(\phi_0) \approx 0$). 
The comparison (see the caption of fig.\,\ref{fig:PotentialFit} for details) suggests that  the potential 
with an approximate stationary infection point is not the best-suited candidate to reproduce our numerical result.

For this reason, we explore an alternative route. 
We note that during the first phase of the dynamics the potential can be computed analytically solving the system in eq.\,(\ref{eq:recPot1}). 
We find 
\begin{align}\label{eq:VSR}
V_{\rm SR}(\phi) = V_{\rm ref}\left[
\frac{
6-2\epsilon_{\rm I} - 2\eta_{\rm I}\sqrt{2\epsilon_{\rm I}}(\phi - \phi_{\rm ref}) - 
\eta_{\rm I}^{2}(\phi - \phi_{\rm ref})^2
}{
2(3-\epsilon_{\rm I})
}
\right]^{1-\frac{3}{\eta_{\rm I}}}\,,
\end{align}
with the subscript $_{\rm SR}$ that indicates that this potential describes the initial slow-roll dynamics. 
{In this analytical derivation, we assumed the linear term in $\epsilon$ 
appearing in eq.~\eqref{eq:HubbleParameters} is negligible, which is justified during the initial slow roll phase.}
In addition, we consider the potential (with the subscript $_{\rm BSR}$ that generically 
indicates that this potential describes the dynamics beyond the initial slow-roll phase)
\begin{align}\label{eq:VBSR}
V_{\rm BSR}(\phi) = \frac{3V_0 - 4\lambda(2+c_4)}{3}  
+ 4\lambda\left(
\frac{\phi}{\phi_0}
\right)^2\left[
1 + c_4\left(
\frac{\phi}{\phi_0}
\right)^2 -
\frac{(1+2c_4)}{3}\left(
\frac{\phi}{\phi_0}
\right)^4
\right]\,,
\end{align}
with, by construction, $V_{\rm BSR}(\phi_0) = V_0$ and  $V_{\rm BSR}^{\prime}(\phi_0) = 0$.

\subsection{Interpretation within multi-field models}

The reconstructed potential suggests that $V(\phi)$ could be obtained by a combination 
of eq.\,(\ref{eq:VSR}) and eq.\,(\ref{eq:VBSR}). 
This is shown in the right panel of fig.\,\ref{fig:PotentialFit}.
In the following, we will make a few comments to motivate this intuition within multi-field models of inflation. We focus for simplicity on two-field models.

\subsubsection{Two-field models: classical dynamics}

Consider a double inflation model with two scalar fields $\phi_{1,2}$ and potential 
$V= V(\phi_1,\phi_2)$. To fix ideas, we can think about the full potential $V$ as the sum of two independent 
contributions, $V = V(\phi_1) + V(\phi_2)$ (even though the following discussion will be valid for a generic $V$).
 The classical equations of motion and the Friedmann equation are (we indicate with $\dot{}$ derivative with respect to the cosmic time $t$)
\begin{align}
\ddot{\phi}_1 + 3H\dot{\phi}_1 + V_{\phi_1} = 0\,,~~~~~~~~~~
\ddot{\phi}_2 + 3H\dot{\phi}_2 + V_{\phi_2} = 0\,,~~~~~~~~~~ H^2= \frac{1}{3}\left(
\frac{1}{2}\dot{\phi}_1^2 + 
\frac{1}{2}\dot{\phi}_2^2 + V
\right)\,,\label{eq:ClassicalTwoFieldSystem}
\end{align}
where we use the short-hand notation $V_x\equiv \partial V/\partial x$. 
Inflation proceeds along some trajectory $\sigma$ in field space that we describe by means of the velocity field 
\begin{align}  
\dot{\sigma} = \dot{\phi}_1 \cos\theta + \dot{\phi}_2 \sin\theta\,,
~~~~~~
\resizebox{45mm}{!}{
\parbox{23mm}{
\begin{tikzpicture}[]
\node (label) at (0,0)[draw=white]{ 
       {\fd{2.75cm}{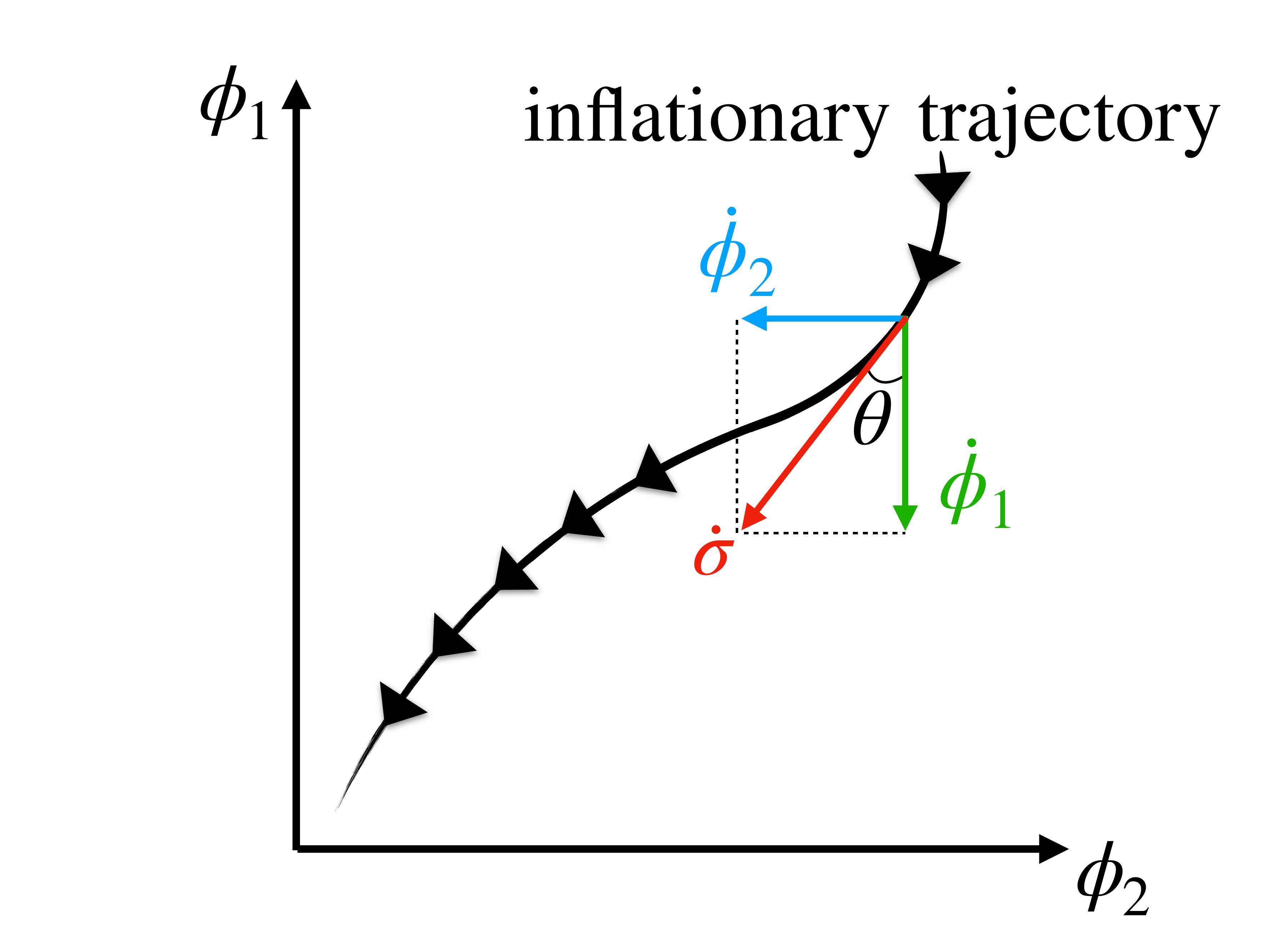}} 
      };
\end{tikzpicture}
}}\label{eq:MainTra}
\end{align}
Since there is no velocity in the transverse direction, we also have $\dot{\phi}_1\sin\theta = \dot{\phi}_2\cos\theta$;
consequently, we find $\dot{\sigma}^2 = \dot{\phi}_1^2 + \dot{\phi}_2^2$.
It should be noted that, in general, the angle $\theta$ depends on time. 
If we combine the time derivative of eq.\,(\ref{eq:MainTra}) with the equations of motion for $\phi_{1,2}$ we find
$\ddot{\sigma} + 3H\dot{\sigma} + V_{\sigma} = 0$,
where $V_{\sigma} = V_{\phi_1}\cos\theta +  V_{\phi_2}\sin\theta$.
All in all, instead of the system in eq.\,(\ref{eq:ClassicalTwoFieldSystem}), it is possible to describe the dynamics from the point of view of the 
effective inflationary trajectory by means of the equations  
\begin{align}
H^2 = \frac{1}{3}\left(
\frac{1}{2}\dot{\sigma}^2 + V
\right)\,,~~~~~~~~~\ddot{\sigma} + 3H\dot{\sigma} + V_{\sigma} = 0\,.
\end{align} 
Given the above expression for $H$, we now compute the Hubble parameters in eq.\,(\ref{eq:HubbleParameters}).
A simple computation shows that (using the number of $e$-folds as time variable)
\begin{align}\label{eq:EffectiveHubble}
\epsilon = \frac{1}{2}\left(\frac{d\sigma}{dN}\right)^2\,,~~~~~~~~~~~\eta = 3 + \frac{V_{\sigma}[6-(d\sigma/dN)^2]}{
2V(d\sigma/dN)
}\,.
\end{align}
We rewrite these equations in the form
\begin{align}
d\sigma = \pm \sqrt{2\epsilon}\,dN\,,~~~~~~~~~~\frac{dV}{V} = \left[\frac{2\epsilon(\eta - 3)}{3-\epsilon}\right]dN\,.
\end{align}
We note that these equations retain precisely the same form compared to eq.\,(\ref{eq:recPot1}).
This means that the reverse engineering approach can be equally well applied to 
the case in which $\phi$ represents the effective inflationary trajectory $\sigma$ of a multi-field model.
In the latter case, the reconstructed potential will be the potential felt by the effective inflationary trajectory.
From this perspective, it is therefore plausible that the reconstructed potential features, as function of 
the effective trajectory, a non-trivial profile like 
the one found in fig.\,\ref{fig:PotentialFit} since inflation could be mostly driven in the first stage by one of the two fields (with potential, say, $V(\phi_1)$) and during a subsequent phase by the other (with potential, say, $V(\phi_2)$).

\subsubsection{Two-field models: perturbations}

The previous discussion was purely classical. 
However, 
the reverse engineering approach requires the solution of the MS equation on the reconstructed potential.
What is the analogue of this part of the analysis in the case of a two-field model? 
To answer this question, we introduce adiabatic ($\delta\sigma$) and entropy ($\delta s$) perturbations 
\begin{align}\label{eq:PerturbationsTwoFields}
\delta\sigma = \delta\phi_1 \cos\theta + \delta\phi_2 \sin\theta\,,~~~~~~~~
\delta s = \delta\phi_2 \cos\theta - \delta\phi_1 \sin\theta\,.
\end{align}
The total comoving curvature perturbation $\mathcal{R}$ takes the form\,\cite{Malik:2004tf}
\begin{align}
\mathcal{R} = \psi - HV = \psi + \frac{H}{(\dot{\phi}_1^2 + \dot{\phi}_2^2)}
\left(
\dot{\phi}_1 \delta\phi_1 + \dot{\phi}_2 \delta\phi_2
\right) = 
\psi + \frac{H}{\dot{\sigma}^2}
\left(
\dot{\sigma}\cos\theta \delta\phi_1 + \dot{\sigma}\sin\theta \delta\phi_2
\right) = 
\psi + \frac{H}{\dot{\sigma}}\delta\sigma\,,
\end{align}
where $\psi$ is the gauge-dependent curvature perturbation and $V$ the total velocity perturbation\,\cite{Malik:2004tf}.  
We note that 
the expression for $\mathcal{R}$, written in terms of the field $\sigma$,
 is identical to that for a single field. 
We now assume the absence of entropy perturbations, $\delta s = 0$.  
From eq.\,(\ref{eq:PerturbationsTwoFields}), it follows that $\delta\phi_2 \cos\theta = \delta\phi_1 \sin\theta$; 
combined with the classical relation $\dot{\phi}_1\sin\theta = \dot{\phi}_2\cos\theta$, it gives the condition 
$\delta\phi_1/\dot{\phi}_1 = \delta\phi_2/\dot{\phi}_2$.
Under this assumption, it is possible to show that 
the equation governing the evolution of adiabatic perturbation is 
the same as that in the single field inflation\,\cite{Malik:2004tf}; in 
Fourier space, it reads
\begin{align}
u_k^{\prime\prime} + \left(
k^2 - \frac{z^{\prime\prime}}{z} 
\right)u_k = 0\,,
\end{align}
where $u_k =-zR_k$, $z=a\dot{\sigma}/H$ and $^{\prime}$ indicates derivative with respect to the conformal time. 
Using the number of $e$-folds as time variable, 
the previous equation takes precisely the same form of the MS equation in eq.\,(\ref{eq:M-S}) but with 
the Hubble parameters $\epsilon$ and $\eta$ given in terms of the effective inflationary trajectory 
as in eq.\,(\ref{eq:EffectiveHubble}). 
The power spectrum of adiabatic perturbation is again given by eq.\,(\ref{eq:PS}).

The conclusion of this brief discussion is the following.
The reverse engineering approach implemented in
the context of single-field inflationary models could be also applied 
in the case of two-field models under the assumption of negligible entropy perturbations. 
The key difference is that the role of the inflaton field $\phi$ is played by 
the effective inflationary trajectory $\sigma$. In such a case,
the reconstructed potential corresponds to the potential along the trajectory $\sigma$; the latter 
could be the result of a non-trivial combination of different potentials along different directions in field space 
as possibly suggested by our numerical result shown in fig.\,\ref{fig:PotentialFit}.

It should be noted that, in general,
entropy perturbations are non-zero and act as an additional source term
in the equation of motion for the adiabatic field perturbation\,\cite{Malik:2004tf}. 
However, there are cases in which their dynamics decouples. 
The equation of motion of $\delta s$
is indeed characterized by an effective mass squared term 
that, if $\gg H^2$, effectively decouples 
entropy from adiabatic perturbations\,\cite{Malik:2004tf} (see also ref.\,\cite{Geller:2022nkr} for a recent discussion of PBH formation in the context of multi-field inflation with non-minimal couplings). 
Moreover, if the trajectory in field space is not strongly curved (that is, more specifically, if $\dot{\theta}^2 \ll H^2$) entropy perturbations also decouple\,\cite{Malik:2004tf}.

Needless to say, the above discussion about inflationary models that fit our numerical results is anything but comprehensive.
On the one hand, keeping the discussion at this level suits the spirit of this paper since the main point of our analysis is precisely that of moving the attention from the details of the inflationary potential to the underlying dynamics. 
On the other one, finding concrete and motivated models that reproduce the reconstructed potential 
plays an important role in our understanding of PBH formation. 
In this sense, our results could stimulate new research in this direction since 
we are not aware of consistent inflationary models that generate a plateau in the power spectrum 
like the one found in our analysis.

\section{Conclusions and Outlook}\label{sec:Conclusions}

{
In this paper, we have discussed the details of the  reverse engineering technique presented in ref.\,\cite{Franciolini:2022pav} for studying the consequences of an USR phase during the inflationary dynamics.}
This approach models the time-evolution of the Hubble parameter $\eta$, eq.\,(\ref{eq:MainEqEta}), and gives as output the power spectrum of curvature perturbation, fig.\,\ref{fig:PS}. 
Our approach makes intuitively clear all features of the power spectrum, and offers a neat connection with a number of key observables related to PBH physics. 

For the first time, we have shown that an USR dynamics consistent with CMB data may generate a raised plateau in the power spectrum of curvature perturbation that can provide a link between three observables: DM made of asteroid-mass PBHs (fig.\,\ref{fig:PBHAbundanceFull}), a detectable stochastic GW signal (fig.\,\ref{fig:GWSignal}) and an observable fraction of solar-mass mergers ascribable to PBHs (fig.\,\ref{fig:PBHAbundanceFull}).
We expect our results to foster new research on consistent inflationary models able to generate a raised plateau in the power spectrum, like the one found in our analysis, giving rise to various interconnected observational signatures of the physics of the early universe. 
{In this respect, it will be important to extend the discussion drafted in section\,\ref{sec:potential} with the goal of finding motivated scalar field potentials that give the dynamics envisaged by our reverse engineering approach.}

In general, USR dynamics is expected to produce a dip in the curvature spectrum, as the one observed in fig.\,\ref{fig:PS} around few$\times 10^3$ Mpc$^{-1}$. This dip may be a complementary probe of this scenario leaving detectable imprints in CMB $\mu$-space distortions \cite{Ozsoy:2021pws} or 21-cm signals \cite{Balaji:2022zur}. 

Furthermore, the  population of PBHs generated within our model  may give rise to detectable events both in the sub-solar range, which is a smoking-gun signature of primordial origin \cite{Franciolini:2021xbq}, and in the purported lower mass gap, predicting a dearth of events within $\approx [2.2\divisionsymbol 6]\,M_\odot$ (see e.g. \cite{LIGOScientific:2021psn,Farah:2021qom}). In particular, it may help explaining some of the special events 
already observed, such as GW190814 \cite{inprepQCD} (see also \cite{Clesse:2020ghq}). 
Within our framework, it is also possible to explain events 
in the upper mass gap, potentially produced by stellar evolution above $\approx 50 M_\odot$ \cite{1967ApJ...148..803R,Barkat:1967zz,1968Ap&SS...2...96F,Woosley:2016hmi,Farmer:2019jed}, such as GW190521 \cite{LIGOScientific:2020iuh,DeLuca:2020sae}; to this end, it is crucial to understand how to properly shape the left-side edge of the plateau in the power spectrum such as to populate the higher-mass region without violating the FIRAS bound, see ref.\,\cite{inprepQCD} for more details.
%, but with more difficulty the one in the %upper mass gap, potentially produced by %stellar evolution above $\approx 50 %M_\odot$ \cite{1967ApJ...148..803R,Barkat:%1967zz,1968Ap&SS...2...96F,Woosley:2016hmi%,Farmer:2019jed}, such as GW190521 %\cite{LIGOScientific:2020iuh,DeLuca:2020sa%e}.
PBH mergers associated to the bulk of the PBH mass distribution in the asteroidal mass range would give rise to GWs at ultra-high frequencies, which may be potentially observed at GW detectors (see ref.\,\cite{Franciolini:2022htd} and refs. therein).

\color{black}
\section{Acknowledgments} 
We thank G.~Ballesteros, V.~De~Luca, I.~Musco, P.~Pani, A.~Riotto, M.~Taoso and H.~Veerm$\ddot{\rm a}$e for discussions.
G.F. acknowledges financial support provided under the European
Union's H2020 ERC, Starting Grant agreement no.~DarkGRA--757480 and under the MIUR PRIN programme, and support from the Amaldi Research Center funded by the MIUR program ``Dipartimento di Eccellenza" (CUP:~B81I18001170001). This work was supported by the EU Horizon 2020 Research and Innovation Programme under the Marie Sklodowska-Curie Grant Agreement No. 101007855.

%%%%%%%%%%%%%%%%%%%%%%%%%%%%%%%%%%%
\bibliography{PlateauUSR}
%%%%%%%%%%%%%%%%%%%%%%%%%%%%%%%%%%%

\end{document}